\documentclass[extra,onecolumn]{gji}

\usepackage{xcolor}
\usepackage{amsmath} 
\usepackage{mathtools}
	\DeclarePairedDelimiter\norm{\lVert}{\rVert}
\usepackage{timet}
\usepackage[utf8]{inputenc}
\usepackage{amssymb,gensymb,dsfont}
\usepackage{upgreek}
\usepackage{tikz}
	\usetikzlibrary{decorations.pathreplacing}
\usepackage{epstopdf}
\usepackage{afterpage}
\usepackage{booktabs}
\usepackage{graphicx}
\usepackage[final]{changes}  
	\colorlet{Changes@Color}{red}
	\setremarkmarkup{\footnote{\textcolor{Changes@Color#1}{#2}}}  

\date{Accepted xxxx.  Received xxxx; in original form xxxx.}
\pagerange{\pageref{firstpage}--\pageref{lastpage}}
\volume{xx}
\pubyear{2018}

\title[Low latitude time-dependent core flow]{Time-dependent low latitude core flow and geomagnetic field acceleration pulses
}

\author[C.\ Kloss and C.\ C.\ Finlay]{
	Clemens Kloss$^{1}$ and Christopher C.\ Finlay$^1$\\
	\vspace*{1mm}
	$^1${DTU Space, National Space Institute, Technical University of Denmark, Centrifugevej 356, 2800 Kgs.\ Lyngby, Denmark.}
}

\newcommand{\ES}{\mathrm{E}^\mathrm{S}}  
\newcommand{\EA}{\mathrm{E}^\mathrm{A}}  
\newcommand{\er}{\mathbf{\hat{r}}}  
\newcommand{\etheta}{\boldsymbol{\hat{\uptheta}}}  
\newcommand{\ephi}{\boldsymbol{\hat{\upphi}}}  
\newcommand{\ez}{\mathbf{\hat{z}}}  
\newcommand{\e}{\mathrm{e}}  
\newcommand{\I}{\mathrm{i}}  

\providecommand*{\diff}%
{\@ifnextchar^{\DIfF}{\DIfF^{}}}
\def\DIfF^#1{%
	\mathop{\mathrm{\mathstrut d}}%
	\nolimits^{#1}\gobblespace}
\def\gobblespace{%
	\futurelet\diffarg\opspace}
\def\opspace{%
	\let\DiffSpace\!%
	\ifx\diffarg(%
	\let\DiffSpace\relax
	\else
	\ifx\diffarg[%
	\let\DiffSpace\relax
	\else
	\ifx\diffarg\{%
	\let\DiffSpace\relax
	\fi\fi\fi\DiffSpace}

\providecommand*{\pderiv}[3][]{%
	\frac{\partial^{#1}#2}%
	{\partial #3^{#1}}} 
\newcommand{\abs}[1]{\left|#1\right|}
\newcommand{\real}[1]{\Re\left[#1\right]}

\usepackage[pdftex,linktoc=all,colorlinks=true,linkcolor=blue,citecolor=blue,pagebackref=false, bookmarksopen=true]{hyperref}
\usepackage[open]{bookmark}

\begin{document}

\label{firstpage}

\maketitle

\begin{summary}
We present a new model of time-dependent flow at low latitudes in the Earth's core between 2000 and 2018, derived from magnetic field measurements made on board the {\it Swarm} and CHAMP satellites and at ground magnetic observatories.  The model, called {\it CoreFlo-LL.1}, consists of a steady background flow without imposed symmetry plus a time-dependent flow that is dominated by geostrophic and quasi-geostrophic components but also allows weak departures from equatorial symmetry. Core flow mode amplitudes are determined at 4-month intervals by a robust least squares fit to ground and satellite secular variation data.  The $l_1$ norm of the square root of geostrophic and inertial mode enstrophies, and the $l_2$ norm of the flow acceleration, are minimised during the inversion procedure.   We find that the equatorial region beneath the core-mantle boundary is a place of vigorous, localised, fluid motions;  time-dependent flow focused at low latitudes close to the core surface is able to reproduce rapid field variations observed at non-polar latitudes at and above Earth's surface.  Magnetic field acceleration pulses are produced by alternating bursts of non-zonal azimuthal flow acceleration in this region.  Such bursts are prominent in the longitudinal sectors from \mbox{80\--130$^\circ$E} and \mbox{60\--100$^\circ$W} throughout the period studied, but are also evident under the equatorial Pacific from 130$^\circ$E to 150$^\circ$W after 2012.  We find a distinctive interannual alternation in the sign of the non-zonal azimuthal flow acceleration at some locations, involving a rapid cross-over between flow acceleration convergence and divergence.  Such acceleration sign changes can occur within a year or less, and when the structures involved are of large spatial scale they can give rise to geomagnetic jerks at the Earth's surface.   For example, in 2014, we find a change in the sign of the non-zonal azimuthal flow acceleration under the equatorial Pacific, as a region of flow acceleration divergence near 130$^\circ$E changes to a region of flow acceleration convergence.  This occurs at a maximum in the amplitude of the time-varying azimuthal flow under the equatorial Pacific and corresponds to a geomagnetic jerk at the Earth's surface.
\end{summary}
   
\begin{keywords}
Core Dynamics, Magnetic Field, Secular variation, Swarm
\end{keywords}
    
\section{Introduction}
\label{sec:intro}
Time variations in the motion of liquid metal in Earth's outer core produce fluctuations in the geomagnetic field.   A striking example of this process is the occurrence of a sharp change in the field acceleration known as a geomagnetic jerk \cite[]{courtillot1984geomagnetic,brown2013jerks,mandea2010geomagnetic}. Improved satellite-data-based models in the late 2000s made it possible for the first time to reliably estimate the global secular acceleration (SA) and its time variations since 2000  \cite[]{olsen2008rapidly,lesur2010second,Chulliat2010}.  Global maps of the field acceleration at the core surface revealed the existence of distinctive  sub-decadal SA pulses, their prominence at low latitudes, their alternating sign, and their relationship with jerks observed at the Earth's surface which can be seen as by-products of successive pulses \cite[]{Chulliat2010,chulliat2014geomagnetic,torta2015evidence}.

An important step towards understanding the underlying physical mechanism producing this phenomenon of SA pulses is to determine how the core flow, and its patterns of acceleration, have altered during these events.  Here, in an effort to clarify this process, we construct a new model of the time-varying core flow spanning the past 18 years, specifically designed to focus on core flow time variations at low latitudes. 


The core flow is linked to geomagnetic field changes through the magnetic induction equation  \cite[]{roberts1965analysis,Bloxham1991,Holme2015}.  Under the approximation that magnetic diffusion plays a secondary role, core flow advects and stretches the magnetic field producing its observed changes.  This enables field variations to be inverted for the responsible core flow \cite[]{kahle1967estimated} but unfortunately this procedure is highly non-unique since flow around contours of $B_r$ produces no observable field change \cite[]{Backus1968}.  The difficulties are exacerbated by the fact that observations constrain only the largest lengthscales of the field \cite[]{roberts2000test,rau2000core,amit2007tests}.  In order to make progress, it is therefore necessary to make assumptions concerning the underlying nature of the flow \cite[]{gubbins1982finding,le1984outer,pais2008quasi,aubert2012flow}, and to account for unobserved small-scale processes \cite[]{hulot1992taking,eymin2005core,Gillet2009,Gillet2015}.

Most studies of core flow time variations have been based on spherical harmonic field models derived primarily from ground observatory data.   In a landmark study, \cite{Jackson1997} constructed time-dependent flows spanning 1840-1990 and demonstrated that time variations were needed to fit observatory data to an acceptable level.   Significant oscillatory motions were evident in these flows, especially at low latitudes (see his Fig. 3 and related discussion). These equatorial flow oscillations have attracted surprisingly little attention in the intervening twenty years.   \cite{le2000time} investigated changes in flow accelerations at the time of geomagnetic jerks, and found similar patterns but with alternating signs, for the 1969, 1979 and 1992 jerks. \cite{bloxham2002origin} argued that time variations in only the equatorially-symmetric, zonal, toroidal, flow components were sufficient to account for geomagnetic jerks.  More recently, using observatory monthly mean data processed to highlight the core field signal, \cite{Whaler2016} showed that such simple time-dependent flows are unable to adequately fit the observations, finding that temporal variability of the non-zonal azimuthal flow was required at low latitudes.

A major observational advance in the past two decades has been the ability to monitor geomagnetic secular variation and secular acceleration from space.   This has enabled studies of core flow time variations based on global data constraints. Building on initial studies of the high resolution flow at a single epoch \cite[e.g.][]{holme2006core},  \cite{olsen2008rapidly} were the first to find evidence for sub-decadal bursts of flow acceleration. In 2003 they inferred a strong equatorward flow acceleration under Asia accompanied by westward acceleration at low latitudes under the Indian ocean.   Also considering flow changes around 2003, \cite{wardinski20082003} found time variations of both the equatorially-symmetric, zonal, toroidal flow coefficients, and that non-zonal toroidal flows with azimuthal wavenumber $m=1$ and $4$ were required. Similarly, \cite{silva2012investigating} concluded that while changes in flow acceleration around 2003 could be purely equatorially-symmetric, they could not be only zonal and toroidal, inferring a change in the sign of non-zonal flow accelerations under India and Indonesia.  Lesur and colleagues \cite[]{lesur2010modelling,wardinski2012extended,lesur2015geomagnetic}  have pursued an approach that involves the co-estimation of field and flow models from satellite data while \cite{Baerenzung2014,Baerenzung2016, baerenzung2017modeling} have developed probabilistic flow inversion methods.  Taken together, these studies suggest that in the absence of strong physical constraints, primarily toroidal flows, with only a small amount of poloidal flow, and a comparatively weak time-dependence, are adequate to explain the observations, a point made early on by \cite{whaler1986geomagnetic}. 

The use of appropriate prior information is essential in mitigating flow non-uniqueness and in making the most of the information provided by geomagnetic observations.  The most important factor controlling the form of fluid motions in Earth's core is its rapid rotation; this results in a leading order geostrophic force balance between Coriolis forces and pressure gradients \cite[e.g.][]{aubert2017spherical}.  \cite{Hide1966free} pointed out that slow flow perturbations about a state of geostrophic balance are expected to take a columnar form, varying only slowly parallel to the rotation axis.  Such motions are commonly referred to as quasi-geostrophic (QG). \cite{busse1970thermal} showed that convection in a rotating spherical shell takes this form.  \cite{jault2008axial} showed that transient flow disturbances are expected to take a QG form even in the presence of strong magnetic fields, provided the inertial wave speed (associated with rotation) is much shorter than the Alfv\'{e}n wave speed.  Further tests using 3D simulations, with strong imposed magnetic fields, were performed by  \cite{gillet2012rationale} who found that QG motions indeed dominate on large spatial scales and short time scales. \cite{schaeffer2011symmetry} further noted that although equatorially symmetric QG flows were dominant with equatorially symmetric forcing, in the presence of equatorially anti-symmetric forcing,  cross-equatorial flow could occur even in the rapidly-rotating regime.  In the past few years, it has become possible to carry out 3D dynamo simulations that are beginning to approach the correct rapidly-rotating regime; \cite{schaeffer2017turbulent} found large-scale fluid motions outside the tangent cylinder remained primarily columnar, while \cite{aubert2018geomagnetic} showed that not only the flow, but also the flow acceleration remained primarily columnar in the rapidly-rotating regime, provided the Alfv\'{e}n wave timescale governing magnetic disturbances was much longer than the timescale of rotation.  \cite{bardsley2016} have also clarified how columnar motions can be built by propagating energy rapidly along the rotation axis at the group velocity of inertial-Alfv\'{e}n waves, in the case that the disturbance wavevector is perpendicular to the rotation axis. \cite{aubert2018geomagnetic} has highlighted that once formed, such QG disturbances can result in non-zonal Alfv\'{e}n waves, due to the heterogeneous nature of a dynamo-generated magnetic field.  Taken together, these studies point to a crucial role for QG motions in time-varying core flow.

Many studies have attempted to incorporate constraints based on rapid rotation into the determination of core flow. \cite{le1984outer} suggested constraining motions to be tangentially geostrophic at the core surface, such that horizontal pressure gradients and the Coriolis force are in balance.  This approach has been since exploited by many workers \cite[e.g.][]{Jackson1997,holme2006core}.  \cite{jault1988westward} and \cite{jackson1993time} went further and extended the zonal, equatorially-symmetric part of such flows throughout the depth of the spherical shell, showing that the resulting motions were in good agreement with observed changes in the length-of-day.  \cite{pais2008quasi} pointed out that QG flows should in addition be equatorially-symmetric.   Enforcing this condition approach enabled them to highlight the presence of an eccentric and planetary scale  anti-cyclonic gyre, consisting of westward flow at mide-to-low latitudes in the Atlantic hemisphere and at high latitudes under the Pacific hemisphere, that has now been broadly supported by many subsequent studies.  \cite{Gillet2009} derived time-dependent QG flows for the periods 1960-2002 and 1997-2008, using an ensemble method to take into account the impact of unresolved scales.  They found the time-varying part of their flow, that involved prominent interannual variations, was dominated by non-zonal flow structures.   \cite{gillet2012rationale} constructed QG and tangentially geostrophic flows spanning 1840-2008 and found that their tangentially geostrophic flows become increasingly equatorially symmetric (i.e. increasingly QG) as the quality of the observational data improved, concluding that QG flows were able to account well for the best quality recent data. \cite{pais2014variability} have studied the mean flow and major empirical modes of flow temporal variability on decade to century timescales within a QG framework.  On top of the planetary scale eccentric gyre, they found that a small number of empirical modes involving large-scale vortices, and zonal motions coupled between low and high latitudes can account for 90\% of the variance in the observed SV, and for decadal changes in the length-of-day.  \cite{Gillet2015} constructed QG flows from the COV-OBS field model, accounting for the first time for time-correlated errors due to unresolved scales, finding non-zonal, time-dependent structures in the azimuthal flow within $10^\circ$ of the equator.  Describing these features as non-zonal equatorial jets, and finding they possessed significant power in periods between 4 and 9.5 years, it was argued that they may be responsible for the low latitude SA pulses.  \cite{Finlay2016} obtained similar results using the same method to invert the higher resolution CHAOS-6 geomagnetic field model spanning 1999-2016. \cite{schaeffer2011symmetry} studied the effects of allowing more power in zonal flows (i.e. anisotropy) and departures from equatorial symmetry, within the framework of predominantly QG flows.  They found a preference for small-scale flow to cross the equator under Indonesia but concluded that the time-dependent large-scale core flow was probably dominated by QG motions on decadal timescales.    \cite{amit2013differences} have investigated differences between tangentially geostrophic and QG (columnar) flows and in all cases found intense patterns of upwelling and downwelling at low latitudes, particularly under the Indian ocean.    \cite{barrois2017contributions, Barrois2018a} recently used a stochastic ensemble Kalman filter algorithm to derive core flows consistent with prior information from a numerical dynamo simulation; the prior flow in this case was predominantly equatorially symmetric but not strictly QG. The resulting flows were again dominated by the planetary scale eccentric gyre, and involved prominent upwelling and downwelling structures in the equatorially region.  In a follow up study  \cite{Barrois2018, Barrois2018a}, applying the same technique to ground and satellite SV data from the past twenty years, found low latitude flow accelerations on interannual time scales, notably under the Pacific.  Despite much progress, an observation-based study of rapid core flow dynamics at low latitudes, taking into account the effects of rapid rotation and using the latest ground and satellite data, with a focus on clarifying the mechanism of SA pulses and geomagnetic jerks, has been lacking. This is the aim of the present study.

We adopt an alternative approach to describing flows expected in a rapidly-rotating spherical geometry.  Solutions to the Navier-Stokes equation in an incompressible fluid sphere, subject to a non-penetration boundary condition, can in general be decomposed into a geostrophic mode plus inertial modes  \cite[]{Zhang2017}.  Analytical solutions for these are available in a full sphere geometry \cite[]{zhang2001inertial,Zhang2017}.  A special class of slow, equatorially-symmetric, inertial modes, that we refer to as QG modes  \cite[]{zhang2001inertial,busse2005,maffei2017characterization}, have been shown to efficiently describe rotating flow in a sphere at the onset of convection \cite[]{zhang2004new,zhang2007asymptotic}, and when combined with the geostrophic mode can also describe weakly-nonlinear convection \cite[]{zhang2004new,liao2012}.  More generally, geostrophic and inertial modes can be used to describe the transient response of rotating spherical systems to a forcing \cite[]{liao2010} and in principle they provide a complete basis for representing flows in a rotating sphere  \cite[]{cui2014completeness,ivers2015enumeration,backus2017completeness}. Of particular interest here is that they are well-suited to  describing motions at low latitudes.  They do not suffer from a formal requirement for the boundary slope to be small as is the case in classical QG theory \cite[e.g.][]{busse1970thermal,pais2008quasi,schaeffer2006quasi}, and they are able to efficiently describe equatorially-trapped inertial wave motions \cite[]{zhang1993equatorially}.

We use a combination of inertial and geostrophic modes to parameterize the core flow.  In particular, our time-dependent flow consists primarily of QG (i.e. almost axially invariant, equatorially-symmetric) inertial modes and a geostrophic mode.  Our hypothesis is that such modes provide a concise description of the spatial structure of core flow.  We therefore minimize the $l_1$ norm of the square-root of the mode enstrophies; this favours a sparse distribution of amplitude amongst the modes, while also penalizing small scales to ensure spectral convergence. In addition, we seek solutions for which the flow acceleration is minimized.  Further details of our parameterization of the flow and the model estimation scheme are given in Section \ref{sec:model_formulation}.  We use the frozen-flux induction equation,  together with a potential field formalism, to link the flow to secular variation time series collected at ground observatories and at satellite altitude, via virtual observatories \cite[]{mandea2006new,olsen2007investigation}.  Details of the data selection and processing are given in section \ref{sec:data}.  Note that this enables us to avoid using spherical harmonic field model based descriptions of the field secular variation (SV) and secular acceleration (SA) that depend heavily on the chosen temporal regularization or window length  for time-averaging, especially above spherical harmonic degree 9.   We account for the influence of unresolved scales of the magnetic field via an augmented state approach \cite[]{Gillet2015,barrois2017contributions} where the small scales are assumed to be time-correlated \cite[]{Gillet2015} and their impact is calculated iteratively using the induction equation.  In section \ref{sec:results} we present a detailed description of our new reference flow model, {\it CoreFlo-LL.1}, including its fit to the data,  time-average structure, and time-dependence.  We find a steady background flow dominated by an eccentric planetary gyre similar to that found in previous studies, and a time-varying flow that is strongly localised in the equatorial region and consists of vigorous non-zonal azimuthal flows.  This flow explains well observed SV and SA at mid and low latitudes. In section \ref{sec:disc} we discuss the origin of SA pulses, which we find are due to intense, alternating, bursts of non-zonal azimuthal flow acceleration.  We find geomagnetic jerks which occur when there is an abrupt sign change or cross-over in the azimuthal flow acceleration of sufficiently large scale.  We go on to investigate variants of our flow model derived using different assumptions for the spatial regularization norm, and regarding the equatorial symmetry. Connections are made to previous studies and implications for the dynamics of the core discussed, before we finish with concluding remarks in section \ref{sec:conc}.

\vfill 

\section{Data}
\label{sec:data}

\subsection{Ground observatory series}
\label{sec:dataGO}
Ground magnetic observatories provide high quality vector observations of the geomagnetic field at a global network of locations on Earth's surface \cite[e.g.][]{Matzka2010}.  We make use of time series of revised monthly means \cite[]{Olsen2014}, based on quality controlled  \cite[]{Macmillan2013} hourly mean magnetic field observations from the BGS database\footnote{{\tt ftp://ftp.nerc-murchison.ac.uk/geomag/Swarm/AUX\_OBS}}, version 0113, that was built from INTERMAGNET and WDC Edinburgh data as available in January 2018. We use data spanning the interval 2000-2018.  Revised monthly means are aimed to isolate as far as possible the core field signal, and are constructed by removing from the hourly mean values (i) estimates of the large-scale magnetospheric field, taken from the CHAOS-6-x7 field model \cite[]{Finlay2016} (ii) estimates of the ionospheric Sq field, and its associated induced field (taken from the CM4 model \cite[]{sabaka2004extending} driven by the latest F10.7 radio flux series), followed by taking a robust mean of the hourly values from all local times, using Huber weights to account for a long-tailed distribution of data.  The resulting revised monthly means were then averaged over 4-month windows (to match the time sampling of the satellite-based virtual observatories described below) and annual differences taken to obtain times series of secular variation (SV) at each observatory location.  

\begin{figure*}
	\centering
	\includegraphics[width=16cm]{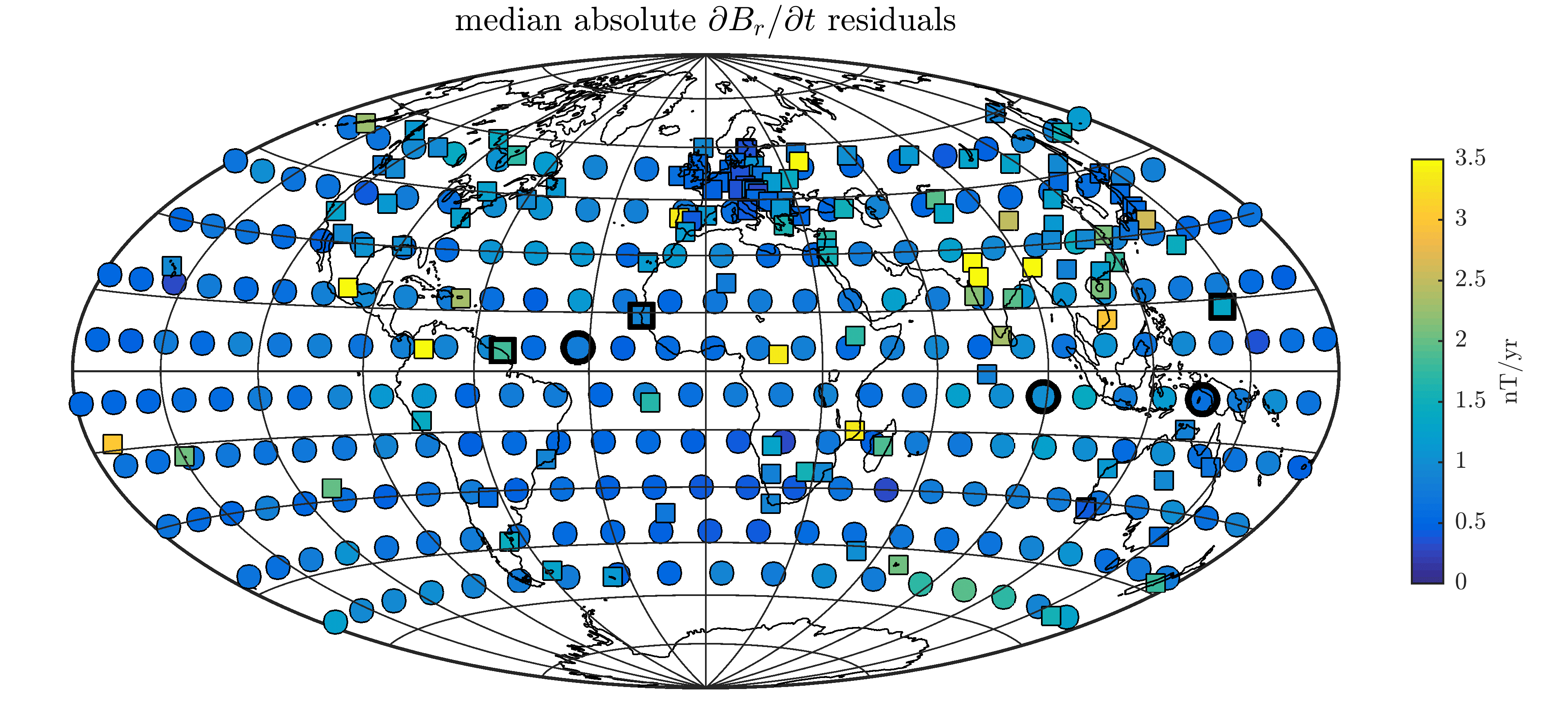}
	\includegraphics[width=16cm]{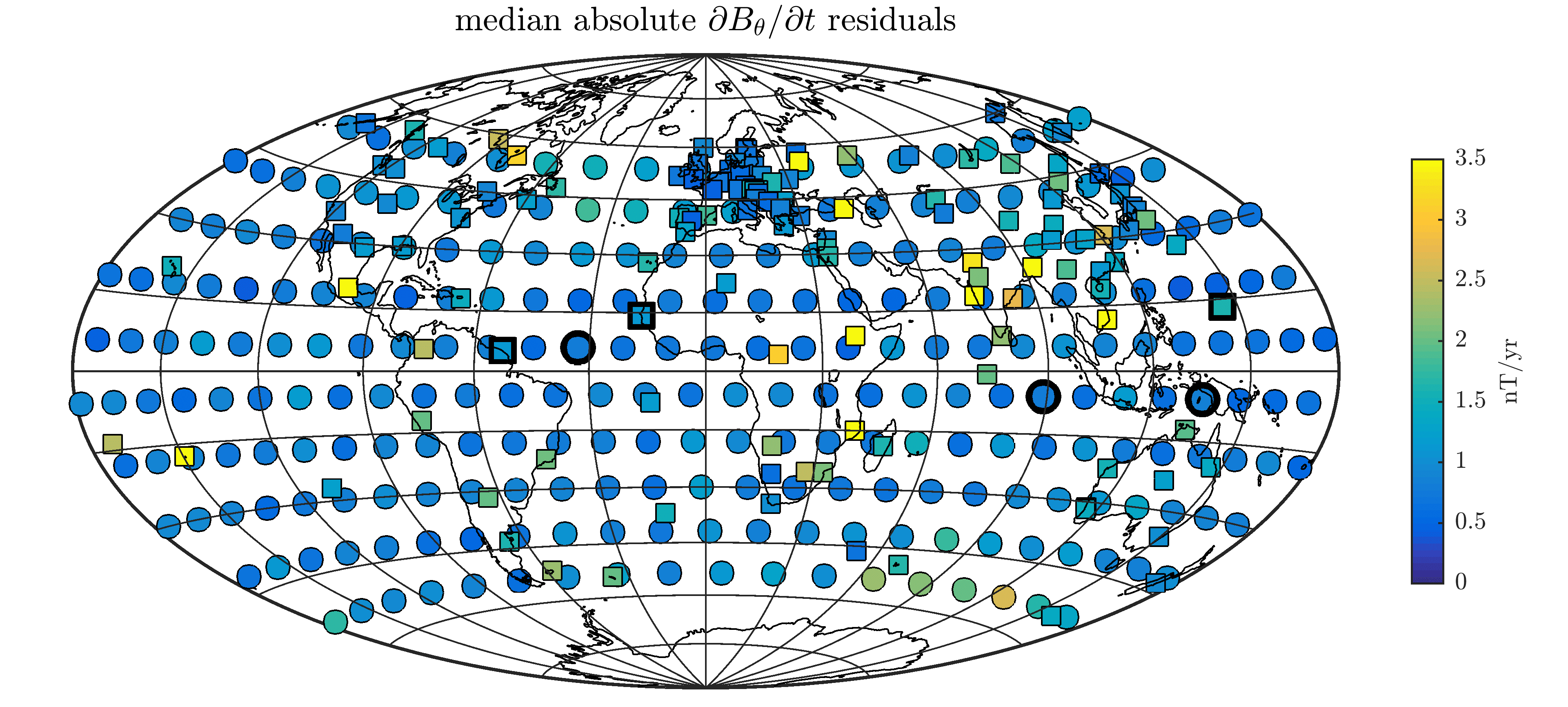}
	\includegraphics[width=16cm]{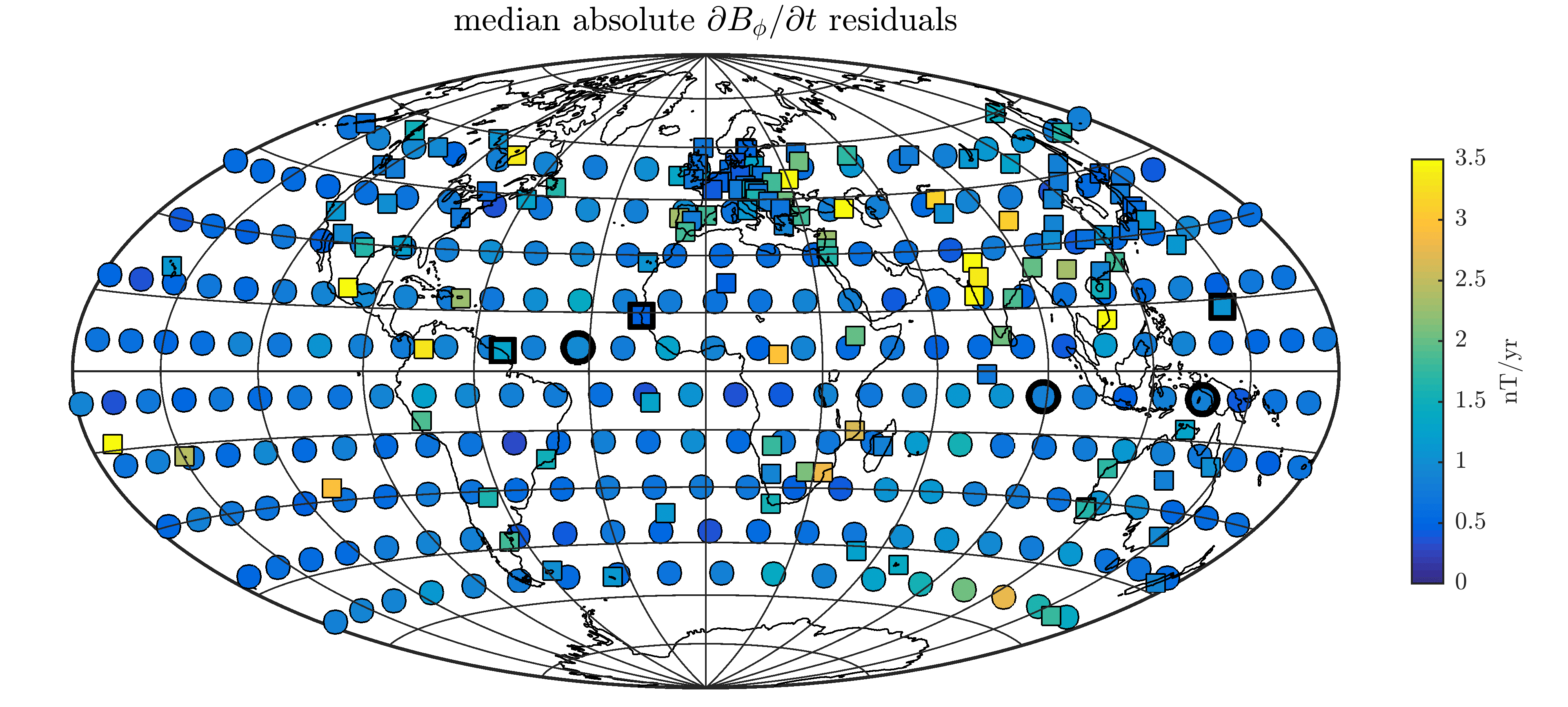}
	\caption{Maps of data locations showing the median absolute residuals from 2000 to 2018 for the three SV components at the virtual (circles) and ground (squares) observatories. Thick markers indicate locations of data shown in Fig.~\ref{fig:timeseries_go_vo}.}
	\label{fig:median_residuals}
\end{figure*}

\begin{figure*}
	\centering
	\includegraphics[width=16cm]{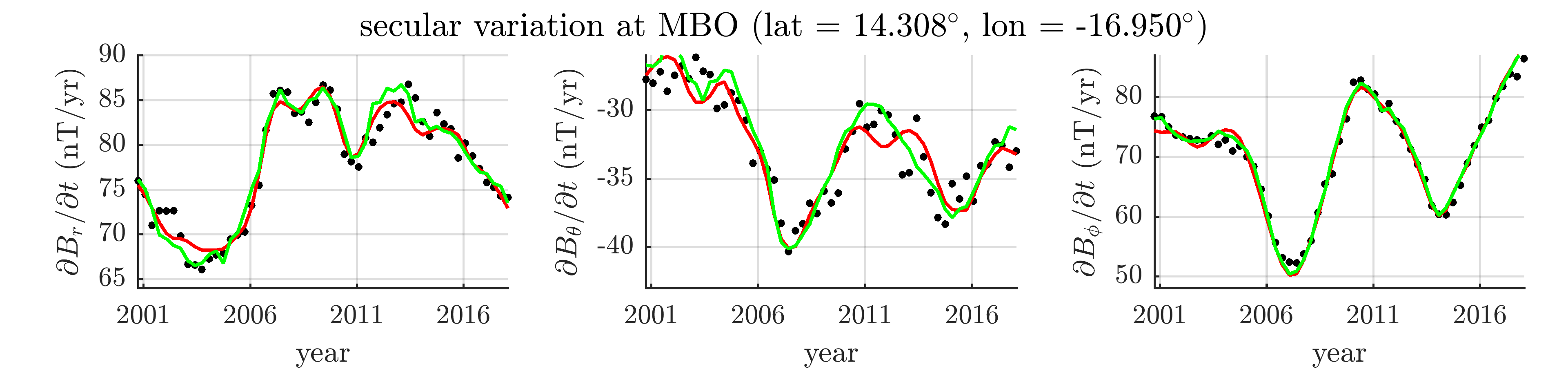}\\[-0.4cm]
	\includegraphics[width=16cm]{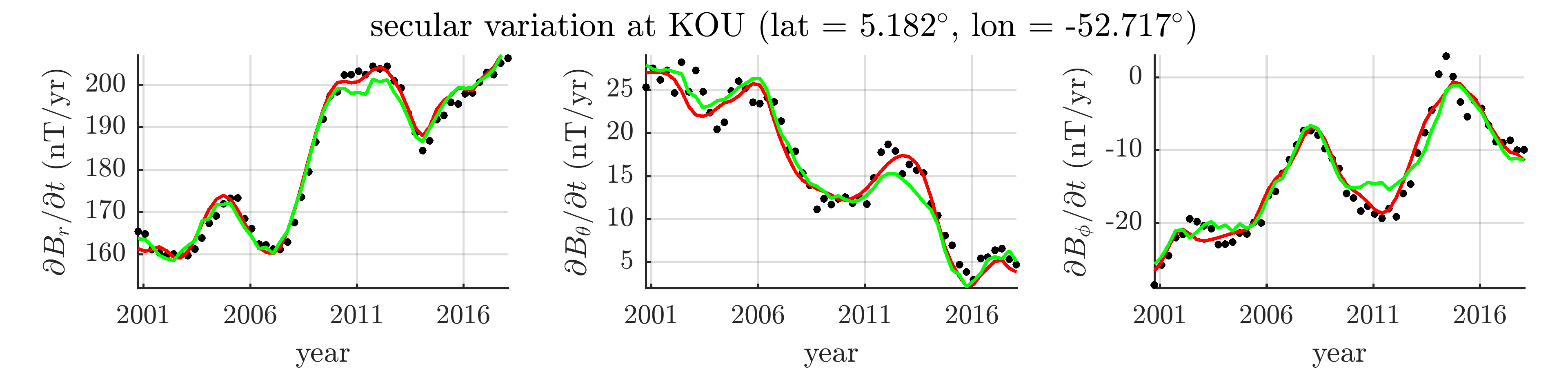}\\[-0.4cm]
	\includegraphics[width=16cm]{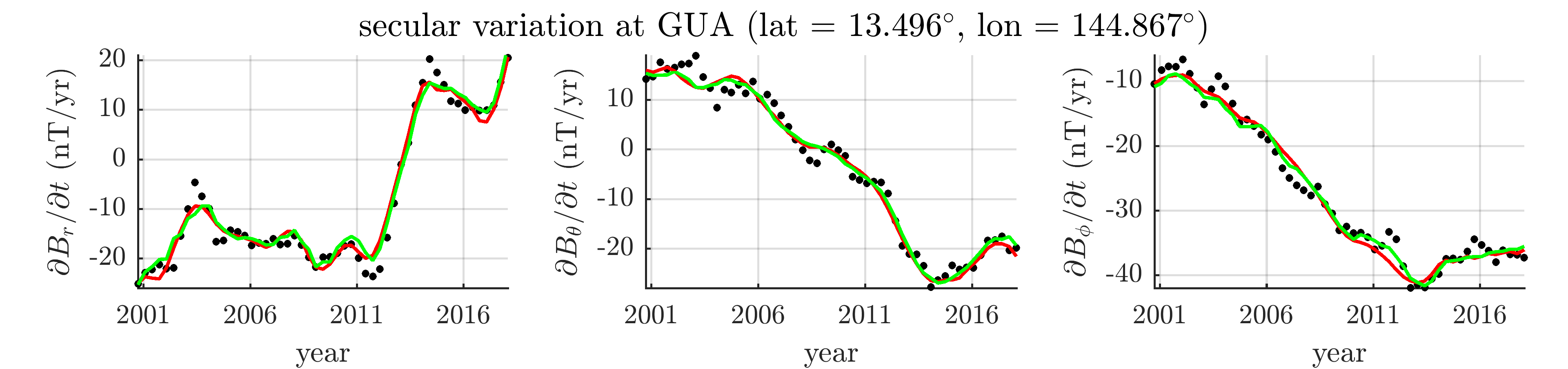}\\[-0.4cm]
	\includegraphics[width=16cm]{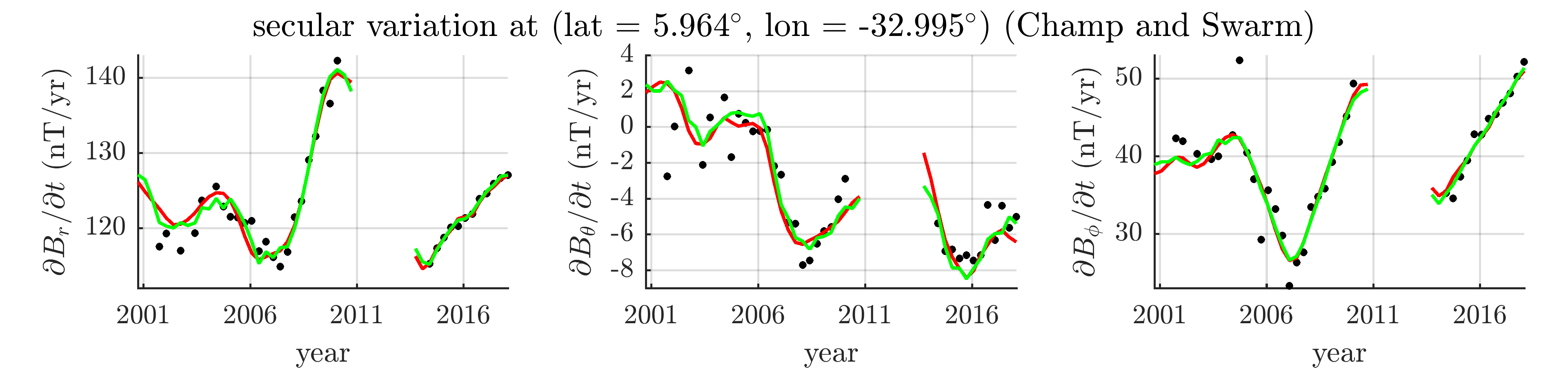}\\[-0.4cm]
	\includegraphics[width=16cm]{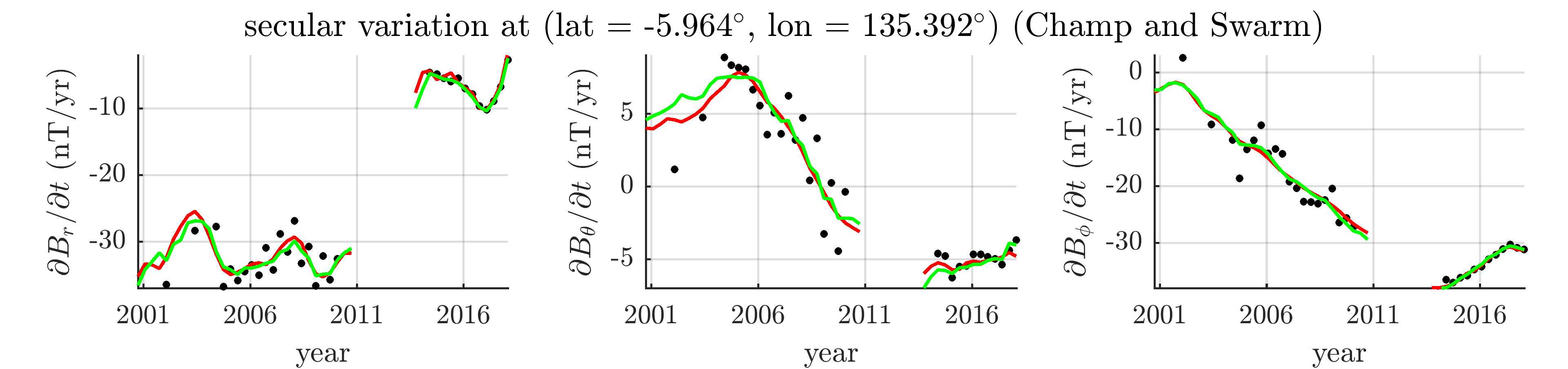}\\[-0.4cm]
	\includegraphics[width=16cm]{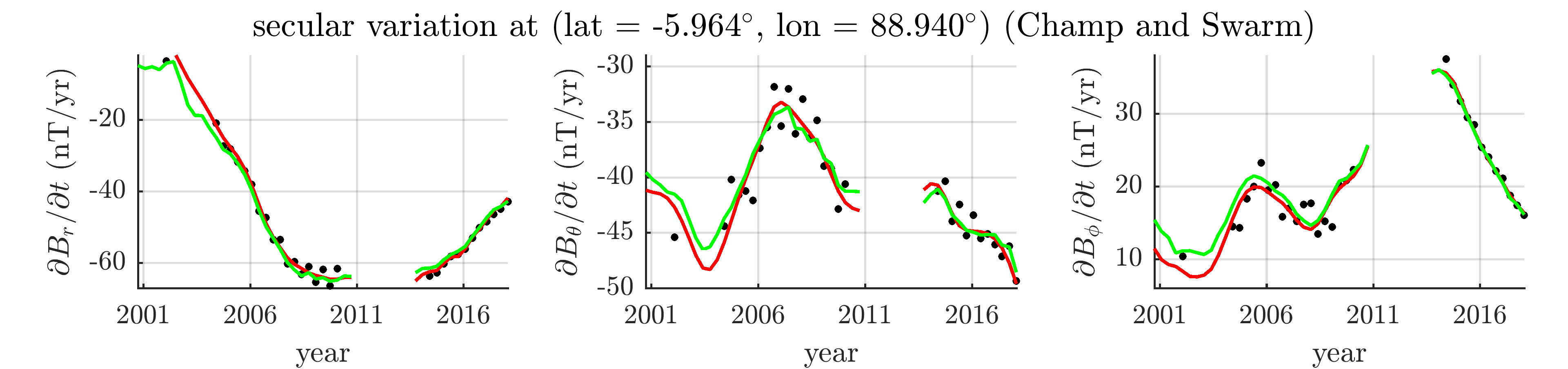}
	\caption{SV time series for the ground observatories in GUA, MBO and KOU, and three virtual observatories, all showing the SV data (black), the CHAOS-6-x7 SV prediction (red), and the SV prediction from the {\it CoreFlo-LL.1} model (green). Note that there is data gap in the time series of the virtual observatories due to lack of satellite data during that time.}
	\label{fig:timeseries_go_vo}
\end{figure*}

In this study, we choose to use only data collected at non-polar latitudes, equatorward of $\pm$60$^\circ$ geographic latitude, since we wish to focus on low latitude flow dynamics and to avoid the complexities associated with magnetosphere-ionosphere coupling currents and related ionospheric currents in the polar regions.   Our final dataset then consists of data from 146 ground observatories; their locations are marked by squares in Fig.~\ref{fig:median_residuals}.  Examples of SV time series from low latitude locations (MBO \-- M'Bour in Senegal, West Africa; KOU \-- Kourou in French Guinea, South America; and GUA \-- Guam, USA, Western Pacific) are shown by the black dots in  Fig.~\ref{fig:timeseries_go_vo}.   

Estimates of the data uncertainty for each component of each observatory were derived as follows.  We calculated a three-by-three covariance matrix  for each observatory from the revised monthly mean time series of the three components, after removing the predictions of the CHAOS-6-x7 field model and de-trending.  The uncertainty estimates were taken to be the square root of the diagonal elements of these covariance matrices.  A similar procedure has been used to determine error estimates for observatory data in recent revisions of the CHAOS model, and in the studies of core flow by \cite{Whaler2016} and \cite{Barrois2018} that also used ground observatory monthly means.


\subsection{Satellite-based virtual observatory series}
\label{sec:dataVO}

Low-Earth-orbit satellites, with dedicated systems for measuring the vector components of the geomagnetic field with absolute accuracy, provide global information on the geomagnetic field within a few weeks.  However, since the position of the satellites are constantly changing, it is difficult to directly assess secular variation from the measurements.  Here, we follow \cite{mandea2006new} and \cite{olsen2007investigation} and derive 'virtual observatory' field estimates at a grid of fixed locations, based on measurements made on board the CHAMP and {\it Swarm} missions within 4-month time windows. Use of 4-month time windows helps reduce local time biases that are found when 1-month windows are used as in the originally proposed version of virtual observatories. Our virtual observatory time series were derived using a modified version of the original procedure \cite[for full details see][]{Barrois2018}.  Here we give only a brief summary of the key steps.
 
We used magnetic field measurements collected by the CHAMP vector field magnetometer between July 2000 and September 2010 and by the \textit{Swarm} vector field magnetometers, on board all three satellites (\textit{Alpha}, \textit{Bravo}, \textit{Charlie}), between January 2014 and July 2018.  Starting from the CHAMP MAG-L3 and \textit{Swarm} Level 1b MAG-L, version 0501, data products, we sub-sampled the data at 15s intervals using the data in the vector field magnetometer (VFM) frame and rotating them into an Earth-Centered Earth-Fixed (ECEF) coordinate frame using the Euler rotation angles estimated in the CHAOS-6-x7  field model.  

We then selected data from dark regions during geomagnetically quiet times according to the criteria  (i) the sun was a maximum of $10^{\circ}$ above the horizon; (ii) geomagnetic activity index \mbox{$\mathrm{K_p}<3^{\mathrm{o}}$}; (iii) the $RC$ disturbance index \cite[]{Olsen2014} had \mbox{$\vert \mathrm{d}\mathrm{RC}/\mathrm{d}t \vert < 3$ nT/h}; (iv) merging electric field at the magnetopause, \mbox{$\mathrm{E_m}\leq 0.8$ mV/m} with $E_m=0.33 v^{4/3} B_t^{2/3} \mathrm{sin}(\vert \Theta \vert/2)$, $v$ the solar wind speed, $\Theta=\mathrm{arctan}(B_y/B_z)$ and $B_t=\sqrt{B_y^2+B_z^2}$ with $B_y$ and $B_z$ the components of the interplanetary magnetic field (IMF), calculated using 2 hourly means of 1-min values of the IMF and solar wind extracted from the OMNI database\footnote{\tt http://omniweb.gsfc.nasa.gov}; (v) IMF \mbox{$B_z>0$ nT} and IMF \mbox{$ \vert B_y \vert<10$ nT}, again based on 2 hourly mean of 1 minute values. Next, estimates of the magnetospheric and its related induced fields, as given by the CHAOS-6-x7 model, the ionospheric and its induced fields, as given by the CM4 model, as well as the static internal field for spherical harmonic degrees \mbox{$n>20$} given by the CHAOS-6 model were removed from data. 

Robust inversions for a static local potential were then carried out in 4-month windows, using all data within $700\,$km of a specified target point.  As a pre-processing step and in order to aid the determination of robust data weights, the time-dependent internal field from CHAOS-6-x7 (for spherical harmonic degrees 1 to 20) was first removed from the data, then a robust least squares inversion for the parameters of the local potential (up to cubic terms) was carried out.  \cite{Barrois2018} present the relevant expressions for the potential.  The three components of the field at the target point was then calculated from the potential, adding this back on to the prediction of CHAOS-6-x7 at the target location.  Note this does not prevent our 4-monthly VO series from departing from CHAOS-6-x7; each estimate is free to depart from the CHAOS model as much as it is required by the data.    

Rather than working directly with the satellite vector field data, we use sums and differences of the vector data, along track (using 15 second differences) and in the east-west direction (from \textit{Swarm} \textit{Alpha} and \textit{Charlie}, constructed by searching for data close in time with smallest differences in latitude).  Use of sum and difference data has been found to be advantageous for studying the internal field \citep{sabaka2015cm5,olsen2015swarm}.  In inverting for the potential we use sums and differences of the matrices linking the vector data at each location to the potential, so no error is made due to the small change in the local coordinate system between the points being differenced.

Times series of such 4-monthly virtual observatory estimates were constructed on an approximately equal area grid of 300 locations, each separated by approximately 12$^\circ$, using a recursive zonal equal area partitioning algorithm \cite[]{leopardi2006}.  The virtual observatories derived from CHAMP data were constructed at $300\,$km altitude, those derived from Swarm data at $500\,$km altitude. This means, on average and only considering the increase in altitude, that the magnitude of the SV for {\it Swarm} is $7\,$nT/yr smaller than for CHAMP. We again chose to use only data series from latitudes equatorward of $\pm$60$^\circ$, so we finally retained data from 258 virtual observatory locations at low and mid latitudes marked using the circles in Fig.~\ref{fig:median_residuals}. Secular variation estimates were computed by taking annual differences of the 4-monthly mean virtual observatory estimates.  Examples of virtual observatory times series at low latitudes are shown in the lower half of  Fig.~\ref{fig:timeseries_go_vo}.  Note the gap in the series between 2010, when the CHAMP mission ended, and 2014 when data was available from {\it Swarm}.

Data uncertainty estimates for the virtual observatory series were derived by a procedure very similar to that used for the ground observatories. For each location, covariances were calculated between the time series of the three components (after removing from each series the predictions of the CHAOS-6-x7 model and de-trending), in order to obtain a three-by-three covariance matrix. A robust procedure for calculating the covariances was employed and the square-root of the diagonal elements of the covariance matrices were taken to be the uncertainty estimates for each series.


\section{Model formulation}
\label{sec:model_formulation}

\subsection{Flow in a rapidly rotating sphere}

In this section, we present details of our parameterization of the core flow.  Our fundamental assumptions are (i) that rotation dominates core dynamics, and (ii) that the core geometry may be approximated by a sphere.  The theory presented below follows closely the notation of \cite{Zhang2017} where more details of the mode-based approach to studying flow in a rotating fluid can be found.  We begin with the Navier-Stokes equations describing fluid motions in an incompressible electrically conducting fluid, uniformly rotating with angular frequency $\mathbf{\Omega}$  \cite[e.g.][]{Gubbins1987magnetohydrodynamics}
\begin{equation}
\label{eq:NavierStokes}
	\begin{aligned}
		\pderiv{\mathbf{u}}{t} + \mathbf{u}\cdot\nabla\mathbf{u} +2\mathbf{\Omega}\times\mathbf{u}&=-\frac{1}{\rho_0}\nabla P+ \mathbf{F}\\
		\nabla\cdot\mathbf{u}&=0,
	\end{aligned}
\end{equation}
where $\mathbf{F}= (\nu\nabla^2\mathbf{u} +\rho'\mathbf{g}+\frac{1}{\rho_0}\mathbf{J}\times\mathbf{B}$) is a forcing/dissipation term, $\rho'$ is the relative departure from the reference density $\rho_0$, $\mathbf{g}$ is the acceleration due to gravity, $\nu$ is the coefficient of kinematic viscosity, $\mathbf{J}$ is the current density and $\mathbf{B}$ is the magnetic field.  $P$ is the departure from the pressure of the reference state which contains both the hydrostatic pressure and the contribution of the centrifugal force.

In order to obtain fundamental solutions representing the basic structure of rotation-dominated flows, we set the forcing and dissipation term $\mathbf{F}=0$, and further neglect the non-linear term $\mathbf{u}\cdot\nabla\mathbf{u}$ by assuming that the flow departs only slightly from rigid body rotation.   For a homogeneous and inviscid fluid confined to a spherical container of radius $r_0$ that is uniformly rotating with $\mathbf{\Omega}=\Omega\ez$, where $\ez$ is the unit vector in the z-direction, and using the inverse of the angular frequency $\Omega^{-1}$ and the radius $r_0$ as reference timescales and lengthscales,  Eq.~\eqref{eq:NavierStokes} reduces in non-dimensional form to
\begin{equation}
\label{eq:CoriolisOperator}
	\begin{aligned}
		\pderiv{\mathbf{u}}{t} +2\mathbf{\hat{z}}\times\mathbf{u}&=-\frac{1}{\rho_0}\nabla P\\
		\nabla\cdot\mathbf{u}&=0,
	\end{aligned}
\end{equation}
while the required boundary condition is that the flow normal to the boundary (in direction $\er$) vanishes at the spherical container at $r=1$,
\begin{equation}
\label{eq:BoundaryCondition}
	\er\cdot\mathbf{u}=0.
\end{equation}

Eqns.~\eqref{eq:CoriolisOperator} and \eqref{eq:BoundaryCondition} together define a boundary value problem with solutions consisting of a single geostrophic mode, corresponding to a steady flow, and an infinite number of time-dependent inertial modes \cite[][]{Zhang2017}. The total flow can be then conveniently expressed as a linear combination of these modes.  Here, we briefly present the key parameters of these modes and refer the reader to \cite{Zhang2017} for a complete description.

\subsubsection{Geostrophic mode}

The geostrophic mode is the solution for a steady flow in a rotating incompressible fluid, that is moving slowly with respect to rigid rotation and when viscosity can be neglected. It follows from the geostrophic balance equation, which is found from Eq.~\eqref{eq:CoriolisOperator} by neglecting time-dependence
\begin{equation}
\label{eq:GeostrophicBalance}
	2\mathbf{\hat{z}}\times\mathbf{u}=-\frac{1}{\rho_0}\nabla P.
\end{equation}
Applying the curl leads to the Taylor-Proudman theorem, which states that
\begin{equation}
	\pderiv{\mathbf{u}(\mathbf{r})}{z}=0.
\end{equation}
Hence, the geostrophic solution describes a flow that is invariant along the rotation axis (i.e. the z-axis). In order to satisfy the boundary condition in a spherical container, Eq.~\eqref{eq:BoundaryCondition}, the geostrophic mode must be axisymmetric and purely azimuthal. The geostrophic mode can therefore be written as an infinite sum of so-called `geostrophic' polynomials $G_{2k-1}$ for integers $k\geq 1$ \cite[]{Zhang2017} of the form
\begin{equation}
\label{eq:geostrophic_modes}
\mathbf{u}^\mathrm{G}(r,\theta)=\sum_{k\geq1}a^\mathrm{G}_k G_{2k-1}(r,\theta)\ephi.
\end{equation}
involving unknown coefficients $a^\mathrm{G}_k$, the radius $r$, the colatitude $\theta$ and the unit vector $\ephi$ in azimuth (see Appendix.~\ref{app:geostrophic_mode} for details of $G_{2k-1}$). Although the geostrophic polynomials depend on the two coordinates $r$ and $\theta$, they take a single argument $s\equiv r\sin\theta$, of which they are odd functions. They are also orthogonal and can be normalized with respect to their mean value over the sphere, i.e.~given a second geostrophic polynomial $G_{2l-1}$ for an integer $l$, they satisfy
\begin{equation}
\label{eq:geostrophic_normalization}
\frac{3}{4\uppi}\int_\mathcal{V}G_{2k-1}G_{2l-1}\mathrm{d}\mathcal{V}=\delta_{kl},
\end{equation}
where $\delta_{kl}$ is the kronecker delta and $\int_{\mathcal{V}}\mathrm{d}\mathcal{V}$ denotes the integration
\begin{equation}
\int_{\mathcal{V}}\mathrm{d}\mathcal{V}\equiv\int_{0}^{2\uppi}\mathrm{d}\phi\int_{0}^{\uppi}\sin\theta\mathrm{d}\theta\int_{0}^{1}r^2\mathrm{d}r,
\end{equation}
with $\phi$ being the longitude. 

\subsubsection{Inertial modes}

Allowing for time-dependent flows, the full solution to Eq.~\eqref{eq:CoriolisOperator} is obtained using the superposition of an infinite number of so-called inertial modes. Following \cite{Zhang2017}, these modes can be specified in terms of three complex-valued velocity components in spherical polar coordinates \mbox{$\mathbf{r}=(r,\theta,\phi)$} as
\begin{equation}
\mathbf{u}(\mathbf{r},t)=[u_r(r,\theta,\phi),u_\theta(r,\theta,\phi),u_\phi(r,\theta,\phi)]\e^{\I2\sigma t},
\end{equation}
where $\I$ is the imaginary number and $\sigma=\tfrac{\omega}{2}$ a half-frequency, which is real-valued and bounded by $0<\abs{\sigma}< 1$.  Each inertial mode has a specific spatial structure that is associated with a specific value of $\sigma$.  

The inertial modes can be divided into two classes according to their symmetry with respect to the equatorial plane. The equatorially symmetric ($\ES$) inertial modes satisfy
\begin{align}
\big(u_r,u_\theta,u_\phi\big)(r,\theta,\phi)&=\big(u_r,-u_\theta,u_\phi\big)(r,\uppi-\theta,\phi),\\
\intertext{whereas the equatorially antisymmetric ($\EA$) inertial modes obey}   
\big(u_r,u_\theta,u_\phi\big)(r,\theta,\phi)&=\big(-u_r,u_\theta,-u_\phi\big)(r,\uppi-\theta,\phi).
\end{align}
We follow the notation of \cite{Zhang2017} and specify the inertial modes by a triple index $(m,n,k)$ with $m,k\geq 0$ and $n\geq 1$. The azimuthal wavenumber $m$ controls the periodic structure in azimuth and is zero in case of axisymmetric modes. The index $n$ determines the degree of complexity in the $z$ direction, along the rotation axis, whereas $k$ indicates the structure along the cylindrical radius perpendicular to the axis of rotation. The inertial modes, hereafter denoted $\mathbf{u}_{mnk}$, are orthogonal when integrated over the full sphere and can therefore be normalized such that
\begin{equation}
\frac{3}{4\uppi}\int_\mathcal{V}\mathbf{u}_{mnk}\cdot\mathbf{u}_{m'n'k'}^*\mathrm{d}\mathcal{V}=\delta_{mm'}\delta_{nn'}\delta_{kk'},
\end{equation}
where $*$ denotes the complex conjugate. Since the expressions for the inertial modes are sinusoidal in azimuth, both the real and the imaginary parts are needed to correctly account for the phase in azimuth. Hence, there are always two coefficients associated with each inertial mode. We introduce $\ES$ and $\EA$ modes separately in the following.

 \paragraph{{\it Equatorially Symmetric Modes}}

The $\ES$ inertial modes $\mathbf{u}^\mathrm{S}_{mnk}$, can be divided into axisymmetric ($m=0$) and non-axisymmetric ($m\geq1$) modes. The half-frequencies $\sigma$ of the axisymmetric $\ES$ modes are the roots of the polynomial \cite[]{Zhang2017}
\begin{equation}
\label{eq:symfrequency1}
0=\sum_{j=0}^{k-1}(-1)^{j}\frac{[2(2k-j)]!}{j!(2k-j)![2(k-j)-1]!}\sigma^{2(k-j)},
\end{equation}
with $k\geq 2$. Hence, for a given $k$, there are \mbox{$k-1$} axisymmetric $\ES$ modes corresponding to the $k-1$ distinct roots.  In case of the non-axisymmetric $\ES$ modes, the half-frequencies satisfy
\begin{equation}
\label{eq:symfrequency2}
0=\sum_{j=0}^{k}(-1)^{j}\frac{[2(2k+m-j)]!}{j!(2k+m-j)![2(k-j)]!}\left[(m+2k-2j)-\frac{2(k-j)}{\sigma}\right]\sigma^{2(k-j)},
\end{equation}
with  $k\geq 1$. There are $2k$ modes for a given $k$ and $m\neq 0$. The absolute value of the roots can be arranged in ascending order $0<\abs{\sigma_1}<\abs{\sigma_2}<\dots<\abs{\sigma_{2k}}$ with the first being the longest period $\ES$ inertial mode for a given $m$ and $k$ (see Appendix~\ref{app:inertial_mode_ES} for explicit expressions).

\paragraph{{\it Quasi-Geostrophic Modes}}

A special subset of the $\ES$ inertial modes are those characterized by having the longest periods. Their force balance is closest to the geostrophic balance in Eq.~\eqref{eq:GeostrophicBalance} which results in a flow that is almost invariant with respect to the rotation axis. Following \cite{Zhang2017}, modes of this subset of $\ES$ inertial modes will therefore be referred to as being quasi-geostrophic (QG). They correspond to the half-frequencies $\sigma^\mathrm{S}_{m1k}$, and are modes with the index $n=1$ in accordance with the previously applied ordering.

\paragraph{{\it Equatorially Antisymmetric Modes}}

The $\EA$ inertial modes are denoted $\mathbf{u}^\mathrm{A}_{mnk}$ and can be again separated into axisymmetric and non-axisymmetric modes. The half-frequencies of the axisymmetric $\EA$ modes satisfy \cite[]{Zhang2017}
\begin{equation}
\label{eq:asymfrequency1}
0=\sum_{j=0}^{k}(-1)^{j}\frac{[2(2k-j+1)]!}{j!(2k-j+1)![2(k-j)]!}\sigma^{2(k-j)},
\end{equation}
with $k\geq 1$,
whereas the half-frequencies of the non-axisymmetric $\EA$ modes are the roots of
\begin{equation}
\label{eq:asymfrequency2}
0=\sum_{j=0}^{k}\frac{(-1)^{j}[2(2k+m-j+1)]!}{j!(2k+m-j+1)![2(k-j)+1]!}\\
\left[(m+2k-2j+1)-\frac{2(k-j)+1}{\sigma}\right]\sigma^{2(k-j)+1},
\end{equation}
with $k\geq 0$. Hence, there are $k$ axisymmetric and $2k+1$ non-axisymmetric $\EA$ modes for a given $k$ (see Appendix~\ref{app:inertial_mode_EA} for explicit expressions).

\begin{figure}
	\centering
	\includegraphics[width=8.5cm]{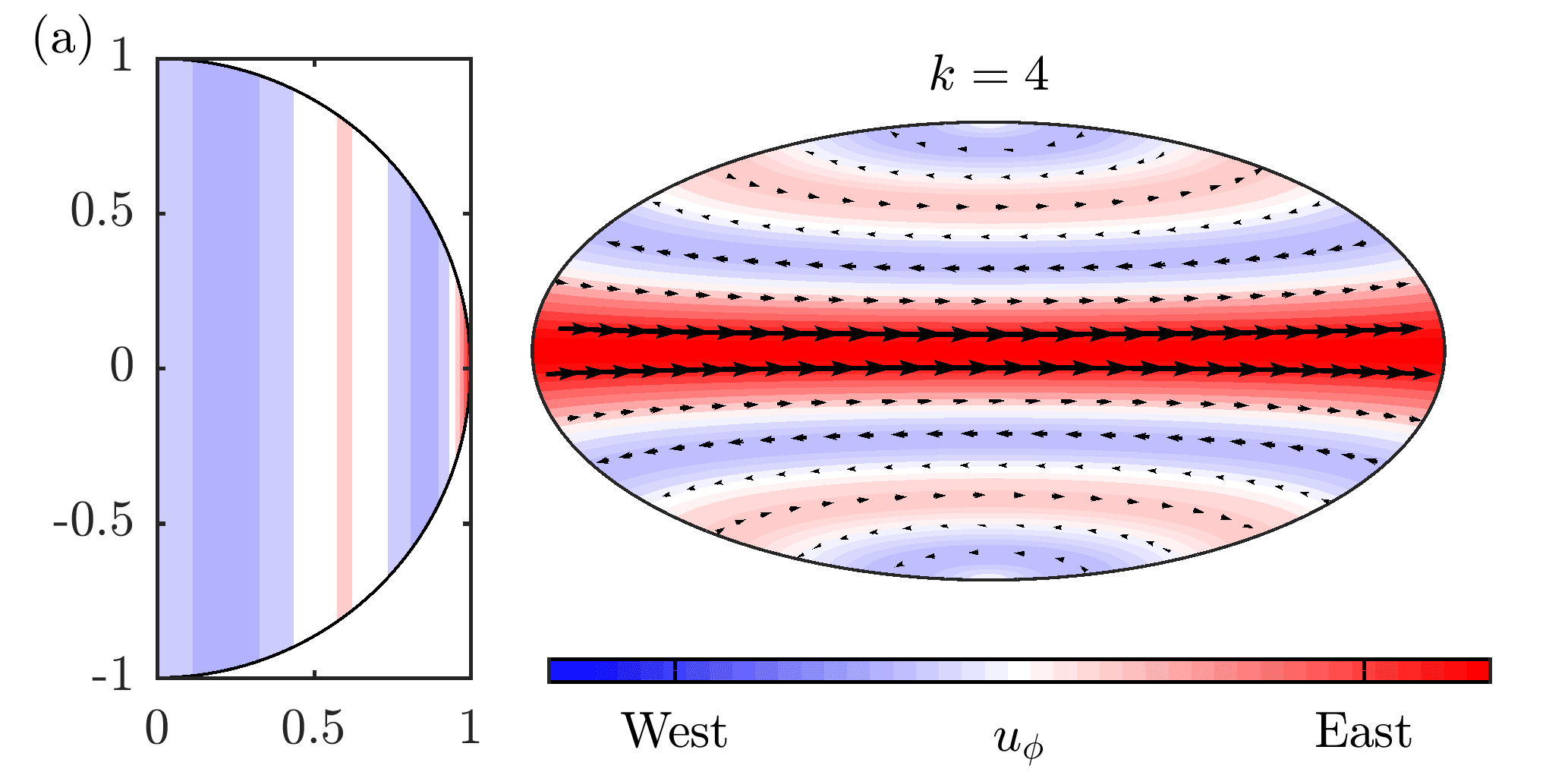}
	\includegraphics[width=8.5cm]{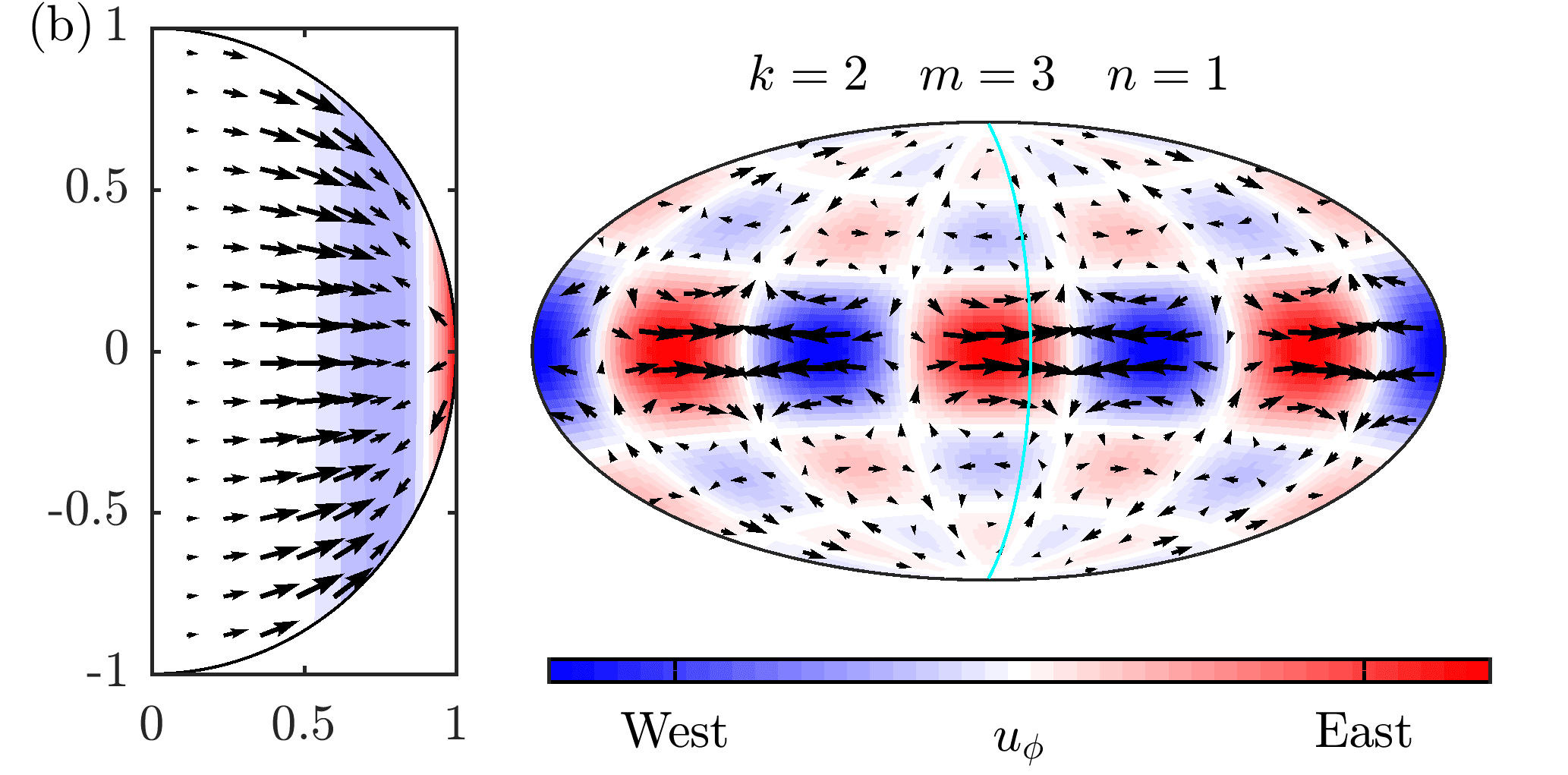}
	\includegraphics[width=8.5cm]{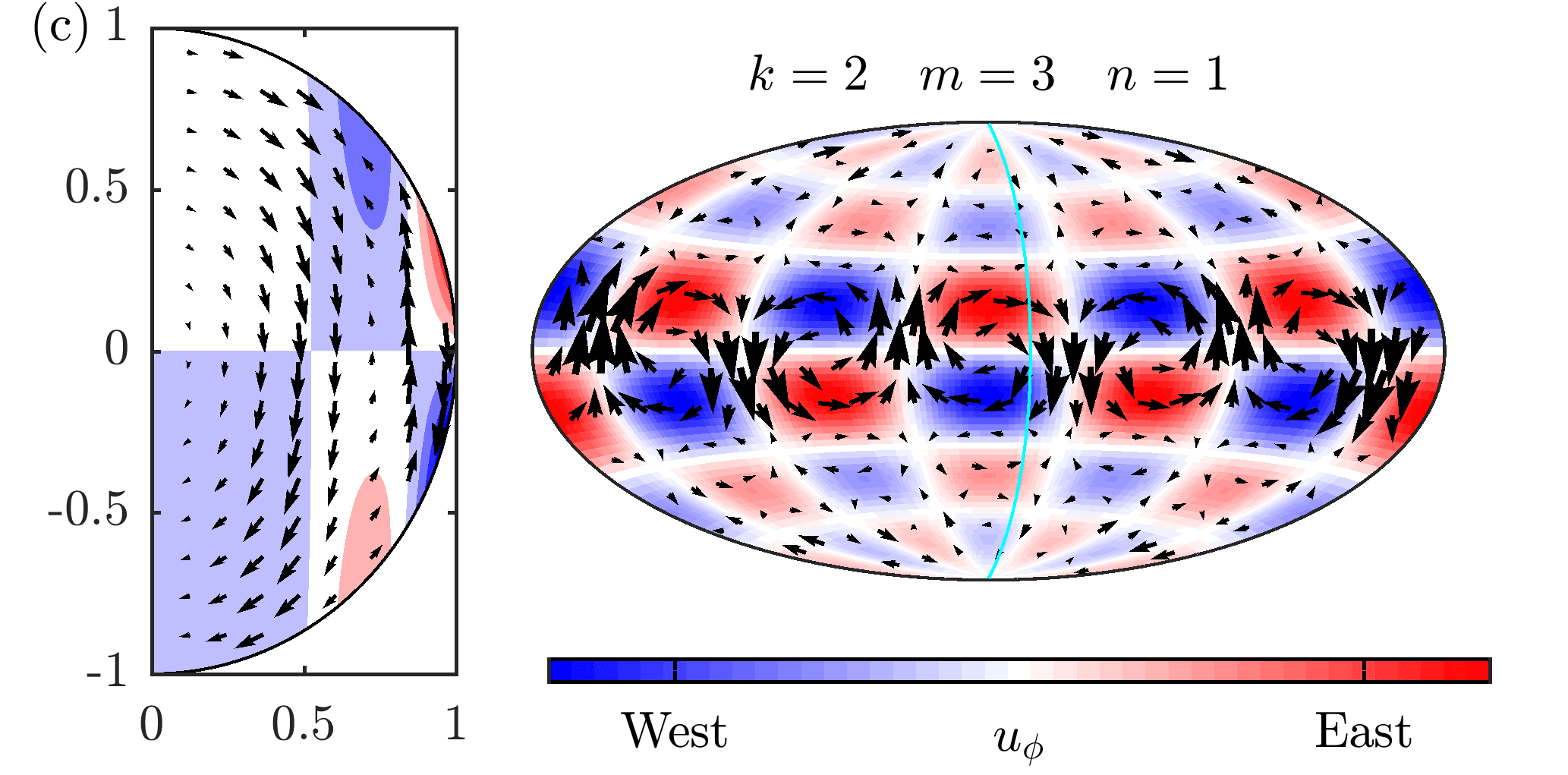}
	\caption{Examples of a Geostrophic polynomial (a) and $\ES$ (b) and $\EA$ (c) inertial modes in a meridional plane (left of each panel) and on the core surface (right). The cyan line indicates the intersection of the meridional plane with the core surface. Note that the Geostrophic polynomials are axisymmetric.}
	\label{fig:mode_examples}
\end{figure}

Fig.~\ref{fig:mode_examples} shows examples of a Geostrophic polynomial, $\ES$ and $\EA$ inertial modes, presenting the structure at the core surface and in meridional sections.  Note that these modes have little energy at small cylindrical radius and are well-suited for describing fluid motions in the equatorial region.

\subsection{Decomposition of the flow into steady and time-dependent parts}

We model the core flow as a linear combination of a finite number of geostrophic polynomials and inertial modes. Further, we divide the flow into two parts, a steady background flow $\mathbf{u}_0$ and a time-dependent flow $\mathbf{u}_t$
\begin{equation}
\label{eq:static_time-dep_flow}
	\mathbf{u}(\mathbf{r},t)=\mathbf{u}_0(\mathbf{r})+\mathbf{u}_t(\mathbf{r},t).
\end{equation}
We allow the geostrophic mode, the QG modes (i.e. $\ES$ modes with $n=1$) and the slowest $\EA$ modes ($n=1$) to be time-dependent, since these are the modes expected to be excited by the very slow (relative to the rotation time scale) forcing processes in the core \cite[e.g.][]{Zhang2017}
\begin{equation}
\label{eq:time-dep_flow}
\begin{aligned}
	\mathbf{u}_t(\mathbf{r},t)=&\sum_{k=1}^{K}a_k^\mathrm{G}(t)G_{2k-1}(\mathbf{r})\ephi + \sum_{k=1}^{K}\sum_{m=1}^{M}a^\mathrm{S}_{m1k}(t)\real{\mathbf{u}^\mathrm{S}_{m1k}(\mathbf{r})}\\
	&+\sum_{k=0}^{K}\sum_{m=1}^{M}a^\mathrm{A}_{m1k}(t)\real{\mathbf{u}^\mathrm{A}_{m1k}(\mathbf{r})} + \text{ imaginary part},
\end{aligned}
\end{equation}
where $\real{\cdot}$ denotes the real part. The steady background consists of the axisymmetric and non-axisymmetric inertial modes with $n\neq 1$, since forcing processes on long timescales, for example inhomogeneous mantle heat flow, is expected to include both equatorial symmetries
\begin{equation}
\label{eq:static_flow}
\begin{aligned}
	\mathbf{u}_0(\mathbf{r})=&\sum_{k=1}^{K}\sum_{m=0}^{M}\sum_{n\neq 1}a^\mathrm{S}_{mnk}\real{\mathbf{u}^\mathrm{S}_{mnk}(\mathbf{r})}\\
&+\sum_{k=0}^{K}\sum_{m=0}^{M}\sum_{n\neq 1}a^\mathrm{A}_{mnk}\real{\mathbf{u}^\mathrm{A}_{mnk}(\mathbf{r})}+\text{ imaginary parts}.
\end{aligned}
\end{equation}
The upper summation limits $K$ and $M$ of the indexes $k$ and $m$ serve as the truncation degrees in this representation. The index $n$ of the inertial modes is automatically bounded by the number of roots of the polynomials in Eqs.~\mbox{\eqref{eq:symfrequency1} - \eqref{eq:asymfrequency2}} for a given $k$. 

It is important to clarify at this point that we do not use the time-dependence of the inertial modes to parameterize the time-dependence of our core flows.  We use only their spatial structure, which we believe to be an efficient way of parameterizing the morphology of rotation-dominated motions at mid and low latitudes.  As discussed in the next section, we solve for the structure of the flow independently at each epoch (i.e. every 4 months), except for a temporal regularization minimizing the flow acceleration.  In effect, we are using the geostrophic and inertial modes to provide a kinematic description of the structure of the core flow epoch by epoch, and do not use them to describe dynamics.  Forcing by buoyancy and magnetic effects, not accounted for in the above framework, will determine the time evolution of each mode, rather than the free oscillation periods of the modes.

\section{Estimation from geomagnetic observations}

\subsection{Induction equation and unresolved scales}

The starting point for inferring the core flow from observations of the magnetic field is the radial component of the induction equation at the core surface, under the frozen-flux approximation \cite[e.g.][]{Bloxham1991}
\begin{equation}
\label{eq:radialfrozenflux}
\pderiv{B_r}{t}=-\nabla_\mathrm{H}\cdot(\mathbf{u}B_r),\qquad \mbox{at } \quad r=1,
\end{equation}
where $\nabla_\mathrm{H}\equiv\nabla-\er\pderiv{}{r}$ is the horizontal gradient. At the core surface, only the large-scale structure of the core field and SV are resolved and can be used to study the core flow. Since all lengthscales of the core field and the flow can interact to give large-scale SV \cite[]{Backus1968}, we decompose the radial core field as $B_r=\overline{B}_r+\tilde{B}_r$, the sum of the resolved large-scale radial core field and an unresolved small-scale part. According to  Eq.~\eqref{eq:radialfrozenflux} the large-scale radial SV at the core surface, denoted $\pderiv{\overline{B}_r}{t}$, is then \cite[]{Gillet2015}
\begin{equation}
\label{eq:radialfrozenflux_largescale}
\pderiv{\overline{B}_r}{t}=-\overline{\nabla_\mathrm{H}\cdot(\mathbf{u}\overline{B}_r)}+e,
\end{equation}
where $e=-\overline{\nabla_\mathrm{H}\cdot(\mathbf{u}\tilde{B}_r)}$ is the radial component of the so-called small-scale error, which represents the contribution of the small-scale radial core field interacting with the flow. In order to use Eq.~\eqref{eq:radialfrozenflux_largescale} in an inversion scheme, we assume that the radial core field and SV can be derived from a scalar potential field which we expand into spherical harmonics (SH). Similarly, we express the flow at the core surface with toroidal and poloidal potentials in terms of an SH expansion. Inserting the SH expansions in Eq.~\eqref{eq:radialfrozenflux_largescale} and numerically integrating over a grid at the core surface \cite[e.g.][]{Lloyd1990,Jackson1997} results in the matrix equation
\begin{equation}
\label{eq:snapshot_equation}
	\mathbf{\dot{b}}=\mathbf{H}_b(\mathbf{b})\mathbf{x}+\mathbf{e}
\end{equation}
where the column vectors $\mathbf{\dot{b}}$, $\mathbf{b}$, $\mathbf{x}$ and $\mathbf{e}$ contain the SH expansions of the SV up to degree $N_{\dot{b}}$, the resolved core field up to degree $N_b$, the toroidal-poloidal flow potentials at the core surface truncated at degree $N_x$ and the small-scale error with truncation degree $N_e$, all evaluated at the core surface. Note that the matrix $\mathbf{H}_b$ depends on the core field $\mathbf{b}$, which we consider here to be known up to the truncation degree and given by a field model; this approximation is often made in such studies \cite[e.g.][]{whaler2015derivation}. Eq.~(22) is a global solution, however here we use only observations from mid and low latitudes that are less contaminated by noise.  Thanks to the form of the Green's function for Laplace's equation subject to Neumann boundary conditions, that links the core surface field to the field at Earth's surface and at satellite altitude, these observations do provide information at all latitudes.  Nevertheless, the data constraint is weaker at polar latitudes and we thus concentrate on time-dependent flows based on inertial modes that are most prominent at mid and low latitudes.

\subsection{Forward problem}

According to Eq.~\eqref{eq:snapshot_equation}, the core surface flow and the small-scale error are connected to the large-scale radial SV at a single epoch. By making use of this equation at discrete times $t_p$, $p\in[1,P]$ spanning the time interval for which observational data is provided, we build a multi-epoch matrix equation to simultaneously compute the SV at all $t_p$ \cite[e.g.][]{Whaler2016}. 

For our formulation of the core flow in terms of modes, the forward problem is completed by including a matrix which allows the computation of the toroidal-poloidal flow expansion from a linear combination of geostrophic polynomials and inertial modes. Hereafter, the column vectors $\mathbf{\dot{b}}_p$, $\mathbf{b}_p$, $\mathbf{x}_p$ and $\mathbf{e}_p$ denote the SH expansions of the SV, the core field, the core surface flow and the small-scale error evaluated at time $t_p$, respectively. We express the SH components of the time-dependent flow as linear combination of geostrophic polynomials, described by coefficients $a^\mathrm{G}_k(t_p)$, QG modes, described by coefficients $a^\mathrm{S}_{m1k}(t_p)$, and the slowest $\EA$ modes, described by the coefficients $a^\mathrm{A}_{m1k}(t_p)$ in a time-dependent column vector $\mathbf{a}_p\equiv\mathbf{a}(t_p)$. Similarly, we store the coefficients $a^\mathrm{S}_{mnk}$ and $a^\mathrm{A}_{mnk}$ with $n\neq1$ in a time-independent column vector $\mathbf{a}_0$ that describes the steady background flow.  The toroidal-poloidal expansion of flow Eq.~\eqref{eq:static_time-dep_flow} at time $t_p$ can therefore be written as
\begin{equation}
\label{eq:FlowExpansion}
\mathbf{x}_p=\mathbf{M}_0\mathbf{a}_0+\mathbf{M}_t\mathbf{a}_{p},
\end{equation}
where $\mathbf{M}_0$ and $\mathbf{M}_t$ are fixed matrices which contain as columns the toroidal-poloidal expansion of the geostrophic and inertial modes matching the ordering of the coefficients in $\mathbf{a}_0$ and $\mathbf{a}_p$. Inserting this expression into Eq.~\eqref{eq:snapshot_equation}, the SH expansion of the SV on the core surface at time $t_p$ becomes
\begin{equation}
\label{eq:sv_coefficients}
	\mathbf{\dot{b}}_p=\mathbf{H}_b(\mathbf{b}_p)\mathbf{x}_p +\mathbf{e}_p=\mathbf{H}_b(\mathbf{b}_p)\bigg[\mathbf{M}_0\mathbf{a}_0+\mathbf{M}_t\mathbf{a}_p\bigg] +\mathbf{e}_p.
\end{equation}

In order to calculate from our model the three vector components of the SV (i.e. $\partial B_r / \partial t$,  $\partial B_\theta / \partial t$ and  $\partial B_\phi / \partial t$) at the known locations of the ground and virtual observatories, listed in the form of a data vector $\mathbf{d}_p$, we multiply the SV expansion $\dot{\mathbf{b}}_p$ in Eq.~\eqref{eq:sv_coefficients} with the matrix $\mathbf{Y}_p$, containing the appropriate spatial derivatives of spherical harmonics, such that
\begin{equation}
\label{eq:single_epoch_forward}
\mathbf{d}_p=\mathbf{Y}_p\mathbf{\dot{b}}_p=\mathbf{Y}_p\mathbf{H}_b(\mathbf{b}_p)\bigg[\mathbf{M}_0\mathbf{a}_0+\mathbf{M}_t\mathbf{a}_p\bigg] +\mathbf{Y}_p\mathbf{e}_p.
\end{equation}
In this step, it is assumed that the mantle conductivity is small and can be neglected. By adjusting the number of rows in $\mathbf{Y}_p$, we can also handle the changing number of SV data over time. 

To summarise, we define our model vector to be \mbox{$\mathbf{m}=[\mathbf{a}_0^\mathrm{T},\mathbf{a}_1^\mathrm{T},\dots,\mathbf{a}_P^\mathrm{T},\mathbf{e}_1^\mathrm{T},\dots,\mathbf{e}_P^\mathrm{T}]^\mathrm{T}$} and combine the $P$ single-epoch expressions in Eq.~\eqref{eq:single_epoch_forward} into one matrix equation of the form
\begin{equation}
\label{eq:forward_problem}
\mathbf{d}=\mathbf{G}\mathbf{m},
\end{equation}
where the forward matrix is
\begin{equation}
	\begin{aligned}
		\mathbf{G}
		=
		&\begin{bmatrix}
			\mathbf{F}_1\mathbf{M}_0 & \mathbf{F}_1\mathbf{M}_t &  &  & \mathbf{Y}_1 &  &  \\ 
			\rotatebox{90}{\dots} &  & \rotatebox{135}{\dots} &  &  & \rotatebox{135}{\dots} &  \\ 
			\mathbf{F}_P\mathbf{M}_0 &  &  & \mathbf{F}_P\mathbf{M}_t &  &  & \mathbf{Y}_P
		\end{bmatrix}\\
	\end{aligned}
\end{equation}
with $\mathbf{F}_p=\mathbf{Y}_p\mathbf{H}_b(\mathbf{b}_p)$. This allows us to calculate model predictions $\mathbf{d}=[\mathbf{d}_1^\mathrm{T},\dots,\mathbf{d}_P^\mathrm{T}]^\mathrm{T}$ at the locations and times of the observational data.

\subsection{Inverse problem}

In order to infer the core flow from the geomagnetic observations stored in $\mathbf{d}^\mathrm{obs}$, we minimize a cost function of the model vector $\mathbf{m}$ of the form
\begin{equation}
\label{eq:cost_function}
	\Phi(\mathbf{m}) = (\mathbf{G}\mathbf{m}-\mathbf{d}^\mathrm{obs})^\mathrm{T}\mathbf{W}_d(\mathbf{G}\mathbf{m}-\mathbf{d}^\mathrm{obs}) +\lambda_0\norm{\mathbf{Q}_0\mathbf{a}_0}_1 + \lambda_t^\mathrm{S}\norm{\mathbf{Q}_t^\mathrm{S}\mathbf{a}_t}_1 + \lambda_t^\mathrm{A}\norm{\mathbf{Q}_t^\mathrm{A}\mathbf{a}_t}_1 + \lambda_a\norm{\mathbf{D}\mathbf{a}_t}^2_2 + \mathbf{e}_t^\mathrm{T}\mathbf{C}^{-1}_e\mathbf{e}_t,
\end{equation}
where $\norm*{\mathbf{x}}_p=(\sum_i\abs{x_i}^p)^{1/p}$ denotes the $l_p$ norm of a vector $\mathbf{x}$, $\mathbf{a}_t\equiv[\mathbf{a}_{1}^\mathrm{T},\dots,\mathbf{a}_{P}^\mathrm{T}]^\mathrm{T}$ is the vector containing all time-dependent flow coefficients, and \mbox{$\mathbf{e}_t\equiv[\mathbf{e}_{1}^\mathrm{T},\dots,\mathbf{e}_{P}^\mathrm{T}]^\mathrm{T}$} is the multi-epoch small-scale error. The strictly positive parameters $\lambda_0$, $\lambda_t^\mathrm{S}$, $\lambda_t^\mathrm{A}$ and $\lambda_a$ control the degree to which the steady background flow, the time-dependent equatorially symmetric and antisymmetric flow, and the flow acceleration are regularized.

The first term on the right side of Eq.~\eqref{eq:cost_function} measures the misfit to the ground and satellite SV data.  It involves the diagonal matrix $\mathbf{W}_d$ which contains the estimated data error variance $\sigma_i^2$ modified following a Tukey biweight scheme in order to downweight outliers \cite[e.g.][]{Constable1988}.
\begin{equation}
\label{eq:tukey-weights}
(\mathbf{W}_d)_{ii} = \left\{
\begin{aligned}
&\frac{\left(1-\big(\frac{r_i}{c\sigma_i}\big)^2\right)^2}{\sigma_i^2}, & &\frac{\abs{r_i}}{\sigma_i} \leq c\\
&0, & &\frac{\abs{r_i}}{\sigma_i} > c
\end{aligned}
\right.
\end{equation}
with breakpoint $c=4.685$ and $r_i$ the residual of the data point $(\mathbf{d}^\mathrm{obs})_i$ relative to the CHAOS-6-x7 geomagnetic field model. The choice of the data error variances $\sigma_i^2$ is described in section \ref{sec:dataVO}.

We regularize the spatial structure of flows by minimizing an $l_1$ norm of the square root of the mode enstrophies calculated with the diagonal matrices $\mathbf{Q}_0$ for the steady part $\mathbf{u}_0$, $\mathbf{Q}_t^\mathrm{S}$ and $\mathbf{Q}_t^\mathrm{A}$ for the time-dependent equatorially symmetric and antisymmetric part of $\mathbf{u}_t$.  We regularize the time-dependence of the flow using an $l_2$ norm of the flow acceleration calculated by applying a time finite-differencing operator $\mathbf{D}$ to the time-dependent part of the model vector.  The small-scale error is regularized using a quadratic form based on the inverse of its multi-epoch covariance matrix $\mathbf{C}_e=\mathbf{C}_e(\mathbf{m})$, which depends on the model vector. We define the matrices $\mathbf{D}$, $\mathbf{Q}_0$, $\mathbf{Q}_t^\mathrm{A}$, $\mathbf{Q}_t^\mathrm{S}$ and $\mathbf{C}_e$ in the next section.

Regarding the minimization of the cost function in Eq.~\eqref{eq:cost_function}, we note that this is a non-linear problem as it involves $l_1$ spatial norms of the flow model and the small-scale error covariance matrix that depends on the flow model. We therefore compute the model solution using an iterative procedure by solving the equation
\begin{equation}
\label{eq:iterativeLS}
	\mathbf{m}_{k+1} = \left(\mathbf{G}^\mathrm{T}\mathbf{W}_d\mathbf{G}+\mathbf{R}(\mathbf{m}_k)\right)^{-1}\mathbf{G}^\mathrm{T}\mathbf{W}_d\mathbf{d}^\mathrm{obs}
\end{equation}
at each iteration step $k$ until convergence criteria are met. We consider a computed model $\mathbf{m}_{k+1}$ to be converged if the relative change between successive iterations of the data misfit \mbox{$(\mathbf{G}\mathbf{m}-\mathbf{d}^\mathrm{obs})^\mathrm{T}\mathbf{W}_d(\mathbf{G}\mathbf{m}-\mathbf{d}^\mathrm{obs})$}, the $l_1$ norm of the two parts of the flow \mbox{$\norm{\mathbf{Q}_t^\mathrm{S}\mathbf{a}_t}_1+\norm{\mathbf{Q}_t^\mathrm{A}\mathbf{a}_t}_1$} and \mbox{$\norm*{\mathbf{Q}_0\mathbf{a}_0}_1$}, and the small-scale error norm \mbox{$\mathbf{e}_t^\mathrm{T}\mathbf{C}^{-1}_e\mathbf{e}_t^\mathrm{\vphantom{T}}$} are each smaller than $10^{-2}$. The regularization matrix $\mathbf{R}$ is block-diagonal and constructed from the model of the previous iterate $\mathbf{m}_k$ according to 
\begin{equation}
\mathbf{R}=
\begin{bmatrix}
\lambda_0\mathbf{W}_0 &  &  \\ 
& \lambda_t^\mathrm{S}\mathbf{W}_t^\mathrm{S}+\lambda_t^\mathrm{A}\mathbf{W}_t^\mathrm{A}+\lambda_a\mathbf{W}_a &  \\ 
&  & \mathbf{C}^{-1}_e
\end{bmatrix}
\end{equation}
where the matrices $\mathbf{W}_0$, $\mathbf{W}_t^\mathrm{S}$ and $\mathbf{W}_t^\mathrm{A}$ iteratively implement the $l_1$ norm, and $\mathbf{W}_a$ the $l_2$ norm. With $\mathbf{R}$ defined in this way, the model parameter estimation usually converged within 20 iterations. The definition of the matrices $\mathbf{W}_0$, $\mathbf{W}_t^\mathrm{S}$, $\mathbf{W}_t^\mathrm{A}$, $\mathbf{W}_a$ and $\mathbf{C}_e$ is given in the following.

\subsection{Regularization norms}

\subsubsection{Spatial regularization: $l_1$ norm based on square root of mode enstrophies}

The flow coefficients $\mathbf{a}_0$ and $\mathbf{a}_t$ are spatially regularized using an $l_1$ norm based on the Ekblom measure \cite[]{Farquharson1998}
\begin{equation}
\mathbf{W}_0=\mathbf{Q}^\mathbf{T}_0\left(\frac{\delta_{ij}}{\sqrt{(\mathbf{Q}_0\mathbf{a}_0)_i^2+\epsilon^2}}\right)\mathbf{Q}^\mathbf{\vphantom{T}}_0,
\end{equation}
for the steady background part and similarly $\mathbf{W}_t^\mathrm{S}$ and $\mathbf{W}_t^\mathrm{A}$ for the time-dependent equatorially symmetric and antisymmetric part of the flow by replacing $\mathbf{Q}_0$ and $\mathbf{a}_0$ with $\mathbf{Q}_t^\mathrm{S}$ and $\mathbf{a}_t$, or $\mathbf{Q}_t^\mathrm{A}$ and $\mathbf{a}_t$, respectively. The constant $\epsilon>0$ is set to a value of $10^{-8}$ in order to prevent numerical problems in case the square root of the enstrophy of a mode approaches zero. Use of an $l_1$ norm promotes a sparse mode representation of the flow, by favouring a scenario where the mode amplitudes are if possible zero, and reducing the amplitude of the remaining modes. In order to also weakly favour large-scale flows (to help the spatial power spectrum of our flow to converge), we apply the $l_1$ norm with weights corresponding to the square root of the enstrophy $Q(\mathbf{u})\geq 0$ for a flow $\mathbf{u}$, with the enstrophy, the square of this quantity, defined as
\begin{equation}
\label{eq:enstrophy}
Q^2(\mathbf{u})=\frac{3}{4\uppi}\int_{\mathcal{V}}\abs{\nabla\times\mathbf{u}}^2\mathrm{d}\mathcal{V}.
\end{equation}
We compute the square root of the enstrophy for every individual inertial mode and geostrophic polynomial, and build the diagonal matrices $\mathbf{Q}_0$ for the steady background, and $\mathbf{Q}_t^\mathrm{S}$ and $\mathbf{Q}_t^\mathrm{A}$ for the time-dependent equatorially symmetric and antisymmetric part of the flow. Expressions of the enstrophy of the modes can be found in the appendix Eq.~\eqref{eq:geostrophic_vorticity}, \eqref{eq:symmetric_vorticity} and \eqref{eq:antisymmetric_vorticity}. The parameter $\lambda_t$ controls the extent to which the time-dependent part of the flow is regularized in space while $\lambda_0$ controls the regularization of the steady background flow.

\subsubsection{Temporal regularization: $l_2$ norm of mode Acceleration}

We regularize the flow in time by minimizing the flow acceleration via an $l_2$ norm based on the matrix $\mathbf{W}_a=\mathbf{D}^\mathrm{T}\mathbf{D}$, where $\mathbf{D}$ is the first order forward time difference operator for equally spaced  times $t_p$
\begin{equation}
\mathbf{D}=\frac{1}{\Delta t}
\begin{bmatrix}
-\mathbf{I} & \mathbf{I} &  &  \\ 
& \rotatebox{135}{\dots}  & \rotatebox{135}{\dots} &  \\ 
&  & -\mathbf{I} & \mathbf{I}
\end{bmatrix},
\end{equation}
where $\Delta t=t_{p+1}-t_p$ is the step-size and $\mathbf{I}$ is the unit matrix that is compatible with the size of $\mathbf{a}_p$. The strength of the regularization of the flow acceleration is controlled by the parameter $\lambda_a$. 

\subsubsection{Regularization of the small-scale error}

We compute the multi-epoch covariance matrix $\mathbf{C}_e$ of the small-scale error by projecting the covariance of the unresolved small-scale core field $\mathbf{C}^\mathrm{\vphantom{T}}_{\tilde{b}}$ of SH degree \mbox{$N_b<n\leq N_{\tilde{b}}$} onto large lengthscales of the SV, via the flow at the previous iteration
\begin{equation}
\label{eq:covariance_projection}
\mathbf{C}_e=\mathbf{H}^\mathrm{\vphantom{T}}\mathbf{C}^\mathrm{\vphantom{T}}_{\tilde{b}}\mathbf{H}^\mathrm{T},
\end{equation}
using the block-diagonal matrix
\begin{equation}
	\mathbf{H}=\mathrm{diag}\big(\mathbf{H}_x(\mathbf{x}_1),\dots,\mathbf{H}_x(\mathbf{x}_P)\big).
\end{equation}
The matrix $\mathbf{H}_x(\mathbf{x}_p)$ is based on the induction equation Eq.~\eqref{eq:radialfrozenflux} and is similar to $\mathbf{H}_b(\mathbf{b}_p)$ but takes the flow as the argument to connect the expansions of the magnetic field and the SV 
\begin{equation}
	\mathbf{\dot{b}}_p=\mathbf{H}_x(\mathbf{x}_p)\mathbf{b}_p.
\end{equation}
We build the covariance matrix for the unresolved small-scale field $\mathbf{C}_{\tilde{b}}$ as a $P\times P$ block matrix where each block $(\mathbf{C}_{\tilde{b}})_{pq}$ for $p,q\in [1,P]$ is equal to the covariance of the small-scale core field of epoch pairs $t_p$ and $t_q$
\begin{equation}
(\mathbf{C}_{\tilde{b}})_{pq}=\mathrm{E}\left[(\tilde{\mathbf{b}}_p)^\mathrm{T}(\tilde{\mathbf{b}}_q)\right]=\underset{n}{\mathrm{diag}}\left(\sigma^2_{\tilde{b}}(n)\left(1+\sqrt{3}\frac{\abs{t_p-t_q}}{\tau_\mathrm{c}(n)}\right)\exp\left(-\sqrt{3}\frac{\abs{t_p-t_q}}{\tau_\mathrm{c}(n)}\right)\right),
\end{equation}
with $\mathrm{E}[\cdot]$ the expectation, $\tau_\mathrm{c}(n)$ the correlation time, $\sigma^2_{\tilde{b}}(n)$ the variance of the small-scale core field coefficient of degree $n$, which we have stored as elements in the vector $\tilde{\mathbf{b}}_p$. Here, we use a Matérn class time correlation function with $\nu=\frac{3}{2}$ \cite[]{Rasmussen2006}. This has previous been used for a similar purpose by \cite{Gillet2015}, it describes an AR-2 process compatible with the observed secular variation spectrum \cite[][]{lesur2017frequency} and allows abrupt slope changes in the SV (jerks). The parameters  $\tau_\mathrm{c}$ and $\sigma^2_{\tilde{b}}$ are estimated for each degree from the field and SV spectra of the CHAOS-6-x7 model in epoch 2015 (see App.~\ref{app:details_covariance} for details). In the computation, we use the time average of the flow over all considered epochs in $\mathbf{H}$ since we expect the covariance of the small-scale error to be only weakly time-dependent when the flow is predominantly steady \cite[]{Gillet2015, Bloxham1992}, this helps improve model convergence.

\section{Results}
\label{sec:results}

In this section, we report the results for our reference flow model  {\it CoreFlo-LL.1}.  For this model the truncation degree of the geostrophic polynomials, the $\ES$ and $\EA$ inertial modes was chosen to be $K=10$ and $M=20$. To ensure an adequate representation of these modes in the computations a SH truncation degree $N_x=60$ was used. The multi-epoch inversion covers the period from September 2000 to January 2018 in 4-month steps, resulting in $P=53$ epochs. The flow is a combination of one geostrophic mode (represented by 10 geostrophic polynomials) and 4720 inertial modes (200 QG modes and  the 220 slowest $\EA$ modes in the time-dependent part, and the remaining 4300 $\ES$ and $\EA$ inertial modes in the steady background). The coefficients of the time-dependent part of the flow together with the small-scale error have to be estimated for every epoch. In total, we estimate 68914 model coefficients: $2\cdot 4300=8600$ determine the background flow, $P\cdot (10+2\cdot 200+2\cdot 220)=45050$ the time-dependent part of the flow and $P\cdot 288=15264$ the SH coefficients of the small-scale error. The truncation degree of the small-scale error $N_e = 16$ is identical to the truncation degree $N_{\dot{b}}$ of the computed SV. In order to compute the SV produced by a given core surface flow, we use the CHAOS-6-x7 model to estimate the core surface field only up to degree $N_b=14$ since higher degrees are contaminated by the lithospheric field \cite[]{Finlay2016}.

In addition to our reference flow model {\it CoreFlo-LL.1}, we computed several other flows that will be discussed further in section \ref{sec:disc}. Table \ref{tab:statistics} summarizes these flows and various statistics (defined below) derived from them. The other flow models were designed to test the robustness of {\it CoreFlo-LL.1}  by changing aspects of the inversion scheme and the model parameterization. In \textit{model 1b}, we constrained the time-dependent part of the flow to be purely equatorially symmetric, whilst continuing to regularize in the same way as \textit{CoreFlo-LL.1}. For \textit{model 1c}, we replaced our spatial flow regularization, based on the $l_1$ norm of the square root of the enstrophy of the modes, with a more traditional approach using an $l_2$ norm that scales with the third power of the spherical harmonic degree in the toroidal-poloidal expansion of the flow.

We characterize and compare the derived flows with the help of the following statistics:
\begin{enumerate}
	\renewcommand{\theenumi}{(\arabic{enumi})}
	\item normalized quadratic measure of the SV misfit between the observational data $\mathbf{d}^\mathrm{obs}$ and our model predictions $\mathbf{d}=\mathbf{G}\mathbf{m}$
	\begin{equation}
		\Phi_\mathrm{SV} = \frac{1}{N_d}(\mathbf{d}-\mathbf{d}^\mathrm{obs})^\mathrm{T}\mathbf{W}_d(\mathbf{d}-\mathbf{d}^\mathrm{obs}),
	\end{equation}
	where $N_d$ is the number of data,
	\item normalized quadratic measure of the SA misfit by computing the central time difference $\mathbf{\dot{d}}$ and $\mathbf{\dot{d}}^\mathrm{obs}$ of the model predictions and data
	\begin{equation}
		\Phi_\mathrm{SA} = \frac{1}{N_{\dot{d}}}(\mathbf{\dot{d}}-\mathbf{\dot{d}}^\mathrm{obs})^\mathrm{T}\mathbf{W}_{\dot{d}}(\mathbf{\dot{d}}-\mathbf{\dot{d}}^\mathrm{obs}),
	\end{equation}
	where $N_{\dot{d}}$ is the number of the SA data,	
	\item length-of-day (LOD) misfit between our model predictions and independent estimates $\mathbf{d}_\gamma$ from \cite{Gillet2015a}
	\begin{equation}
		\Phi_\mathrm{LOD} = \frac{1}{\sigma_\gamma^2N_\gamma}(\mathbf{G}_\gamma\mathbf{m}-\mathbf{d}_\gamma)^\mathrm{T}(\mathbf{G}_\gamma\mathbf{m}-\mathbf{d}_\gamma),
	\end{equation}
	where $\sigma_\gamma=0.24\,\mathrm{ms}$ is the standard deviation of the LOD over the considered time period and $\mathbf{G}_\gamma$ is the matrix to compute the LOD from our modeled core flow \cite[e.g.][]{Holme2015},
	\item $l_1$ norm of the square root of the mode enstrophy as a measure of spatial complexity
	\begin{equation}
		\Phi_\mathrm{E} = \frac{1}{N}(\|\mathbf{Q}_t^\mathrm{S}\mathbf{a}_t\|_1+\|\mathbf{Q}_t^\mathrm{A}\mathbf{a}_t\|_1+\|\mathbf{Q}_0\mathbf{a}_0\|_1),
	\end{equation}
	where $N$ is the total number of flow coefficients,
	\item fraction $f_\mathrm{S}$ of the total time-averaged flow power that is contributed by the equatorially symmetric flow $\mathbf{u}^\mathrm{S}$ (geostrophic mode and $\ES$ inertial modes) integrated over the core surface
	\begin{equation}
		f_\mathrm{S} = \frac{\left\langle\int_{\Omega}\abs{\mathbf{u}^\mathrm{S}}^2\mathrm{d}\Omega\right\rangle}{\left\langle\int_{\Omega}\abs{\mathbf{u}}^2\mathrm{d}\Omega\right\rangle},
	\end{equation}
	where $\langle\cdot\rangle$ stands for the time-average, and
	\item fraction $f_t$ of the total time-averaged flow power that is contributed by the time-dependent part of the flow $\mathbf{u}_t$ (geostrophic mode, QG inertial modes, and the slowest $\EA$ inertial modes for {\it CoreFlo-LL.1}), integrated over the core surface.
	\begin{equation}
		f_\mathrm{t} = \frac{\left\langle\int_{\Omega}\abs{\mathbf{u}_t}^2\mathrm{d}\Omega\right\rangle}{\left\langle\int_{\Omega}\abs{\mathbf{u}}^2\mathrm{d}\Omega\right\rangle}
	\end{equation}
\end{enumerate}
\begin{table*}
	\caption{Summary of the computed flow models. \textit{Model 1} is our reference solution {\it CoreFlo-LL.1}, which is derived using an $l_1$ norm of the square root of the mode enstrophies, and including time-dependent geostrophic, QG and the slowest $\EA$ modes (G+QG+$\EA$). In \textit{model 1b}, we use the same $l_1$ regularization norm as for \textit{model 1}, but restrict the time-dependence to the geostrophic mode and QG inertial modes (G+QG). For \textit{model 1c}, which uses the same parameterization as \textit{model 1b}, we use as spatial regularization an $l_2$ norm of the toroidal-poloidal flow expansion that scales as $n^3$ \cite[]{Gillet2009}. Definitions of the diagnostics are given in the text.}
	\label{tab:statistics}
	\centering
	\begin{tabular}{cccccccccccc}
		\toprule
		model & regularization norm & $\lambda_t^\mathrm{S}$ & $\lambda_t^\mathrm{A}$ & $\lambda_0$ & $\lambda_a$ & $\Phi_\mathrm{SV}$ & $\Phi_\mathrm{SA}$ & $\Phi_\mathrm{LOD}$ & $\Phi_\mathrm{E}$ & $f_\mathrm{S}$ & $f_t$ \\
		& (time-dep. modes) & (km$\cdot$yr) & (km$\cdot$yr) & (km$\cdot$yr) & (km/yr$^2$)$^{-2}$ & & & & (km$\cdot$yr)$^{-1}$ & & \\
		\midrule
		1 & $l_1$ (G+QG+$\EA$) & 0.60 & 3.0 & 5.0 & 167 & 0.77 & 0.58 & 0.55 & 0.57 & 0.82 & 0.17\\
		1b & $l_1$ (G+QG) & 0.25 & - & 6.3 & 167 & 0.76 & 0.57 & 1.7 & 0.45 & 0.83 & 0.41\\
		1c & $l_2$ (G+QG)& 0.16\footnotemark[1] & - & 1.6\footnotemark[1] & 167 & 0.77 & 0.58 & 0.67 & $9.1\cdot 10^6$ & 0.90 & 0.16\\
		CHAOS-6-x7 & - & - & - & - & - & 0.80 & 0.64 & - & - & - & -\\
		\bottomrule
	\end{tabular}
\end{table*}\footnotetext[1]{Units are (km/yr)$^{-2}$.}

\subsection{Fit to observations}
In Fig.~\ref{fig:median_residuals} we present the median absolute deviation between {\it CoreFlo-LL.1} model predictions and the observed SV at ground and virtual observatories as a function of location, for the radial, north-south and east-west SV components, respectively.  The data at the majority of the sites is fit to better than $2\,\textrm{nT/yr}$, with the exception of some ground stations, for example those in India, where the mean residuals are somewhat larger.  There is however no obvious systematic trend in the amplitude of the residuals, and the data at both low and mid latitudes is overall fit well, suggesting that {\it CoreFlo-LL.1} provides a reasonable explanation of field variations at these latitudes.   
\begin{figure}
	\centering
	\includegraphics[width=8.5cm]{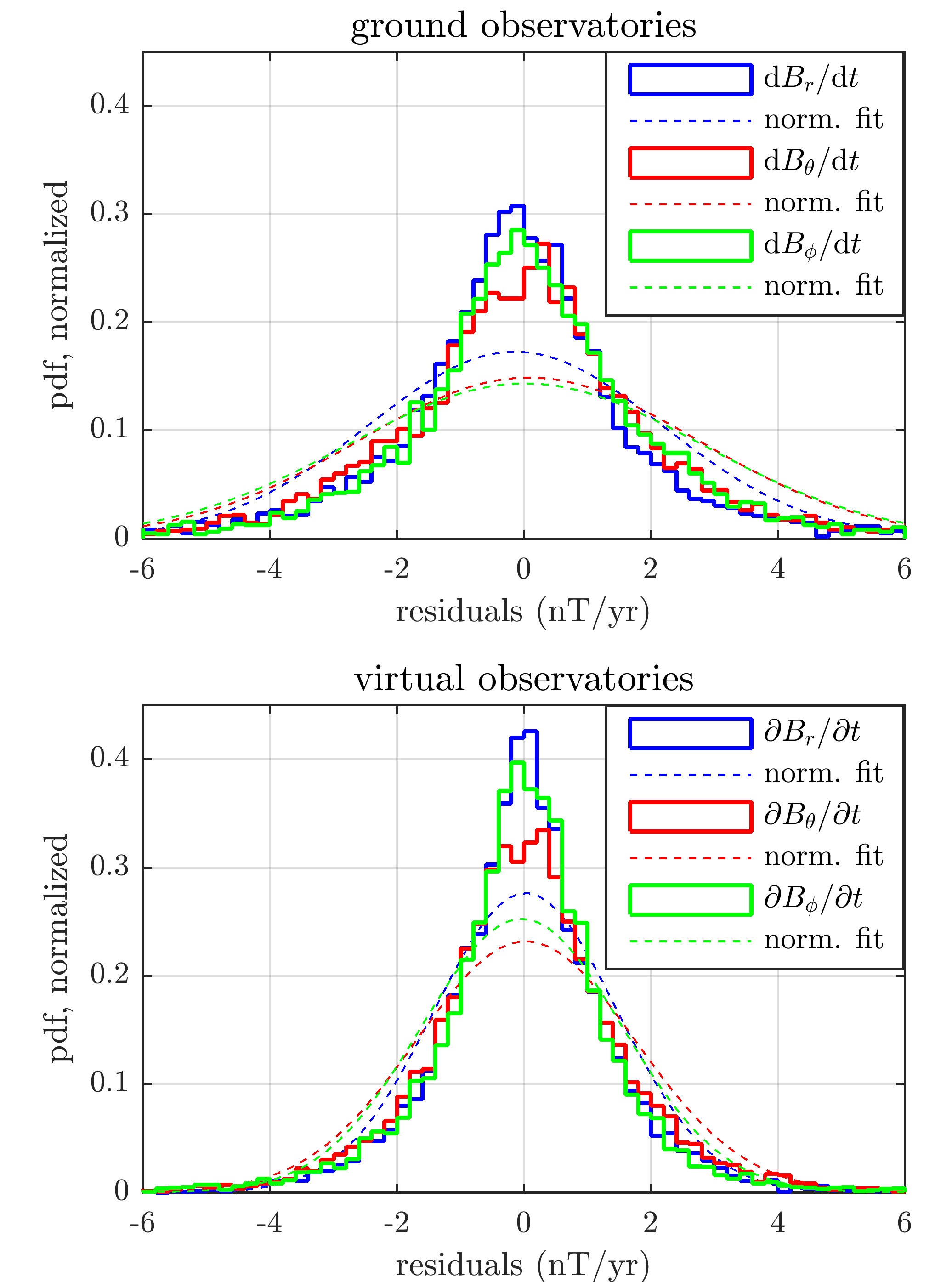}
	\caption{Histograms of the residuals between predictions of the {\it CoreFlo-LL.1} model and the SV at ground and satellite altitude. For comparison, the dashed curves show the probability density function of Gaussian distributed residuals given the respective mean values and variances.}
	\label{fig:hist_residuals}
\end{figure}

Fig.~\ref{fig:timeseries_go_vo} shows time series of ground and SV data from example low latitude sites, compared with the predictions from the CHAOS-6-x7 field model (red) and {\it CoreFlo-LL.1} (green).    {\it CoreFlo-LL.1} fits the data as well as CHAOS-6-x7, and shows remarkably similar patterns of SV.  There is no evidence for systematic offsets in the  {\it CoreFlo-LL.1} series, and it successed in capturing 'V' shaped structures in the SV associated with geomagnetic jerks, see for example the $\phi$-component at MBO in 2007 \cite[e.g.][]{Chulliat2010} and the VO at latitude 5.964$^\circ$ degrees, longitude -32.995$^\circ$.  The time-dependence in our SV predictions comes almost entirely from changes in the time-dependent part of our flow, with the small-scale error term simply adding an almost constant offset to the prediction of the steady part of the flow at a given location.  Thus in {\it CoreFlo-LL.1} changes of the SV (field accelerations) are largely explained by changes (accelerations) in the flow interacting with the large-scale core field.   We come back to this point in section \ref{sec:disc_pulses}.

Histograms of the residuals between the SV data and {\it CoreFlo-LL.1} predictions, over the entire time period, are presented in Fig.~\ref{fig:hist_residuals}. We separate ground and virtual observatories as well as the corresponding vector components. The distributions are mostly symmetric but more long-tailed than for a Gaussian distribution of errors; this has been accounted for during the model estimation by use of a robust Tukey biweight scheme in the misfit function. The radial, north-south and east-west SV residuals have rms values of 2.07, 2.30 and 2.30\,nT/yr at the ground observatories and 1.20, 1.42 and 1.21\,nT/yr at the virtual observatories. These histograms provide further evidence that 
{\it CoreFlo-LL.1} is describing well the observed SV at mid and low latitudes.

\begin{figure}
	\centering
	\includegraphics[width=10cm]{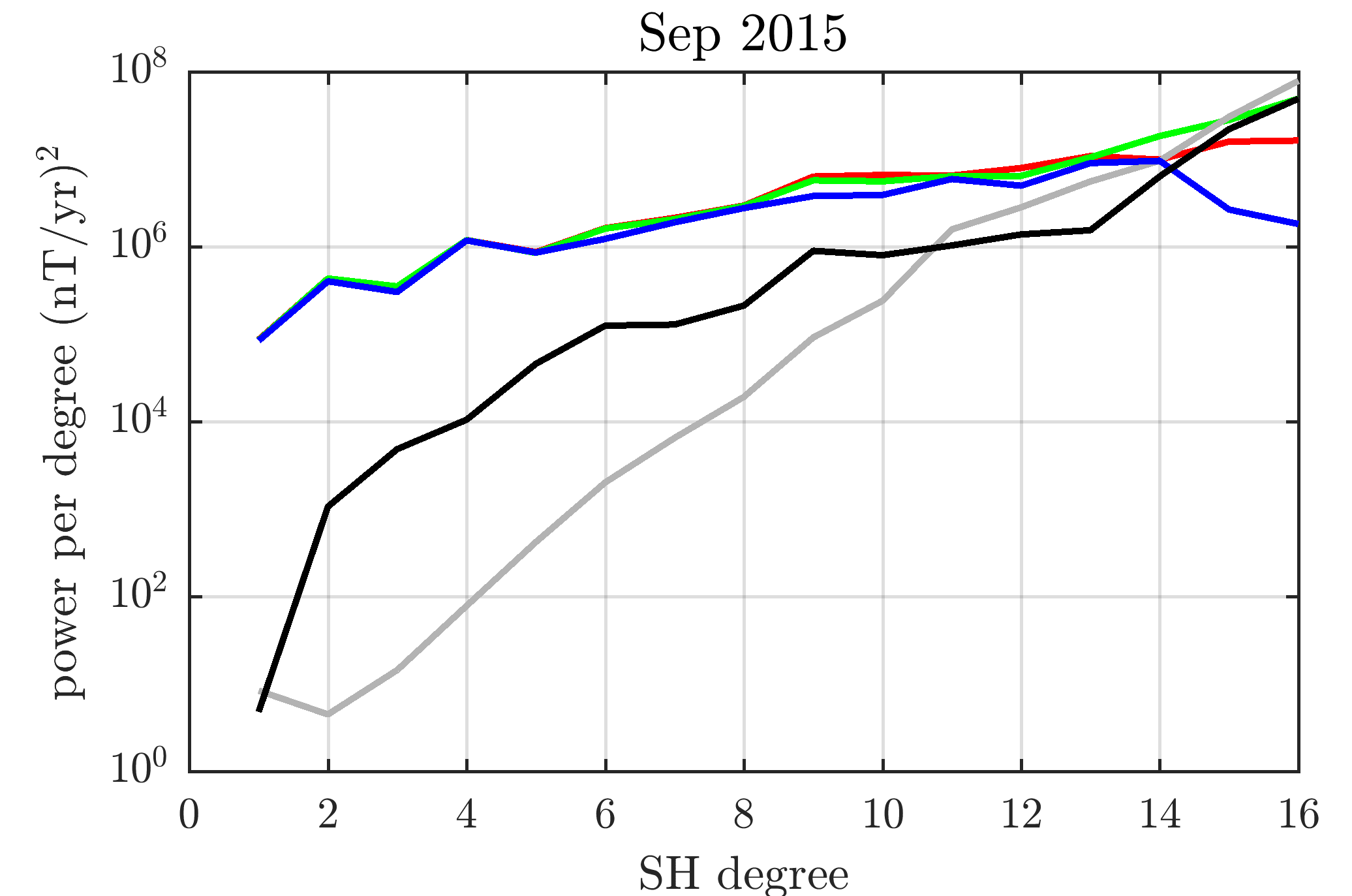}
	\caption{Power spectrum at the core surface in September 2015 of the SV computed from {\it CoreFlo-LL.1} (green), the SV generated by the modeled flow alone, i.e.~excluding the small-scale error (blue), the small-scale error (black), the CHAOS-6-x7 model (red), and the difference between our prediction and the CHAOS-6-x7 model (grey).}
	\label{fig:sv_spectrum}
\end{figure}

In Fig.~\ref{fig:sv_spectrum}, we show the SV power spectrum, taking September 2015 as an example, a time well within a period of good data coverage provided by the {\it Swarm} satellites.
The SV spectrum at the core surface estimated in the CHAOS-6-x7 model (red) is well reproduced by {\it CoreFlo-LL.1} (green), up until SH degree 11 where the power of the difference (grey) becomes of the same order of magnitude as the SV itself. The power in the small-scale error (black) is consistently less than the power in the flow-generated SV (blue), which confirms that the majority of the observed SV is reproduced by the {\it CoreFlo-LL.1} flow interacting with the large-scale core field. 
\begin{figure*}
	\centering
	\includegraphics[width=16cm]{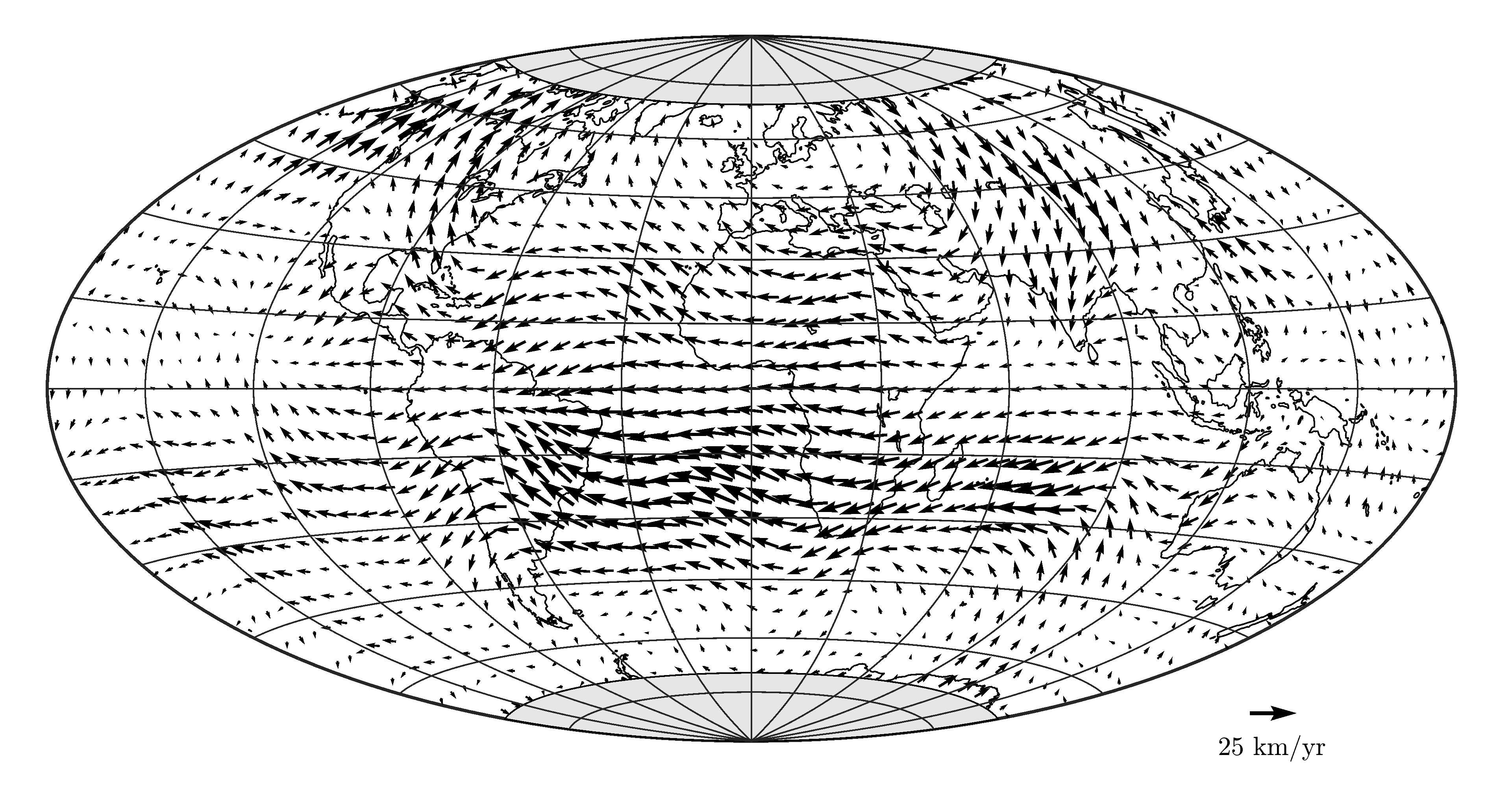}
	\caption{Time-averaged flow at the core surface. The projection is Hammer-Aitoff. The grey area masks the region inside the tangent cylinder where our flow parameterization is inappropriate.}
	\label{fig:flow_taverage}
\end{figure*}

\subsection{Time-averaged flow}
\begin{figure}
	\centering
	\includegraphics[width=10cm]{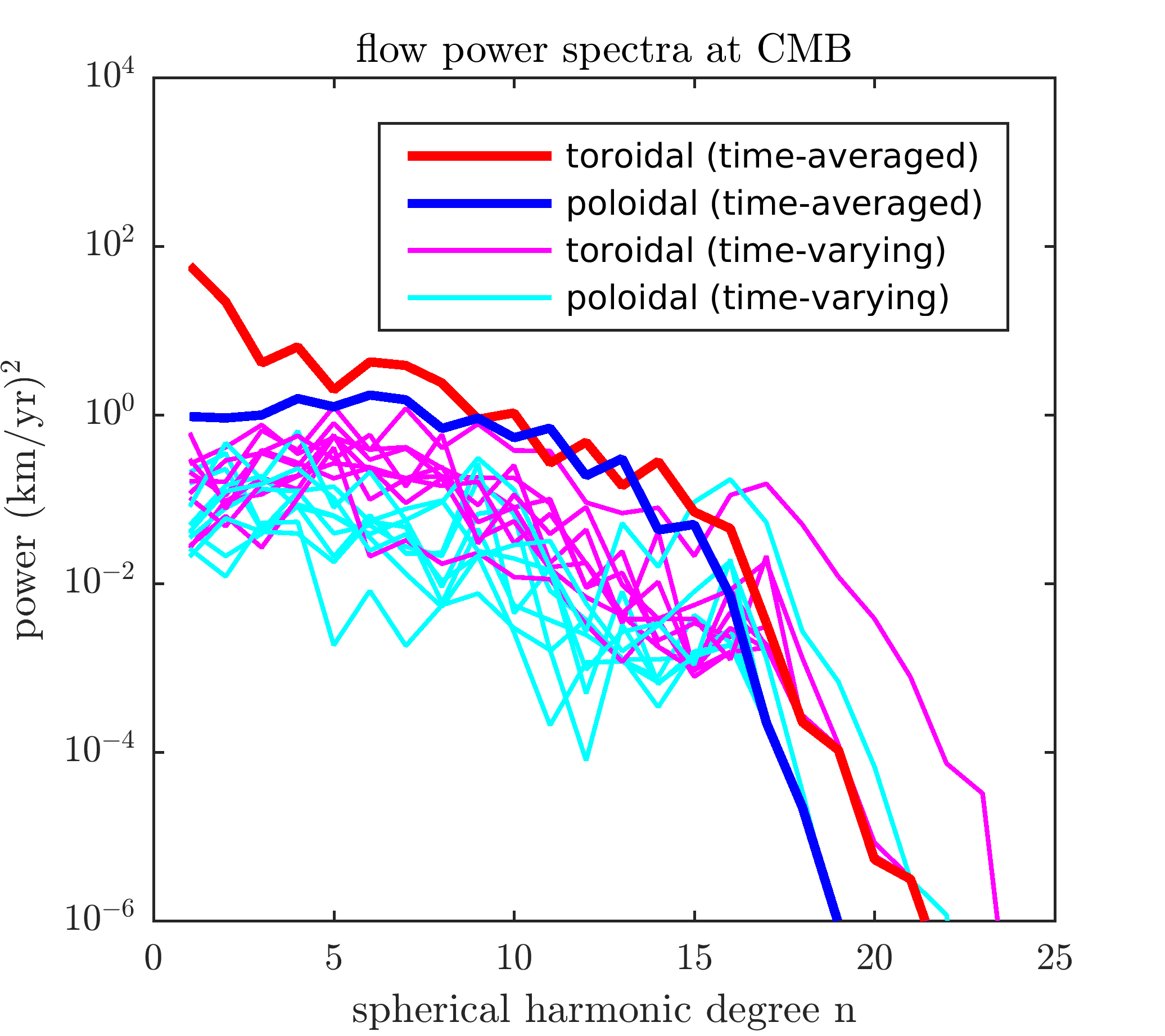}
	\caption{Toroidal-poloidal power spectrum of the time-averaged flow (thick lines) and time-varying flow (thin lines, every 2-years starting in September 2000).}
	\label{fig:torpol_spectrum}
\end{figure}

 In Fig.~\ref{fig:flow_taverage} we present the flow from {\it CoreFlo-LL.1}  time-averaged over the modeled time interval. The regions poleward of the tangent cylinder in this and subsequent plots have been masked in grey to emphasise that the determined flows are not reliable in this region. The basic time-averaged flow structure is that of a planetary-scale eccentric gyre familiar from previous studies \cite[e.g.][]{pais2008quasi}, with equatorward flow around $100^\circ$E, then strong and large-scale westward flow at mid and low latitudes under the Atlantic, slightly stronger in the southern hemisphere in agreement with previous findings \cite[][]{amit2013differences,Baerenzung2016}, and poleward flow under North America.   Note that because our background flow is described by both $\ES$ and $\EA$ flow modes, departures from equatorial symmetry, including equator crossing, are allowed.  Nonetheless the basic gyre structure, including its meridional flows, are recovered despite the fact that we only use SV data from mid and low latitudes.  We find that $82\,\%$ of the time-averaged power at the core surface results from equatorially symmetric flow, which is remarkably close to the value of $82\,\%$ found by \cite{Baerenzung2014} via a probabilistic inversion scheme.  Fig.~\ref{fig:torpol_spectrum} shows the toroidal-poloidal SH power spectrum of the time-averaged flow as well as several examples of the time-varying flow.  The toroidal spectrum decreases gradually and the poloidal spectrum is relatively flat up to degree 12, both decrease more rapidly beyond degree 14.

\subsection{Time-varying flow}

By removing the time-average from the instantaneous flow, we extract the time-varying part of the flow, which is by construction primarily equatorially symmetric in {\it CoreFlo-LL.1}. Looking at the amplitudes of the modes representing our time-dependent flow we find that the large-scale geostrophic polynomials (small $k$) and large-scale QG modes (small $m$ and $k$) contribute most to the time-dependent part of the flow (0.93 of its time-averaged power due to equatorially symmetric flow). More complex modes (high $m$ and $k$) are in contrast less pronounced; this is at least partly a consequence of the applied spatial regularization.

\begin{figure*}
	\centering
	\includegraphics[width=8.75cm]{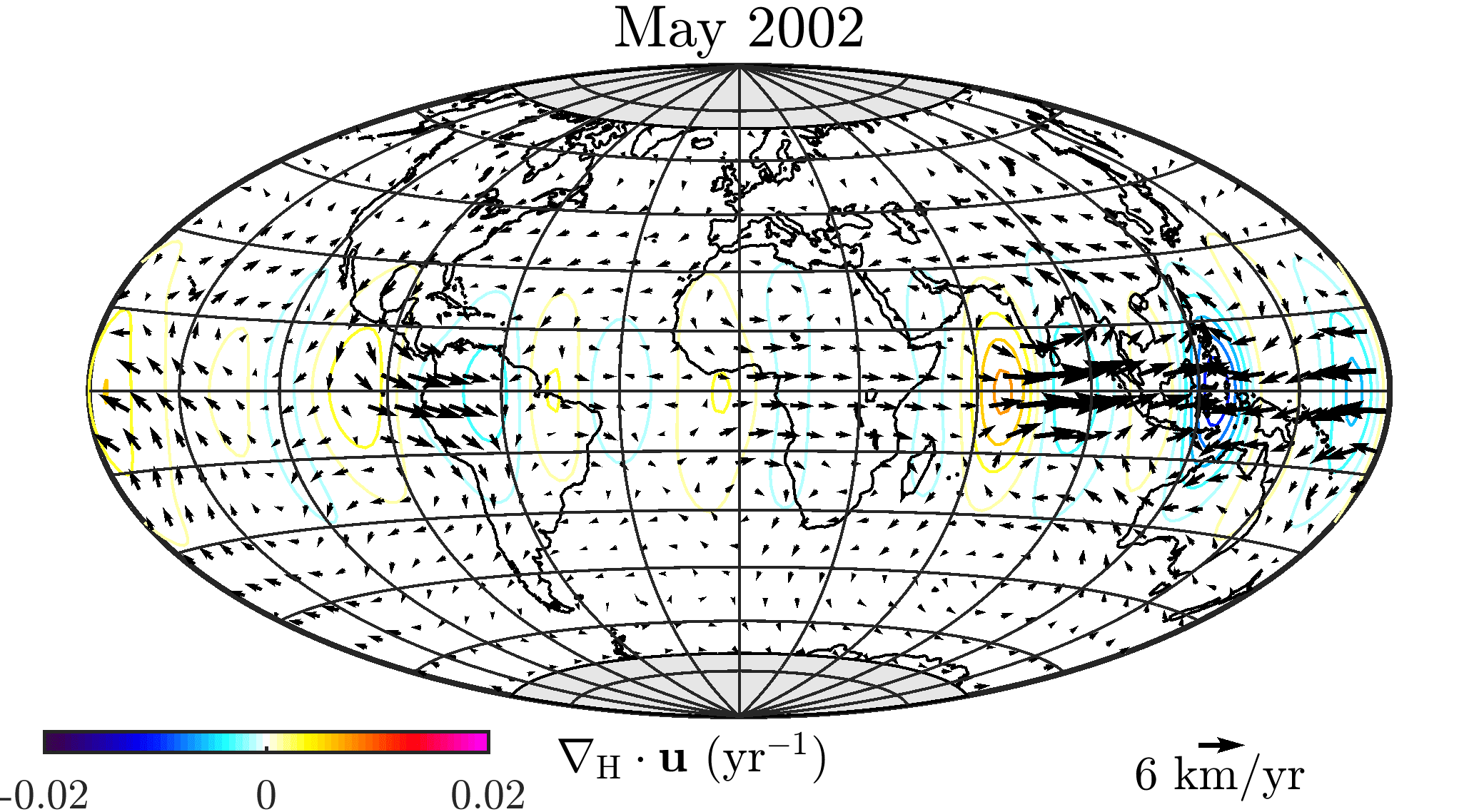}
	\includegraphics[width=8.75cm]{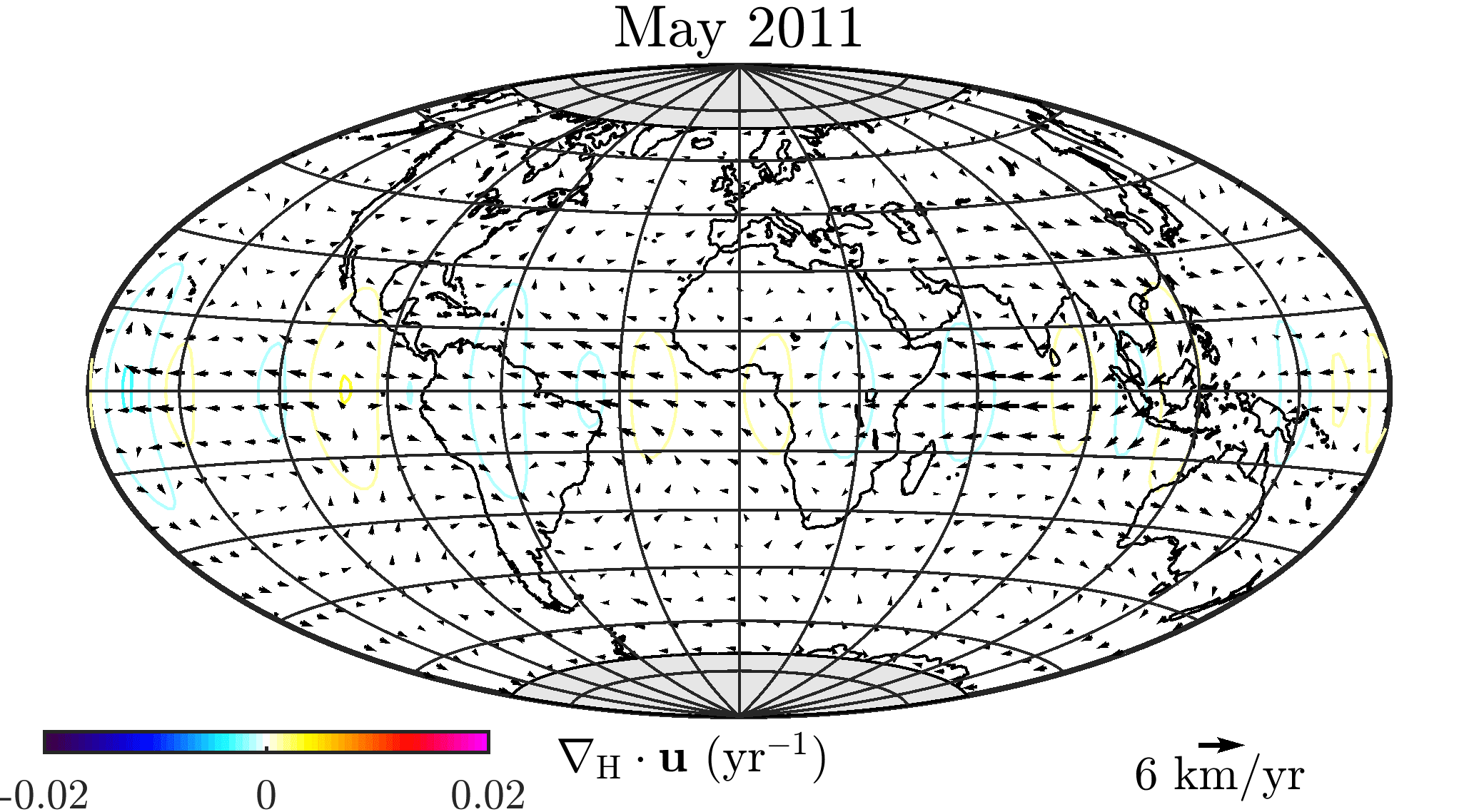}\\	
	\vspace{0.2cm}
	\includegraphics[width=8.75cm]{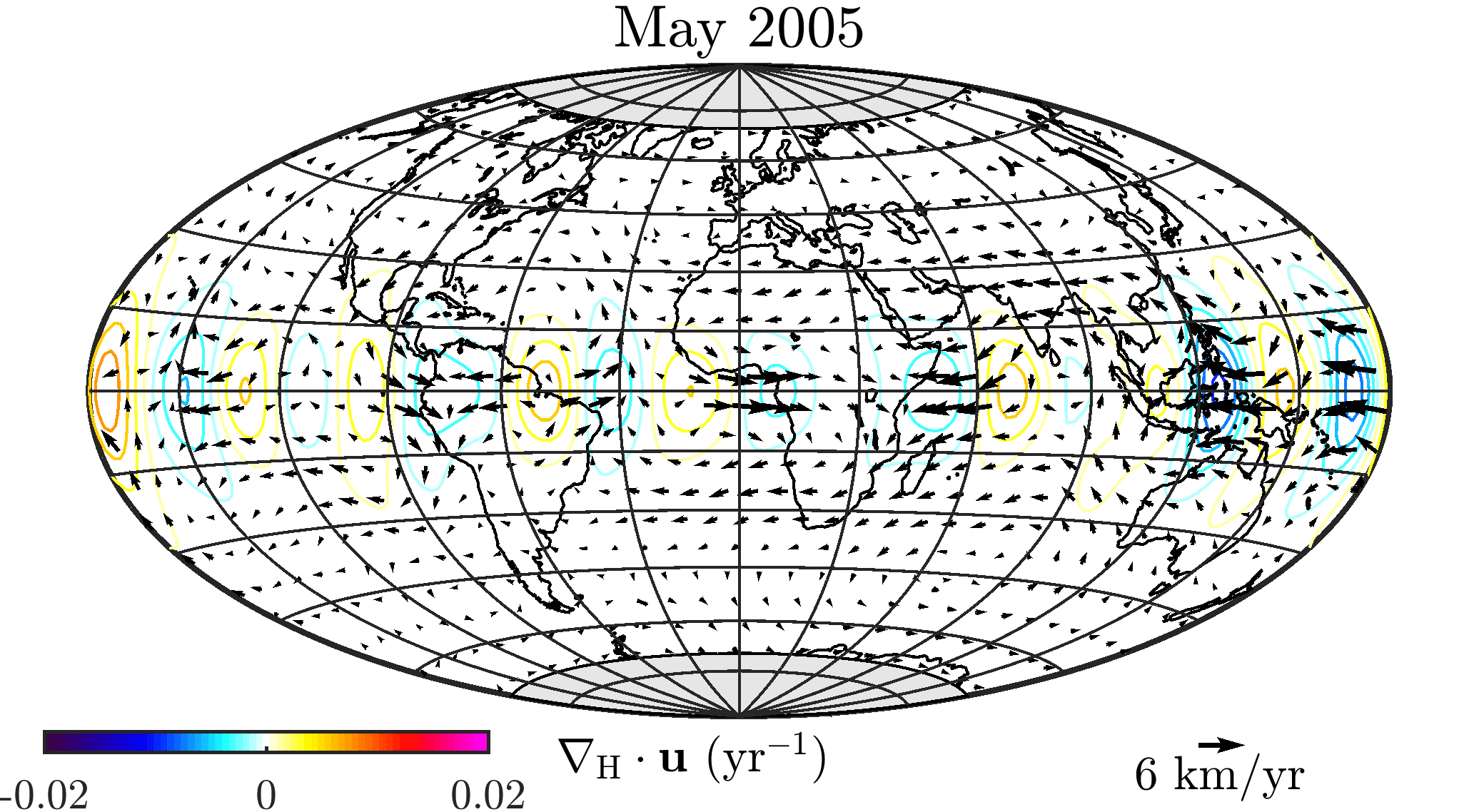}
	\includegraphics[width=8.75cm]{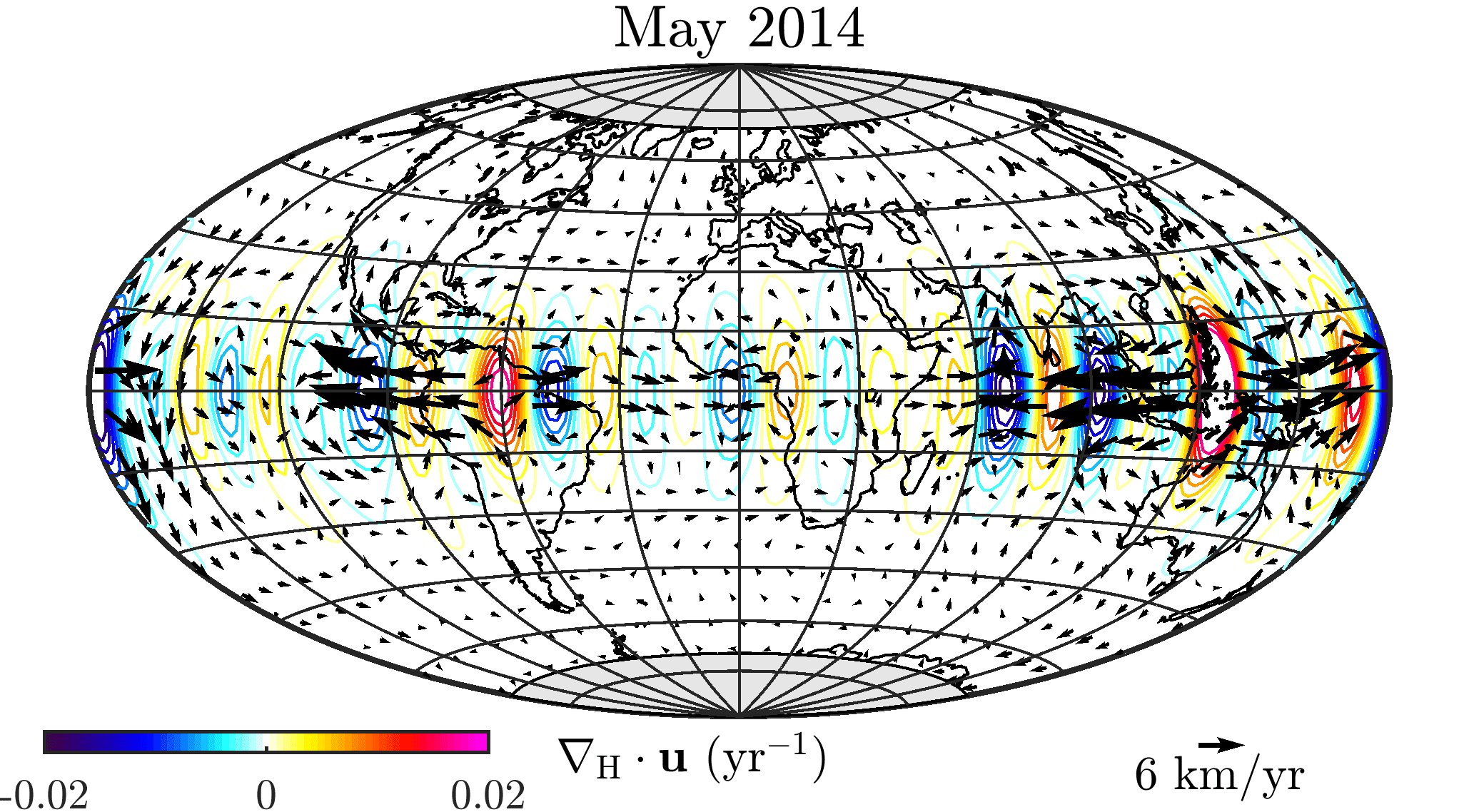}\\	
	\vspace{0.2cm}
	\includegraphics[width=8.75cm]{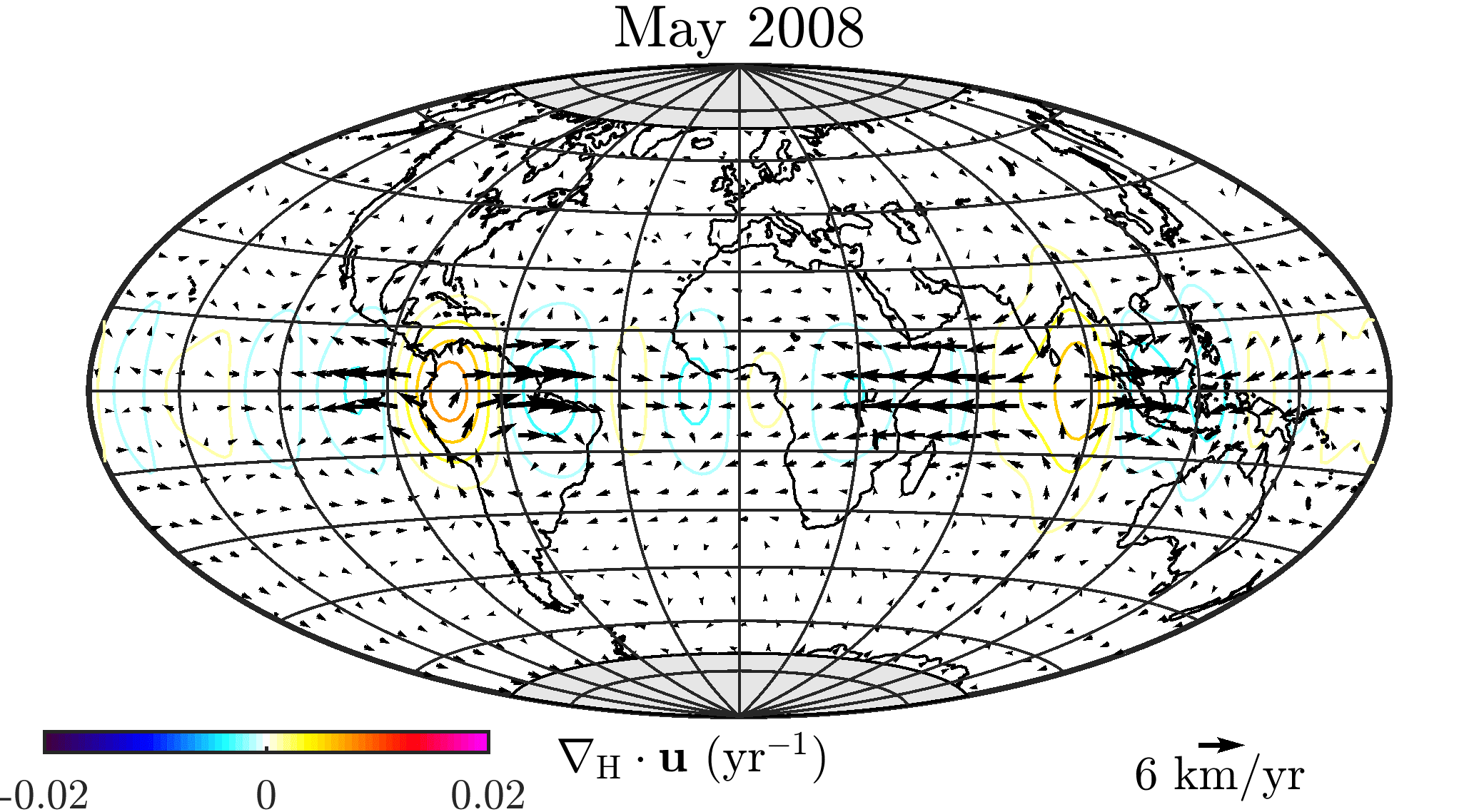}
	\includegraphics[width=8.75cm]{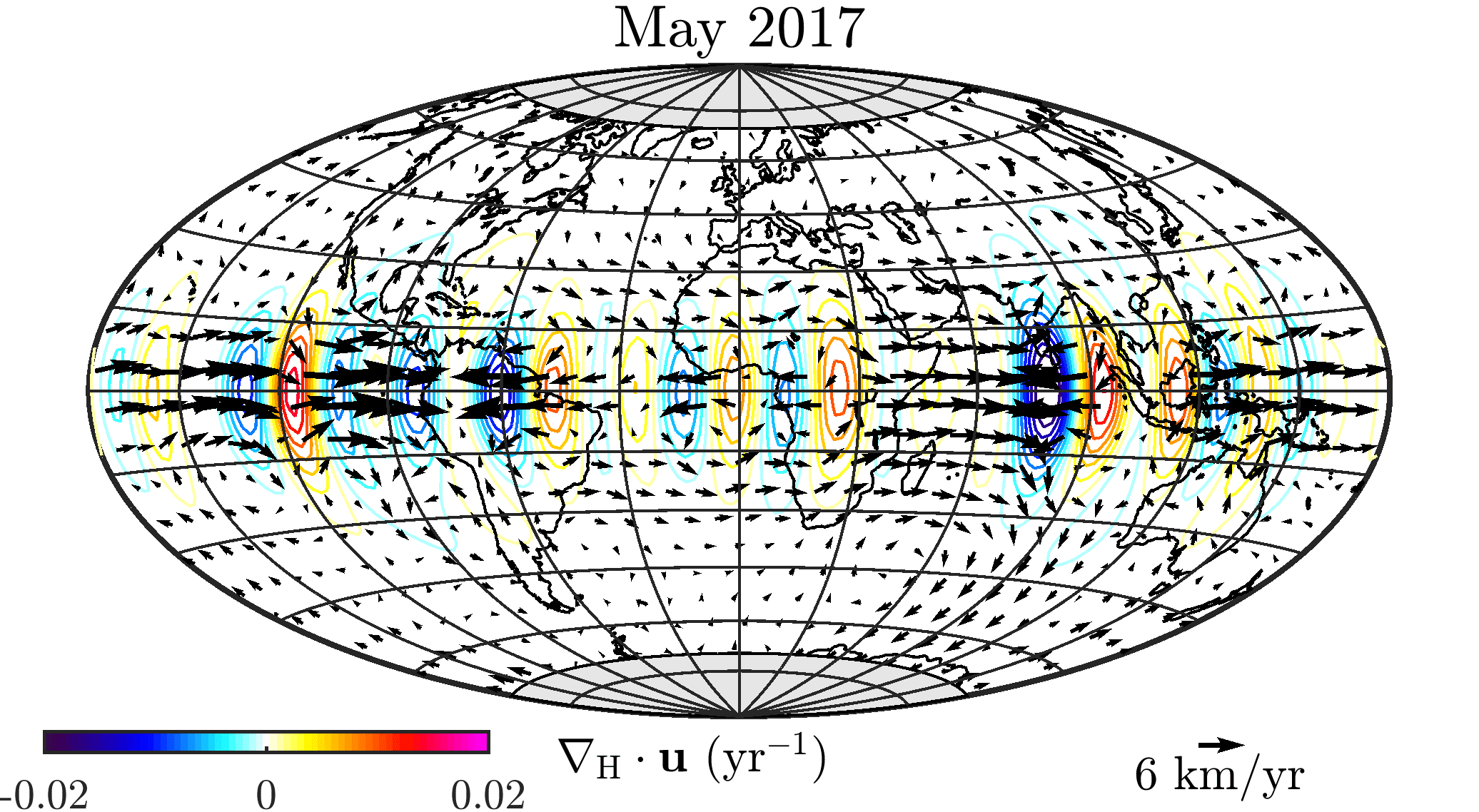}
	\caption{Snapshots of the time-varying core surface flow (arrows) and the related horizontal divergence (contours) during the observation period from 2000 to 2018. The projection is Hammer-Aitoff and the grey area masks the region inside the tangent cylinder.}
	\label{fig:flow_tdep}
\end{figure*}

In Fig.~\ref{fig:flow_tdep} snapshots of the time-varying flow at three year intervals are presented to illustrate how the flow changes over the studied time interval.
We find strong non-zonal azimuthal flow in the equatorial region between latitudes $15^\circ\text{N and S}$, that is particularly enhanced around Indonesia ($90^\circ\text{E}$) and the northern part of South America ($90^\circ\text{W}$). The overall amplitude of the time-varying flow apparently increases in the snapshots after 2014.  This is a data-driven effect, since after 2014 is when high quality data from the {\it Swarm} satellite mission are available. Also noteworthy are the time alternating patterns of converging and diverging azimuthal flow around the equator region that are associated with regions of downwelling and upwelling, respectively. For example, the time-varying flow under northern South America was converging in 2004, diverging in 2008, weakly converging in 2011, and then more strongly diverging in 2014 and strongly converging in 2017.

Next, as an independent test of the time-variations of the geostrophic part of {\it CoreFlo-LL.1}, we computed the predicted change in the LOD and compared with geodetic observations\footnote{{\tt https://www.iers.org/IERS/EN/DataProducts/EarthOrientationData/eop.html}}, that have been corrected for known atmospheric and oceanic signals filtered using a one year moving average \cite[as in][]{Gillet2015a}.
We applied a moving-average to the LOD changes predicted by our model to be able to compare with the observations. We find a trend of generally increasing LOD from 2003 to 2006 in agreement with the trend in the observations. About this trend our flow predicts a maximum in the change of LOD in 2007 and again in 2016, similar to the observations, but with lower amplitude especially in 2007.

\begin{figure*}
	\centering
	\includegraphics[width=16.0cm]{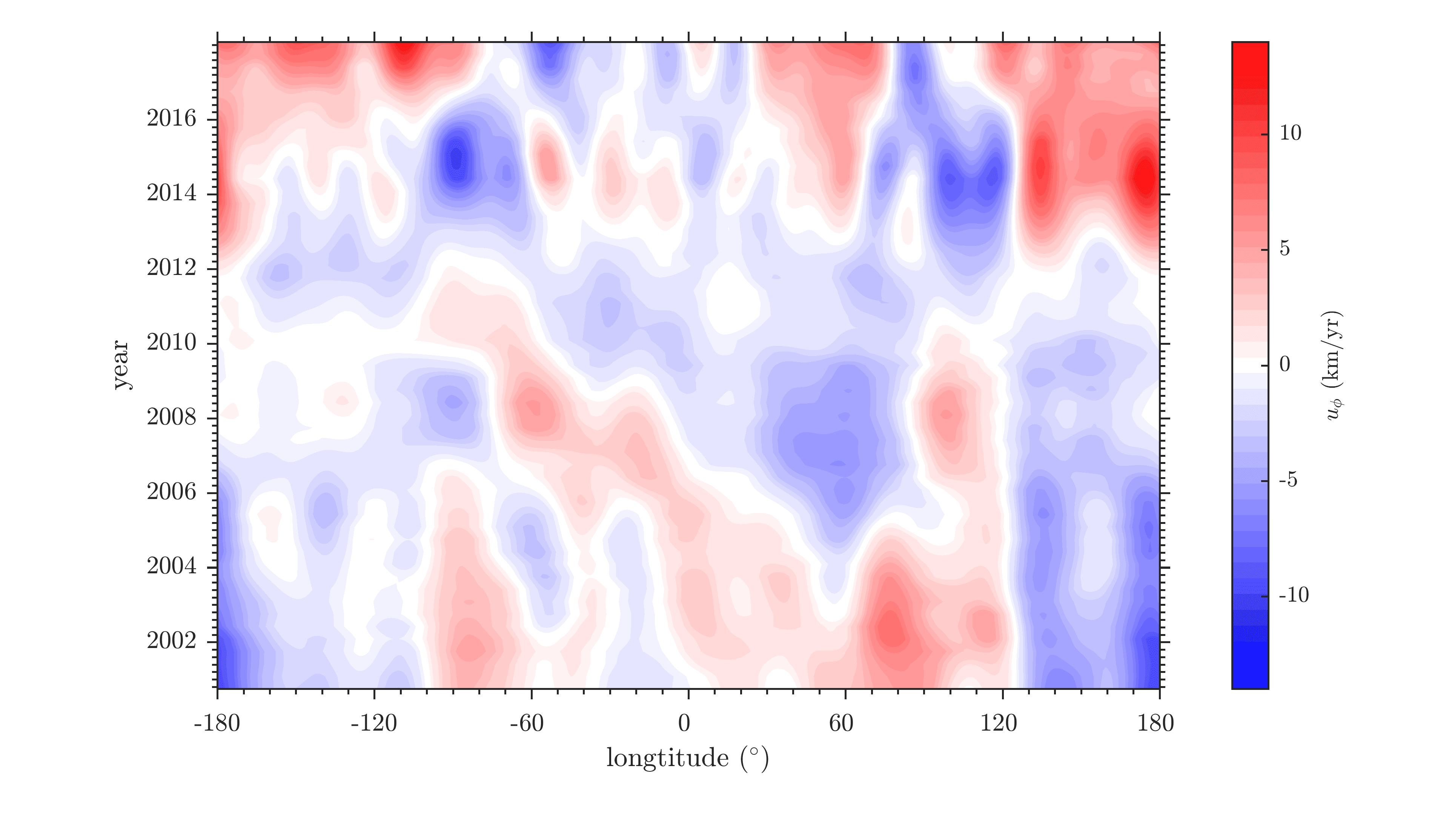}
	\caption{Time varying part of the azimuthal flow averaged at low latitude as a function of longitude and time. The azimuthal flow was averaged over latitudes $15^\circ\text{N, S}$ of the equator in approx. $1^\circ$-bins of longitude.}
	\label{fig:flow_tdep_av_eqline}
\end{figure*}

In order to more clearly visualize the interesting converging and diverging patterns of azimuthal flow at low latitude, we finally plot the average of the azimuthal component of the time-varying flow over $15^\circ\text{N, S}$ latitude in Fig.~\ref{fig:flow_tdep_av_eqline} as a function of longitude and time.
Around 130$^\circ$E, a region of initially converging azimuthal flow changes sign to become an azimuthal flow divergence around 2011. The azimuthal flow around northern South America ($90^\circ$W) changes its direction more frequently in time at around a relatively fixed location on the equator. There is some indication of westward propagation of azimuthal flow features with alternating signs between  $90^\circ$W and $45^\circ$E especially from 2005 to 2015. In the next section we discuss in detail how this time varying flow gives rise to secular acceleration pulses and geomagnetic jerk events.

\section{Discussion}
\label{sec:disc}

\subsection{The origin of core surface field acceleration pulses}
\label{sec:disc_pulses}

\begin{figure*}
	\centering
	\includegraphics[width=16cm]{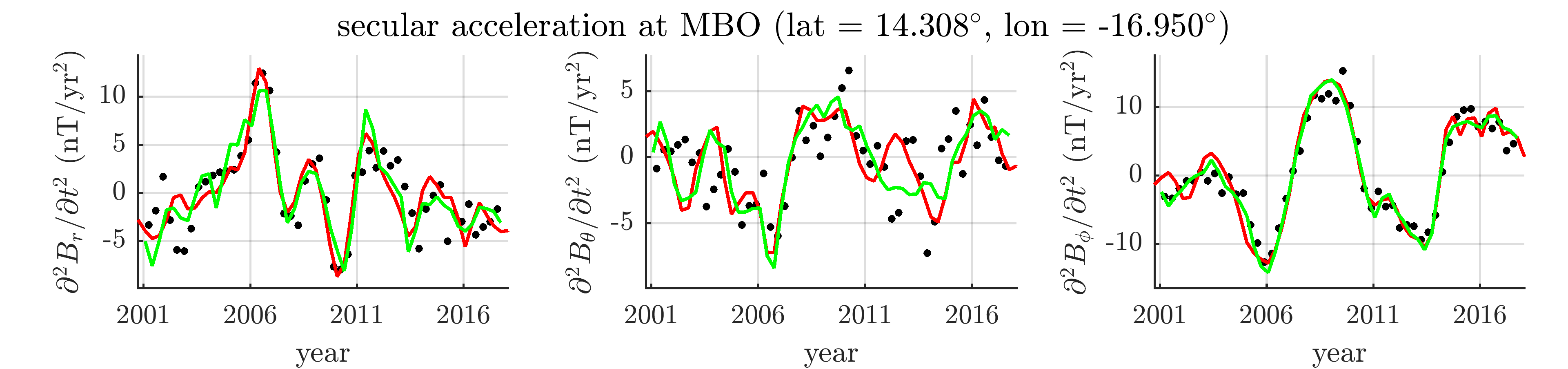}\\[-0.4cm]
	\includegraphics[width=16cm]{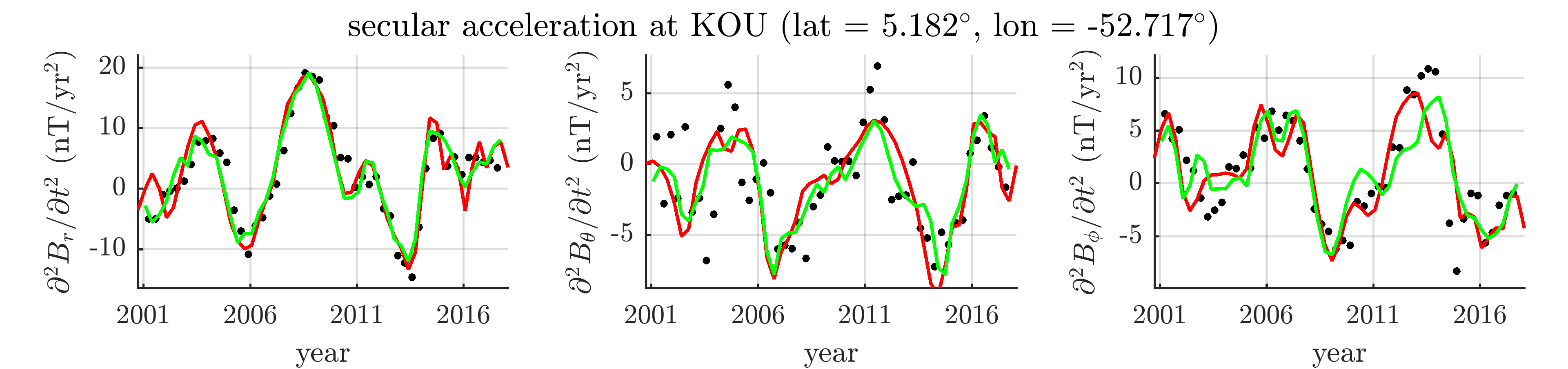}\\[-0.4cm]
	\includegraphics[width=16cm]{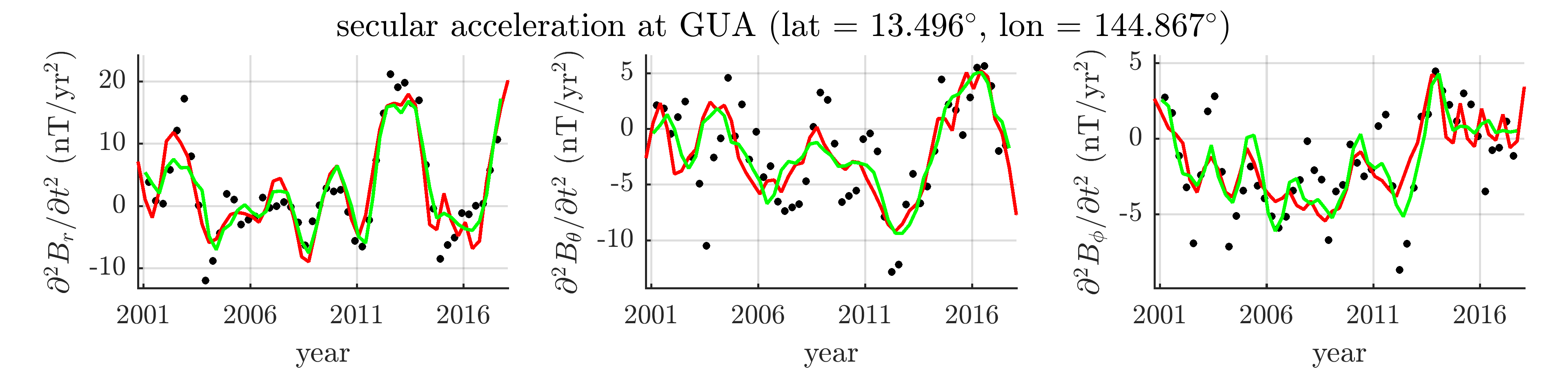}\\[-0.4cm]
	\includegraphics[width=16cm]{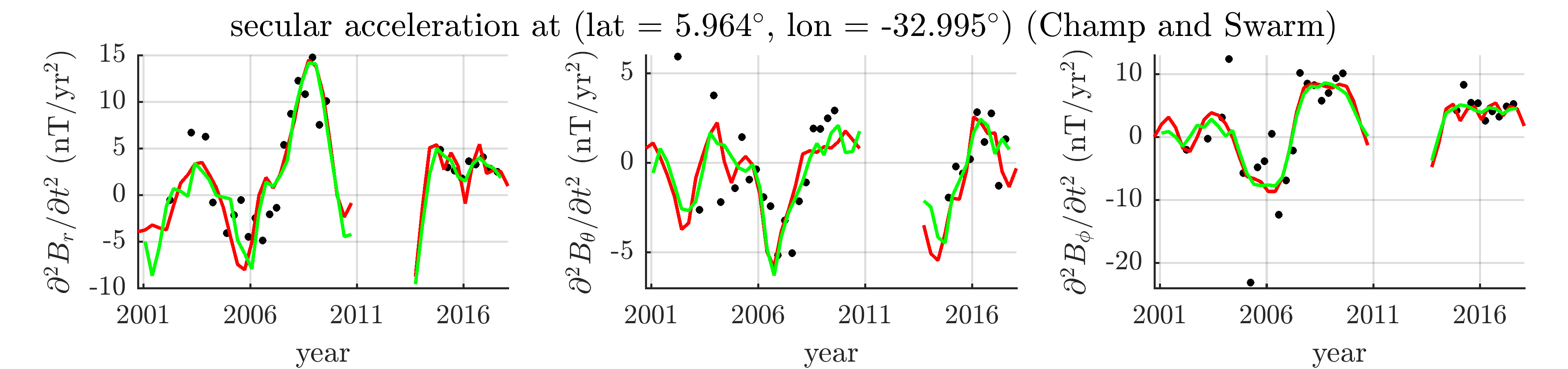}\\[-0.4cm]
	\includegraphics[width=16cm]{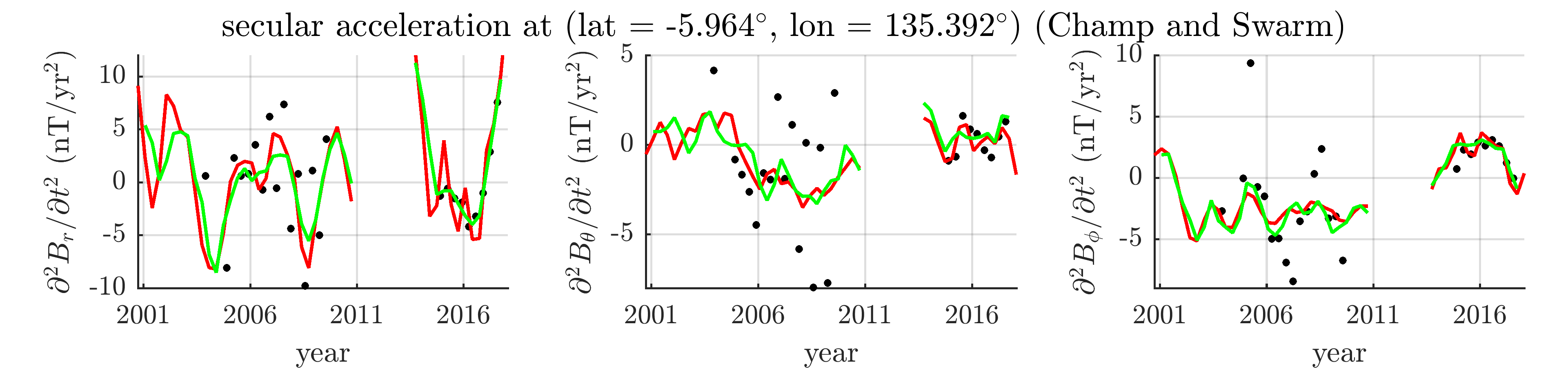}\\[-0.4cm]
	\includegraphics[width=16cm]{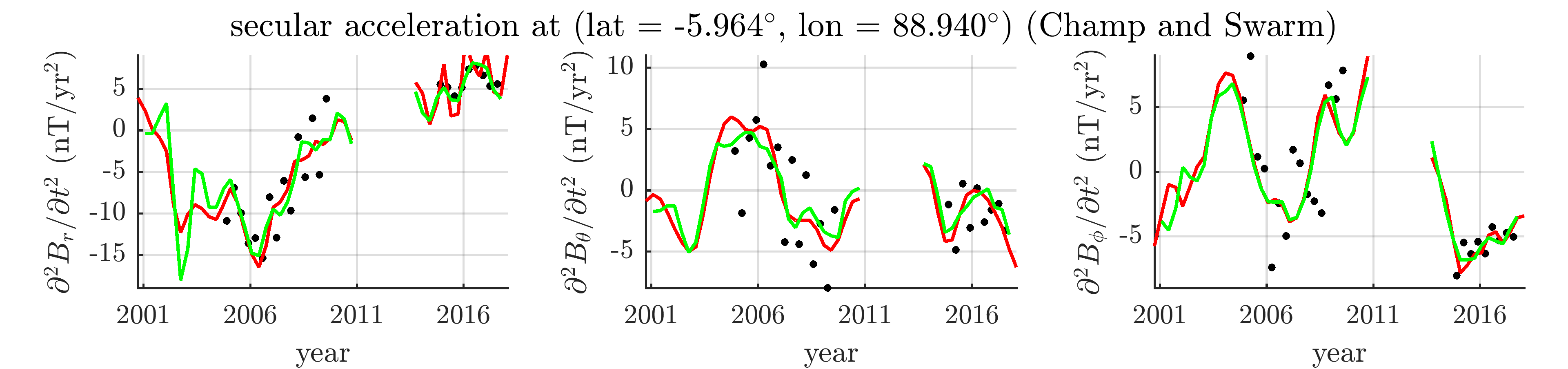}
	\caption{SA time series for the ground and virtual observatories in Fig.~\ref{fig:timeseries_go_vo}, showing the observed SA (black), the CHAOS-6-x7 SA prediction (red) and our model's SA prediction (green), both truncated at spherical harmonic degree 9.}
	\label{fig:sa_timeseries_go_vo}
\end{figure*}

The future evolution of the geomagnetic field is difficult to predict due to intermittent pulses of field acceleration that cause linear extrapolations to fail.  It is therefore of significant interest to investigate  whether our time-dependent flows can reproduce observed pulses of field acceleration, and if so by what process.  In this respect it is important to recall that our flows were not derived from SV computed from spherical harmonic field models (e.g.~CHAOS or GRIMM type models where the applied regularization plays an important role in the time-dependence of the core surface field acceleration), but directly from ground and satellite-based SV time series.   Fig.~\ref{fig:sa_timeseries_go_vo} presents the observed field accelerations at Earth's surface and at satellite altitude, calculated by taking annual differences of the SV series from Fig.~\ref{fig:timeseries_go_vo}, together with the field accelerations predicted by the {\it CoreFlo-LL.1} model and the CHAOS-6-x7 model.  The {\it CoreFlo-LL.1} predictions match the observations well with a weighted rms misfit of 2.75, 3.42 and 3.42\,nT/yr$^2$ to the accelerations in $B_r$, $B_\theta$ and $B_\phi $, compared to 3.00, 3.49, 3.50\,nT/yr$^2$ for CHAOS-6-x7. Notable pulses of high amplitude SA are seen in $\partial^2 B_\phi / \partial^2 t$ at MBO in 2006 and 2009, and in $\partial^2 B_r / \partial^2 t$ in GUA near 2013.  The SA observed at the satellite-based virtual observatories, including those at low latitudes, as shown for example in Fig.~\ref{fig:sa_timeseries_go_vo}, are also generally well fit.  From this we conclude that {\it CoreFlo-LL.1} accounts well for mid and low latitude SA observed at ground and satellite altitude.

\begin{figure*}
	\centering
	\includegraphics[height=6cm]{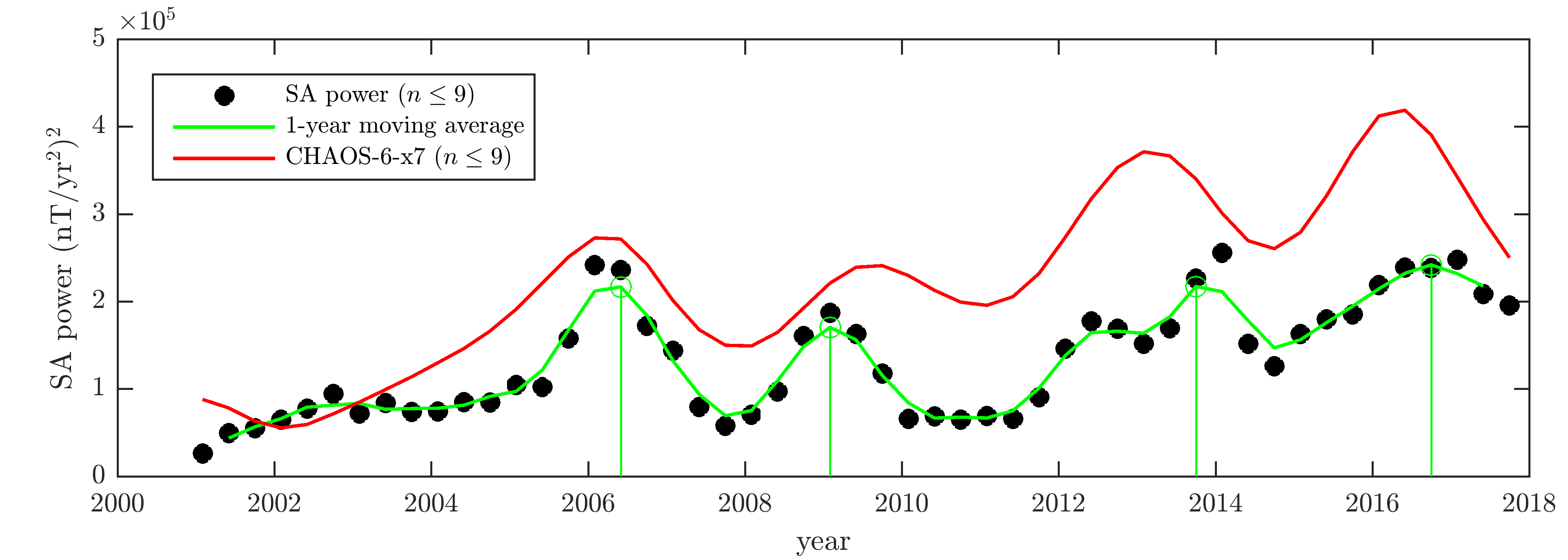}
	\caption{SA power for SH degrees $n\leq 9$ predicted by the {\it CoreFlo-LL.1}  flow at the core surface as a function of time (black), 1-year moving average (green) and CHAOS-6-x7 prediction (red). The vertical lines indicate times of enhanced SA power, for which we plot model snapshots in Fig.~\ref{fig:flow_acc}.}
	\label{fig:sa_norm}
\end{figure*}

Next, we examine the occurrence and form of the SA pulses at the core surface, comparing those predicted by our flow model with those found in CHAOS-6-x7. Fig.~\ref{fig:sa_norm} presents the time evolution of the SA power integrated over the core surface \cite[e.g.][]{Finlay2015}, calculated up to SH degree 9 from our modeled flow alone (i.e. without small-scale error contribution) and, for comparison, the same quantity from the CHAOS-6-x7 model.   Peaks of the SA power around January 2006, January 2009, September 2013 and September 2016 correspond to well-known SA pulses  at the core surface \cite[]{Chulliat2010,Finlay2015,chulliat2014geomagnetic}. There are however differences in the times of the pulse peaks of up to a year compared to CHAOS-6-x7. This may be, at least in part, due to the strong temporal regularization applied when constructing CHAOS-6-x7 (and similar models) which results in an averaging in time of the true field that depends on the spherical harmonic degree \cite[]{olsen2009chaos}, or because our flow does not predict well the SA at high latitudes that will contribute to the SA power in the case of the field model. 

\begin{figure*}
	\centering
	\includegraphics[width=8.25cm]{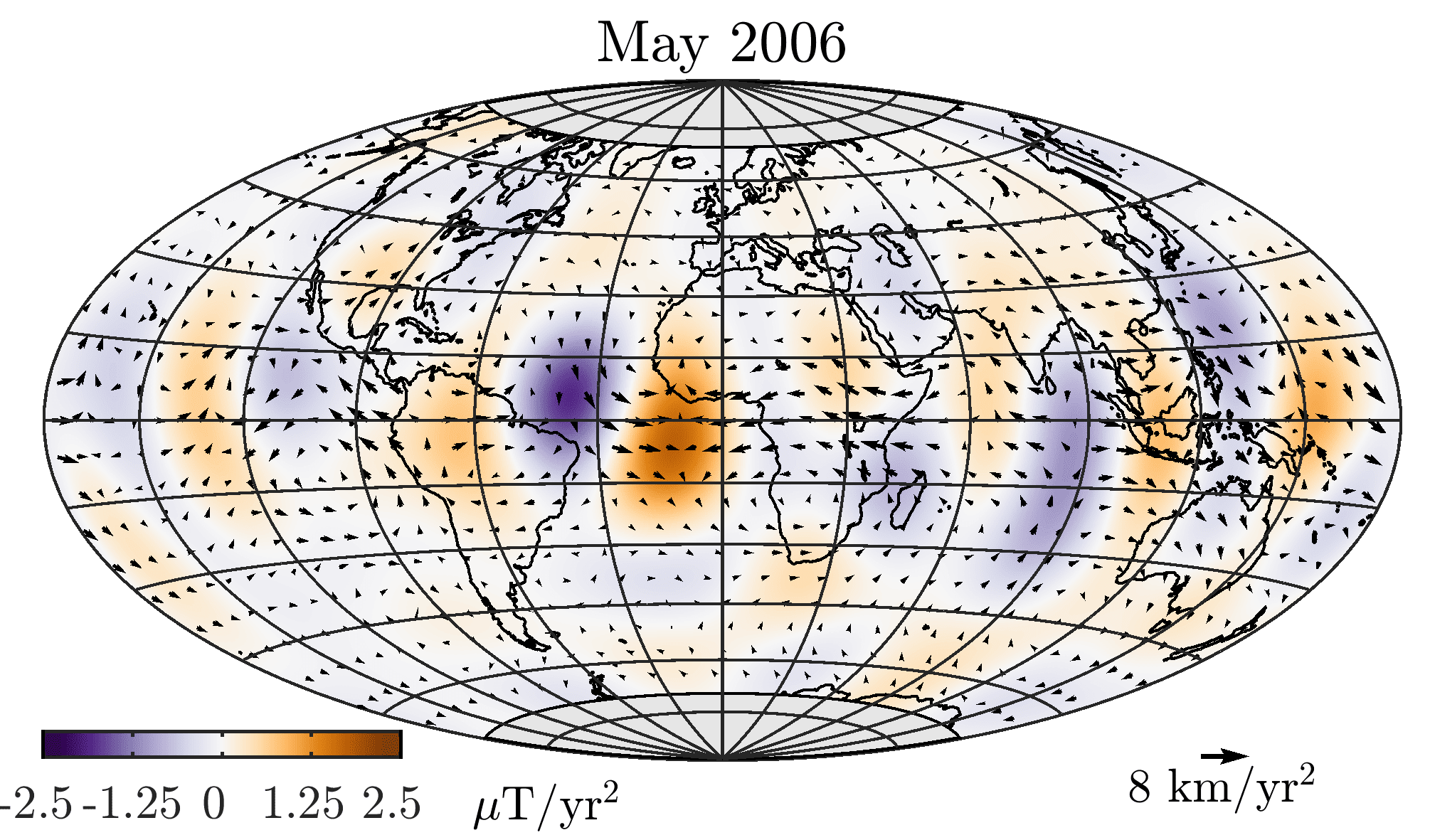}
	\includegraphics[width=8.25cm]{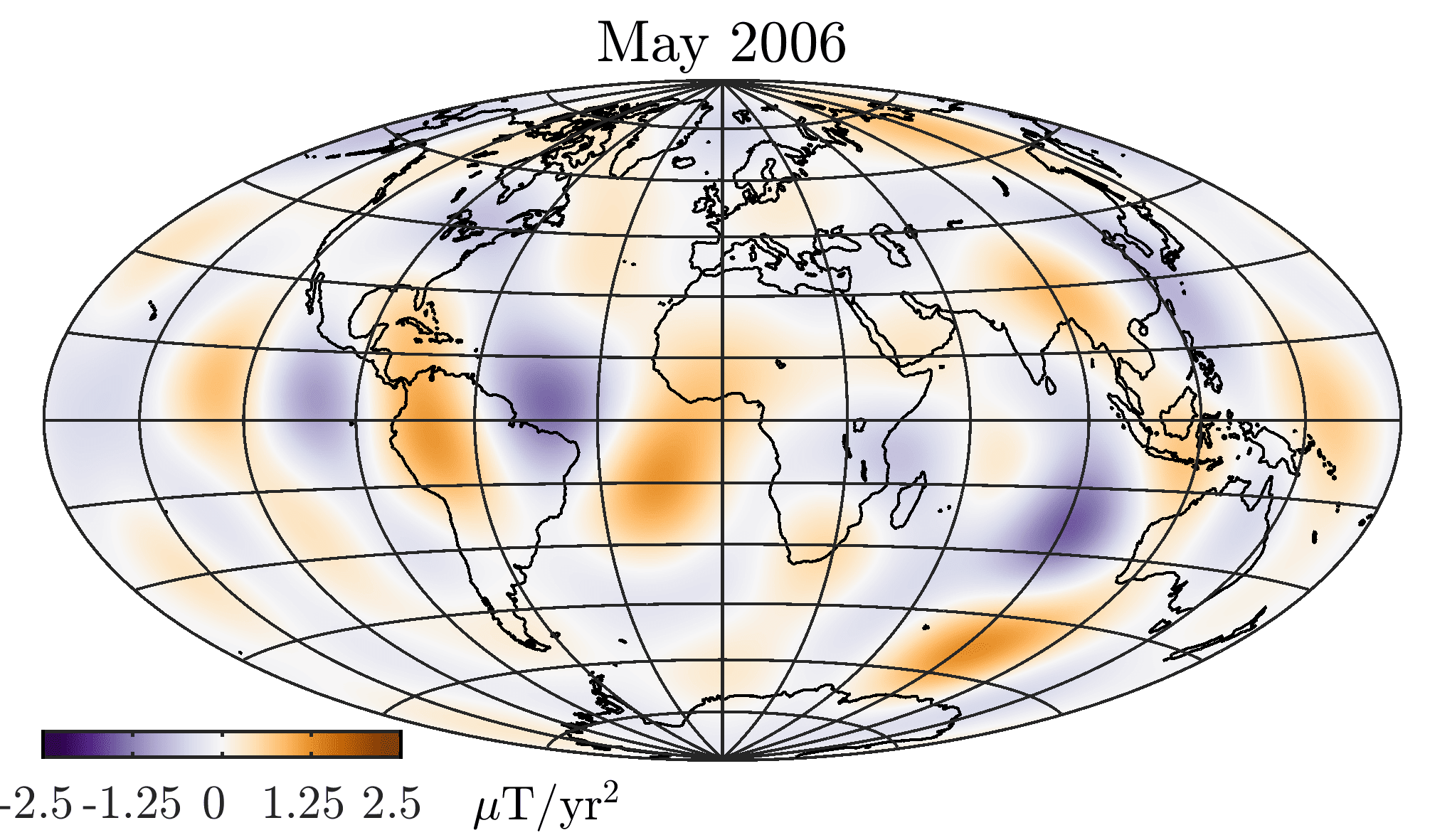}\\
	\vspace{0.5cm}
	\includegraphics[width=8.25cm]{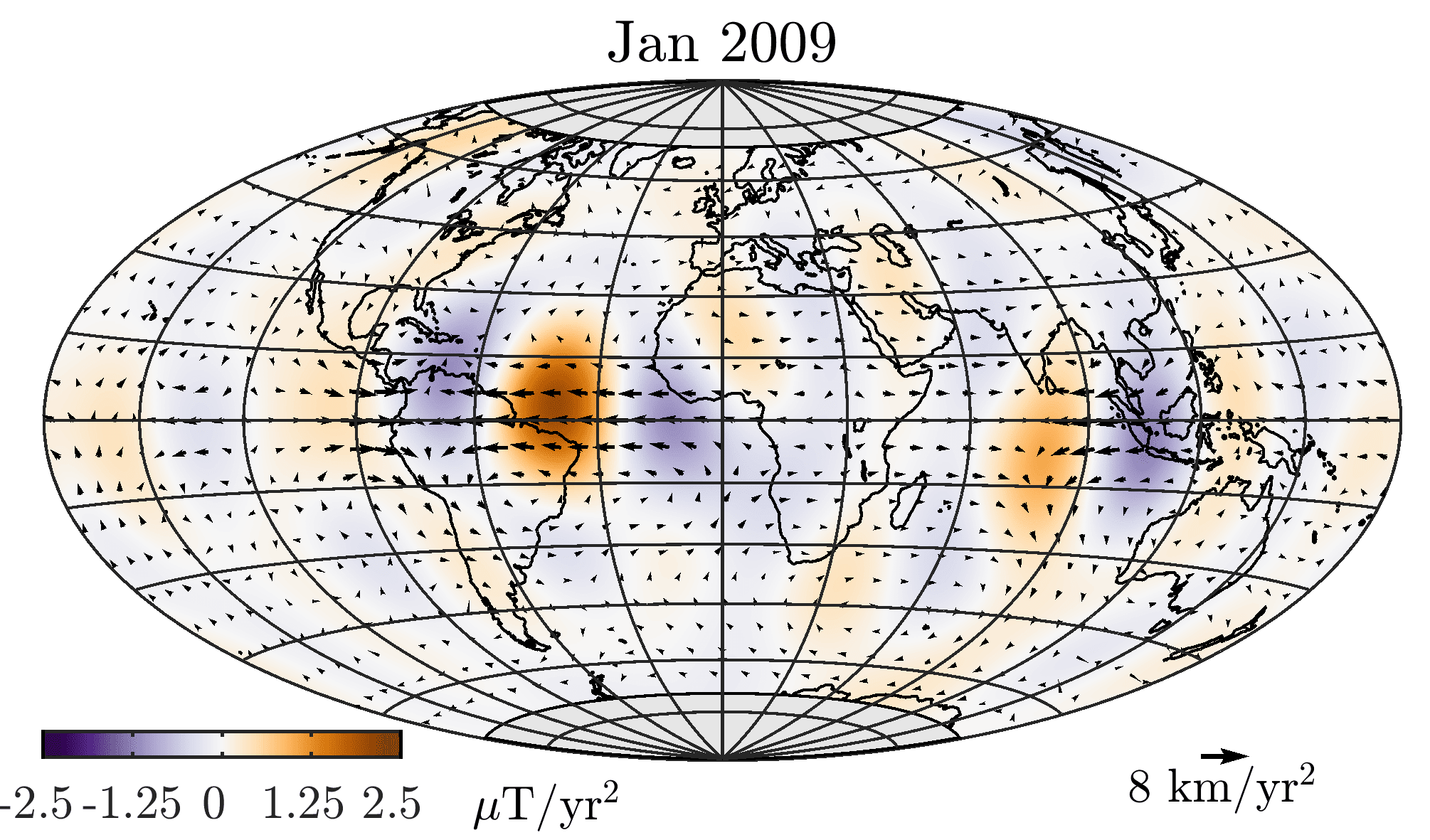}
	\includegraphics[width=8.25cm]{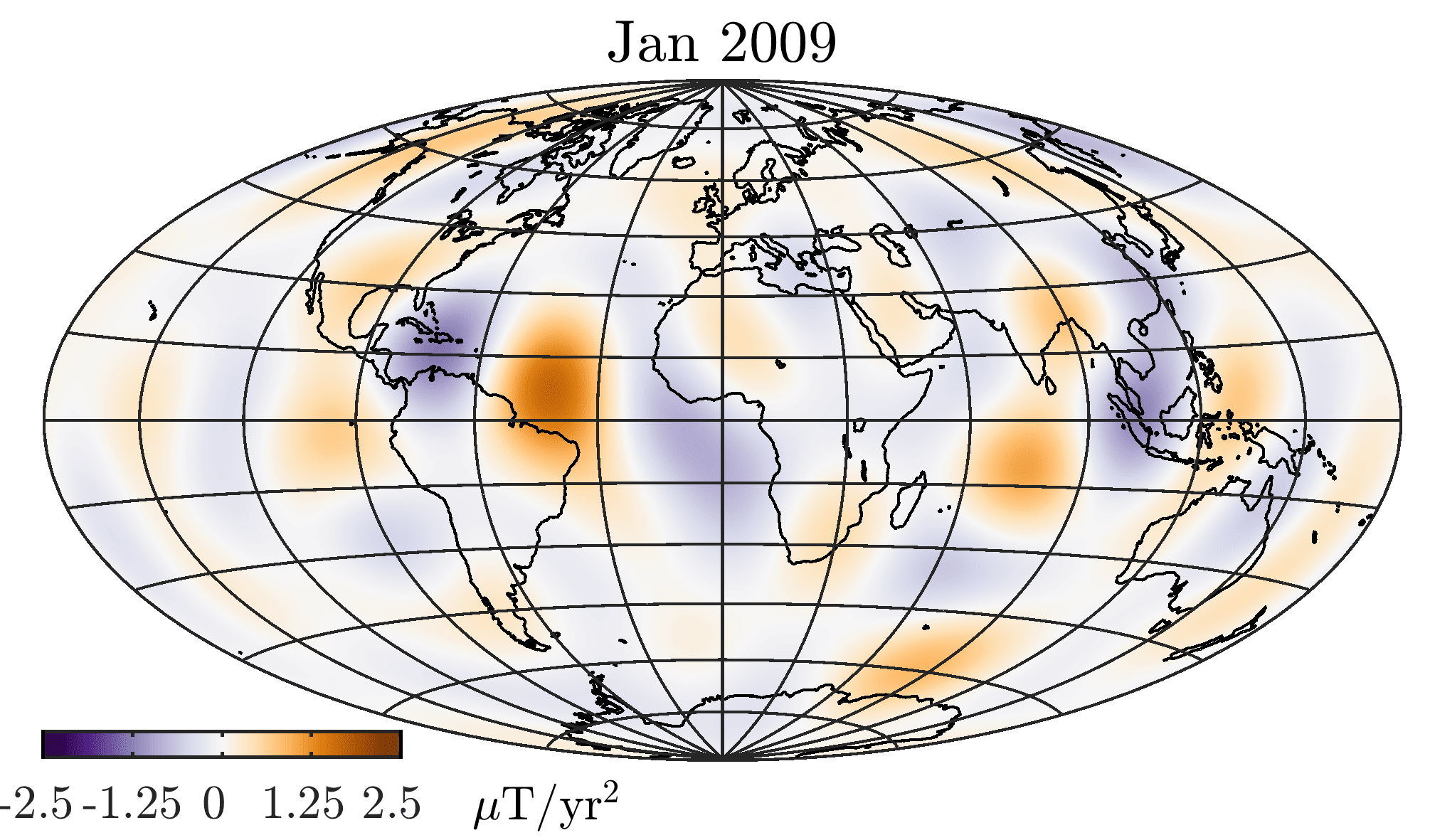}\\
	\vspace{0.5cm}
	\includegraphics[width=8.25cm]{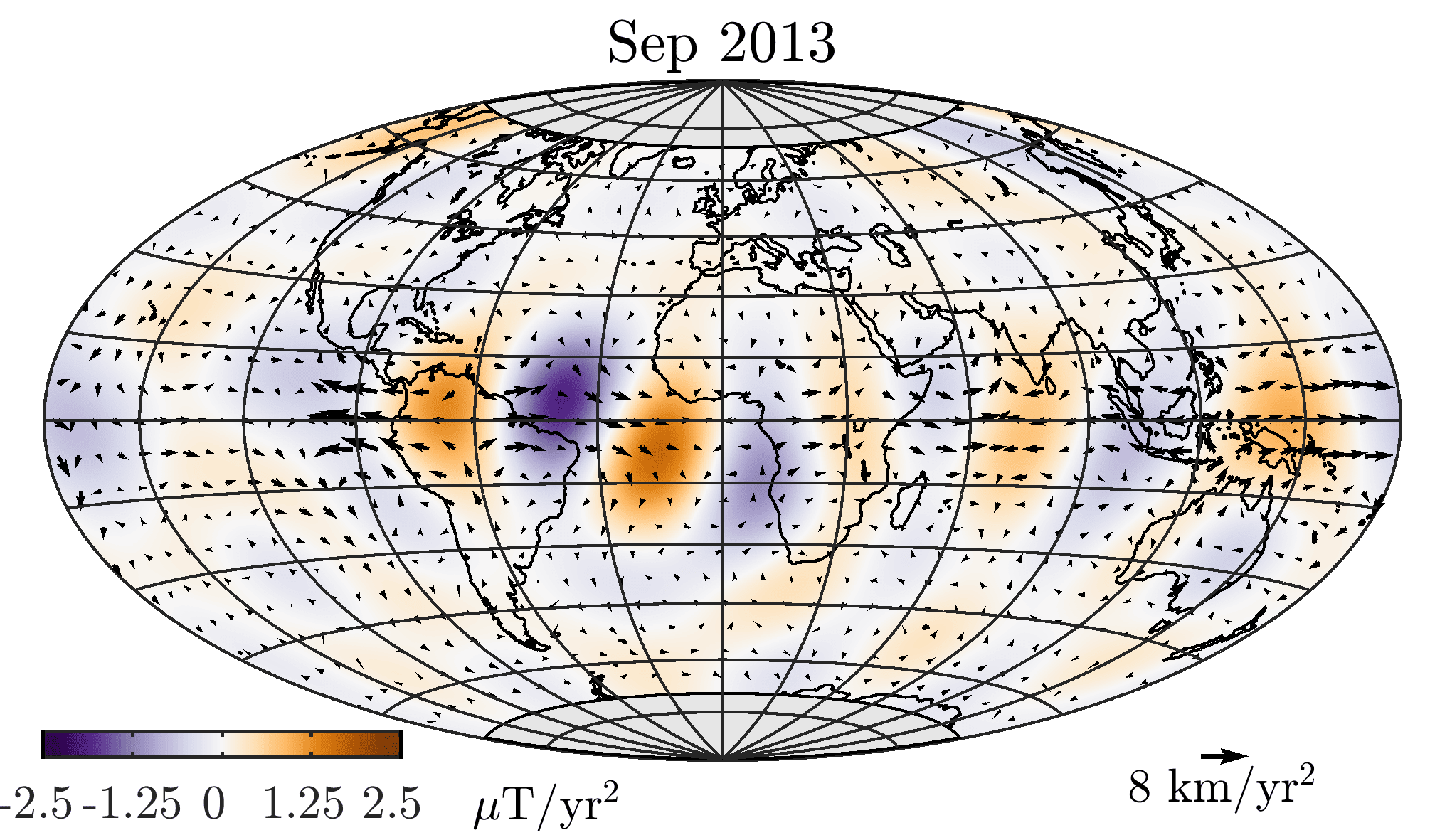}
	\includegraphics[width=8.25cm]{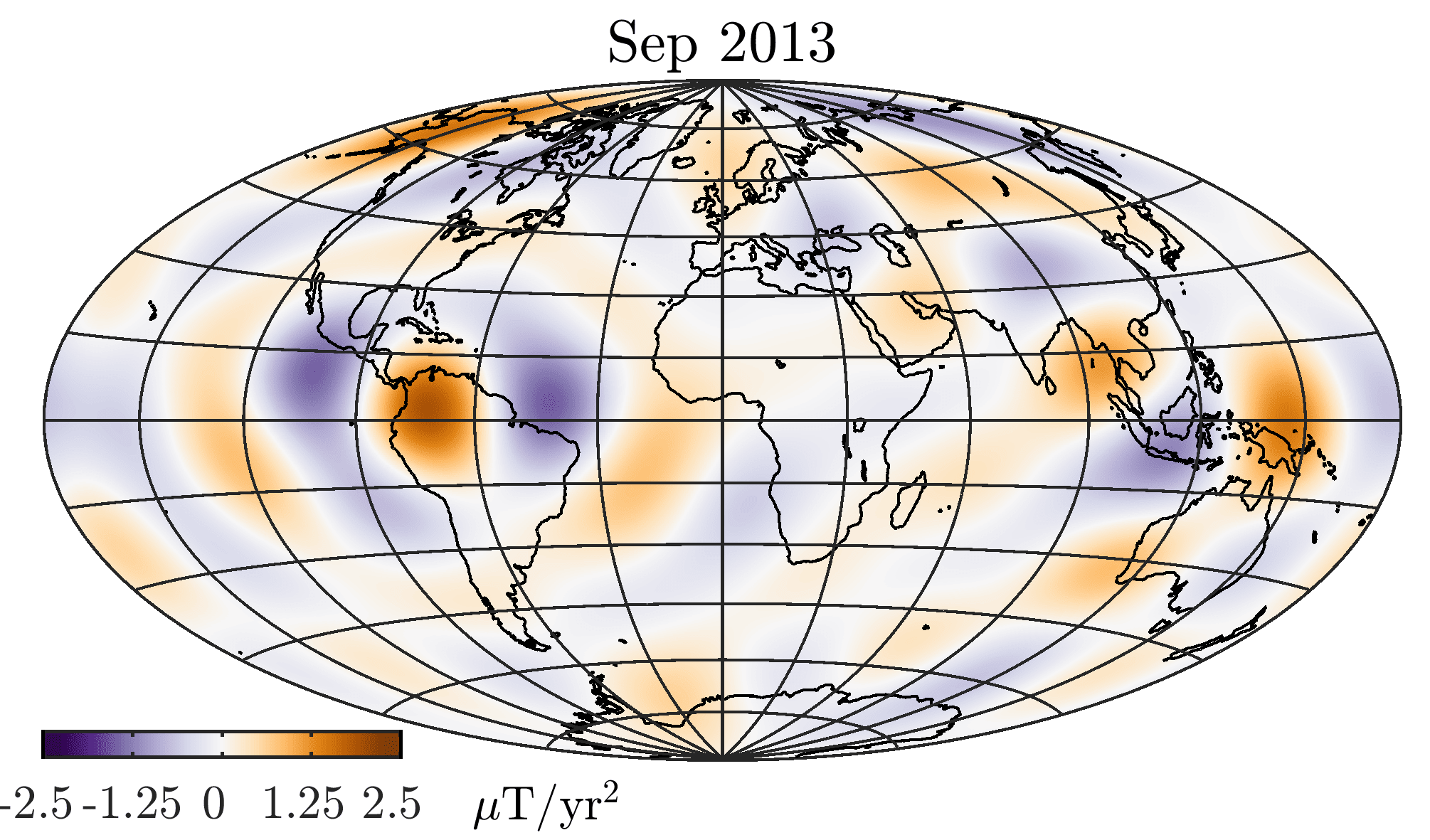}\\
	\vspace{0.5cm}
	\includegraphics[width=8.25cm]{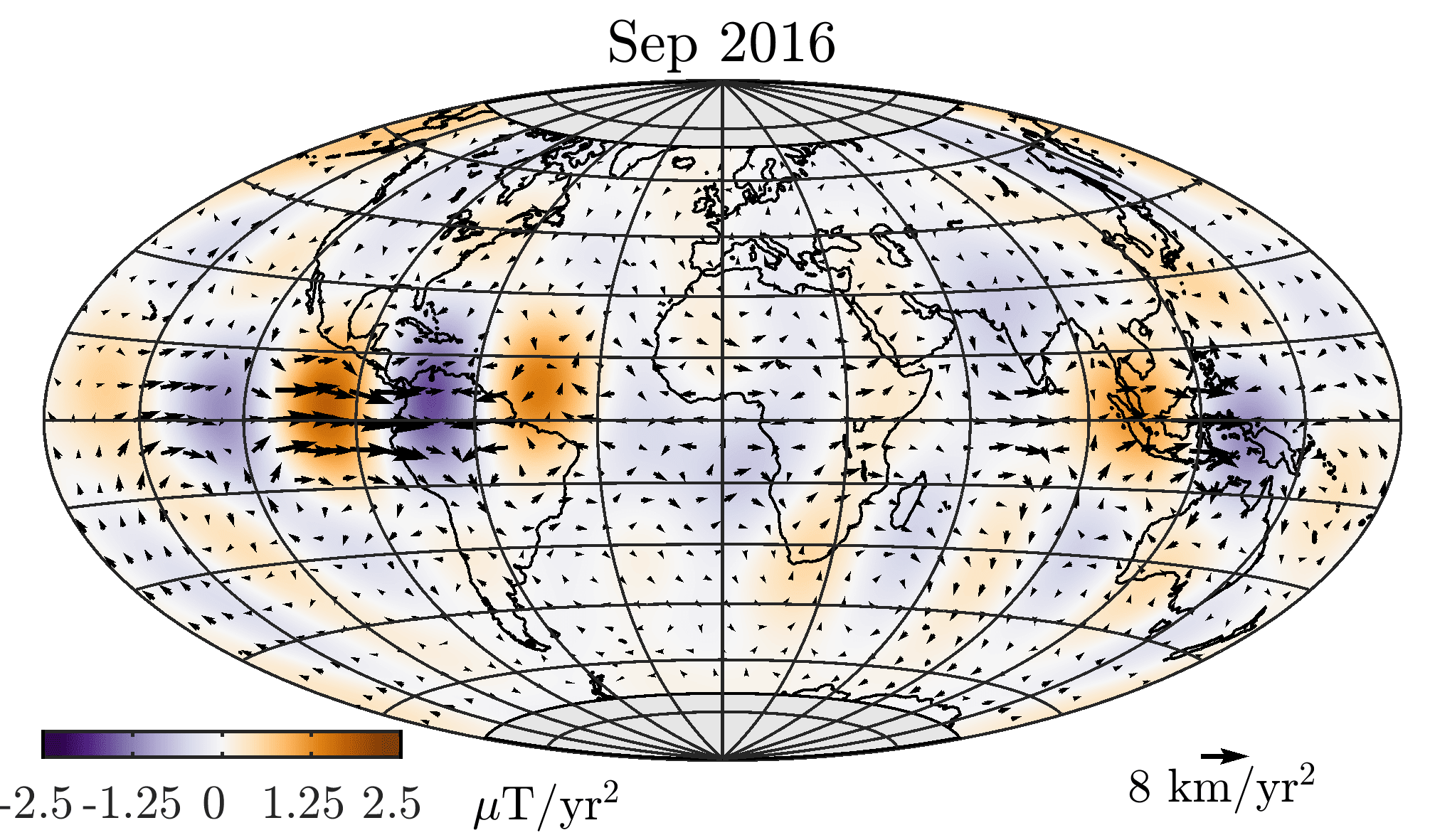}
	\includegraphics[width=8.25cm]{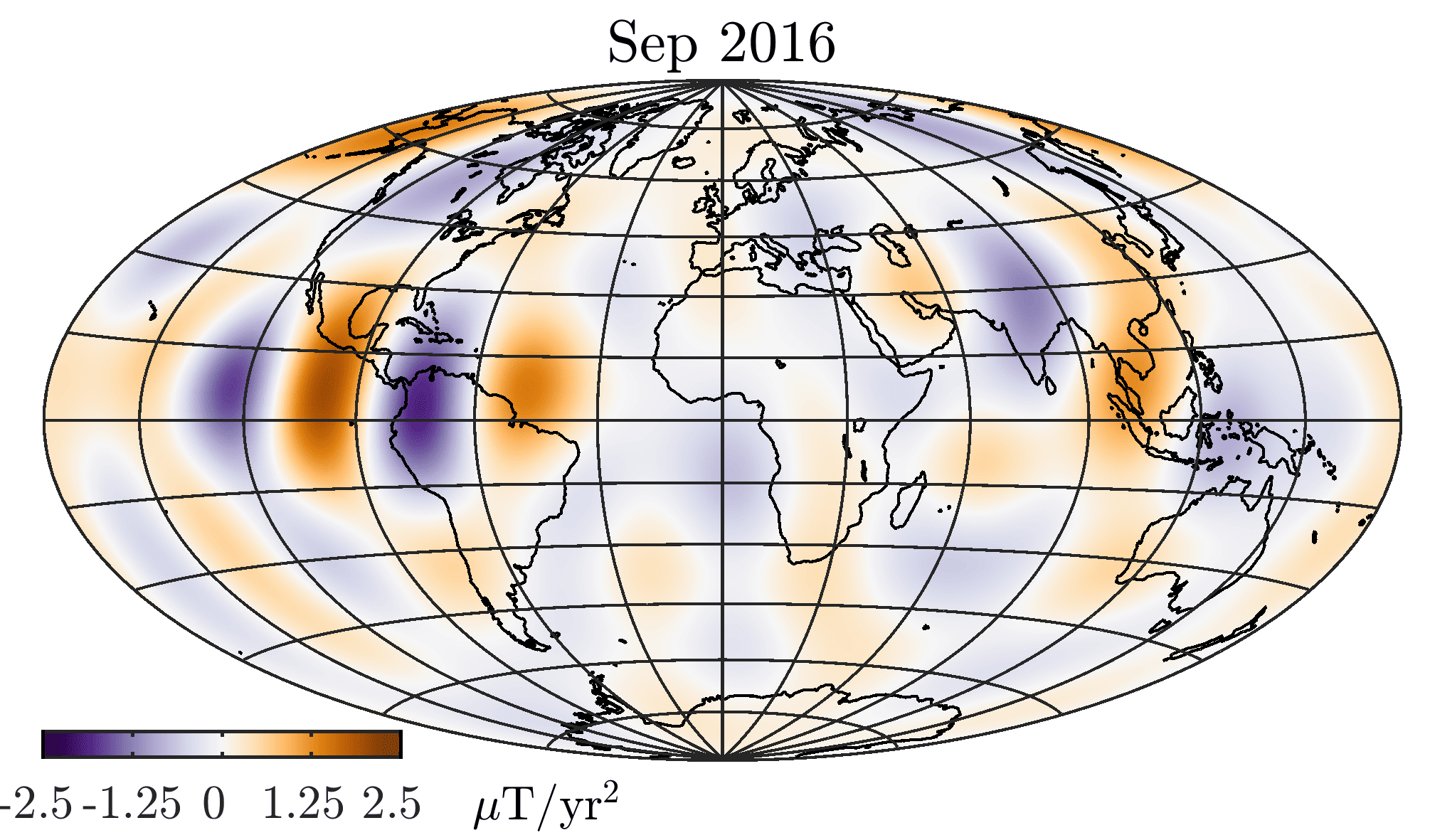}
	\caption{Snapshots of the flow acceleration and predicted SA from {\it CoreFlo-LL.1}  (left) during increased SA power as indicated in Fig.~\ref{fig:sa_norm}. CHAOS-6-x7 prediction of the SA is also shown for comparison at the same time (right). SA has been truncated at spherical harmonic degree 9 in each case. The grey area masks the region inside the tangent cylinder.}
	\label{fig:flow_acc}
\end{figure*}

Fig.~\ref{fig:flow_acc} presents the corresponding maps of the core surface SA, plotted together with the core surface flow acceleration from {\it CoreFlo-LL.1} at pulse times.  The SA has again been truncated here at spherical harmonic degree 9 in order to facilitate comparisons with CHAOS-6-x7 field model, shown in the right column.  {\it CoreFlo-LL.1} generates localised patches of strong SA in the equatorial region, for example around South America, which change polarity over time in agreement with the patterns seen in CHAOS-6-x7.  We find that pulses of SA power in Fig.~\ref{fig:sa_norm} correspond to times of high amplitude equatorial SA patches, that alternate in sign and are located in specific longitudinal sectors.  There are however differences in the patterns of core surface SA predicted by  {\it CoreFlo-LL.1} and those found in the CHAOS-6-x7 model, despite that both explain well the observed SA at Earth's surface and at satellite altitude (recall Fig.~\ref{fig:sa_timeseries_go_vo}).  This suggests care should perhaps be taken interpreting the details of the core surface SA in field models, since a variety of possibilities exist for adequately explaining the observations, depending on the chosen  regularization and other constraints imposed during model construction.  Our time-dependent flow is by construction primarily equatorially symmetric \-- this leads to core surface SA patterns that are more equatorially symmetric than those seen in CHAOS-6-x7.

\begin{figure*}
	\centering
	\includegraphics[width=11cm]{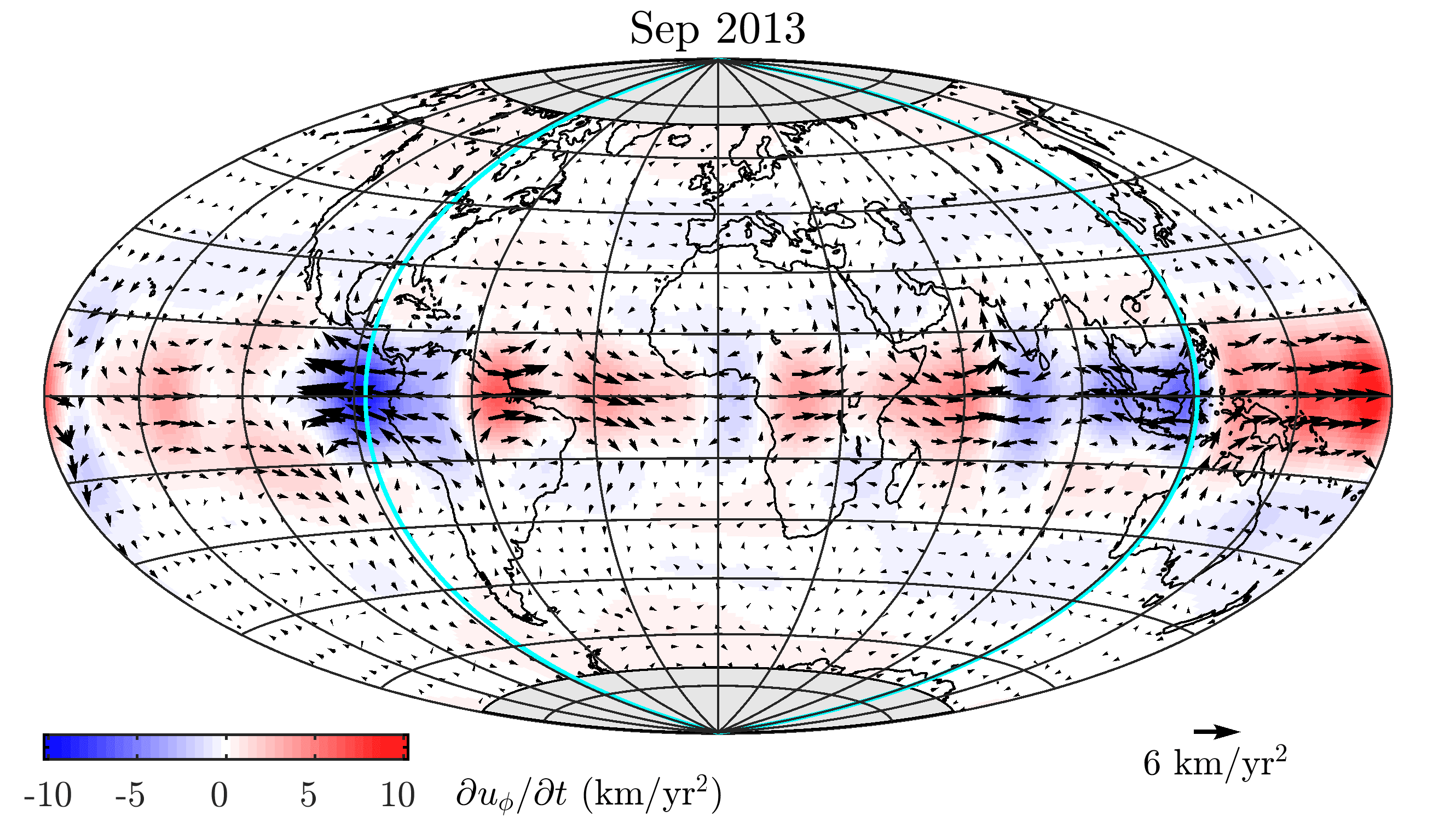}\\
	\includegraphics[width=4.5cm]{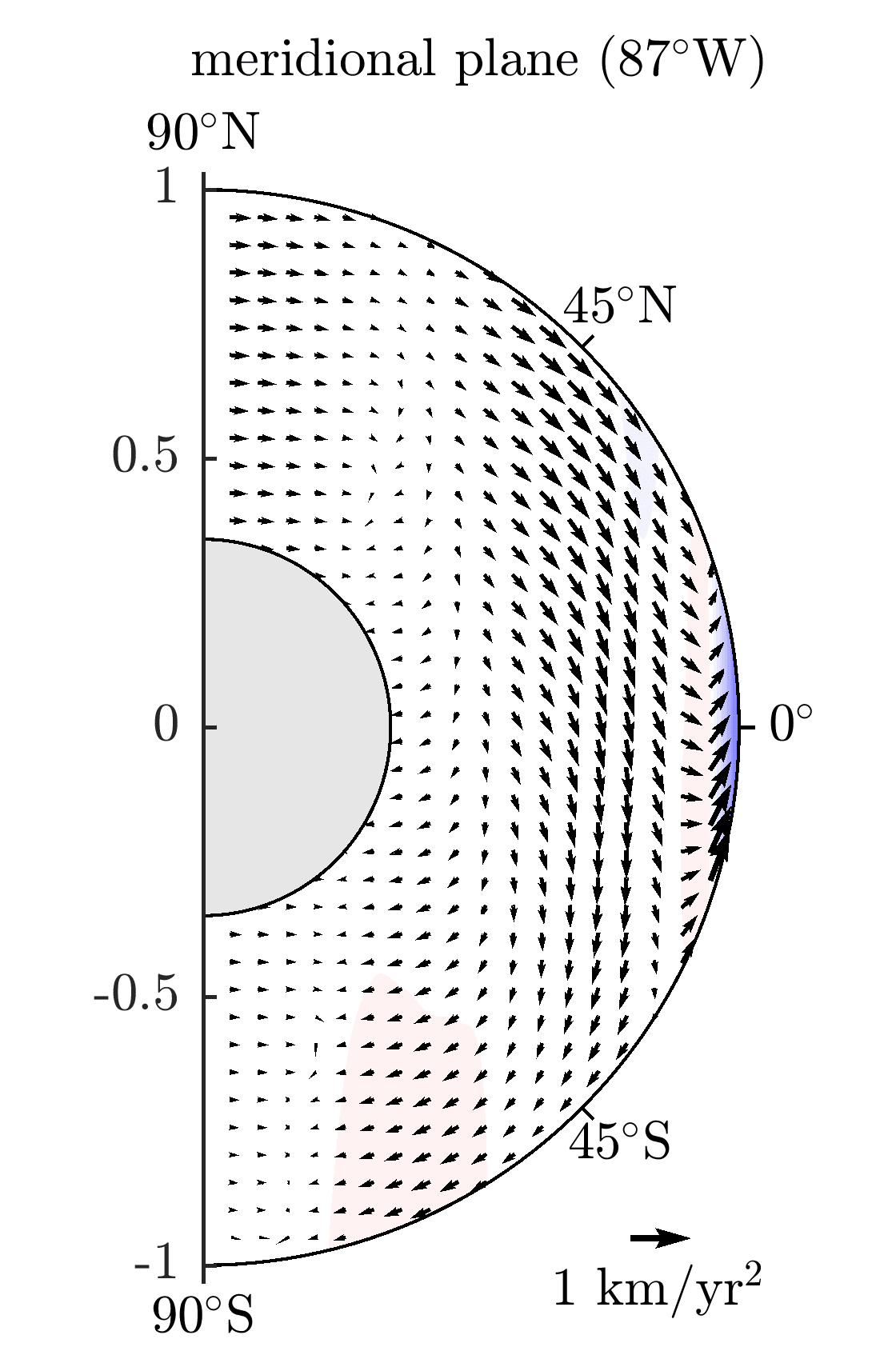}
	\includegraphics[width=4.5cm]{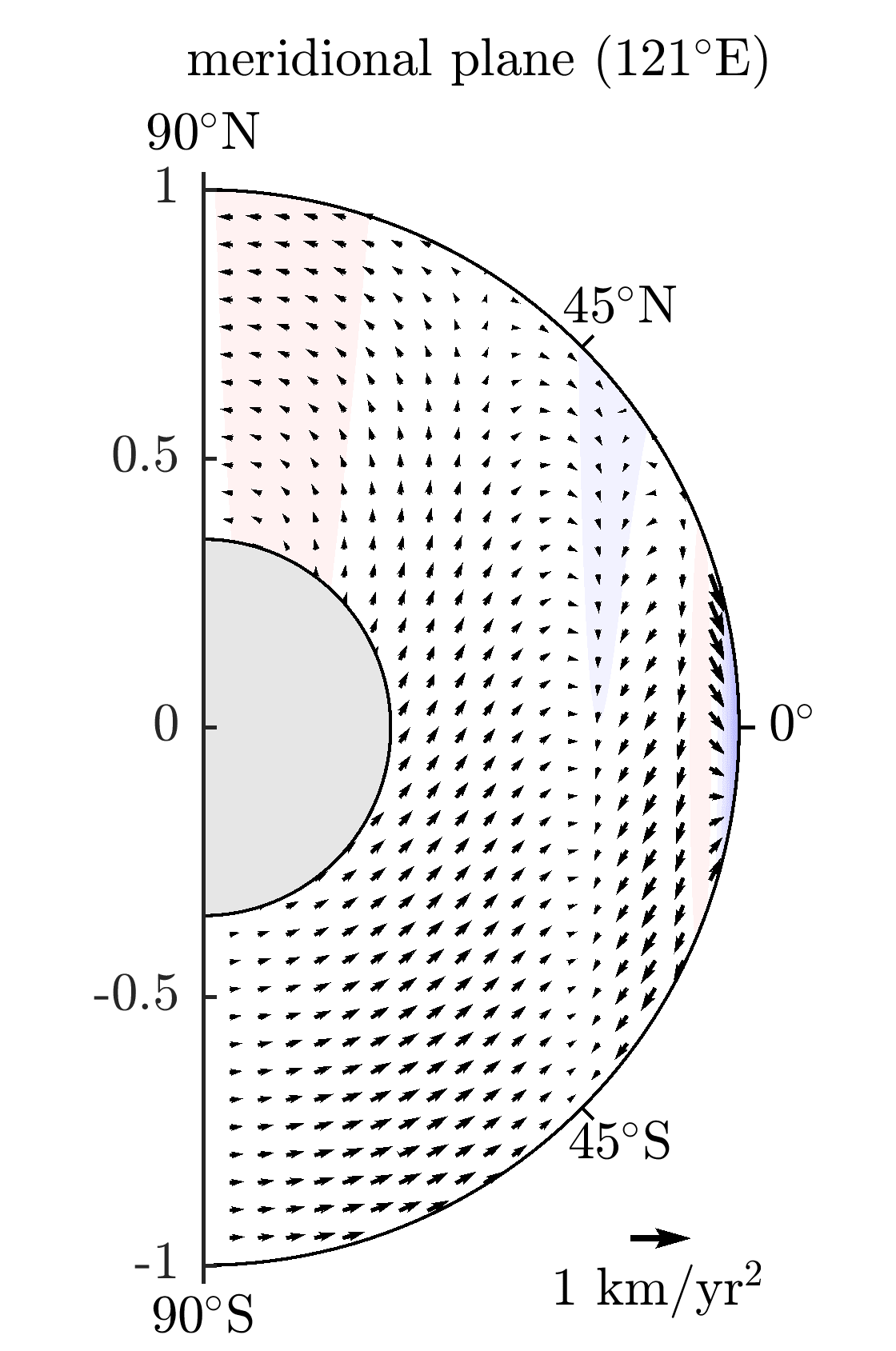}
	\includegraphics[width=7.25cm]{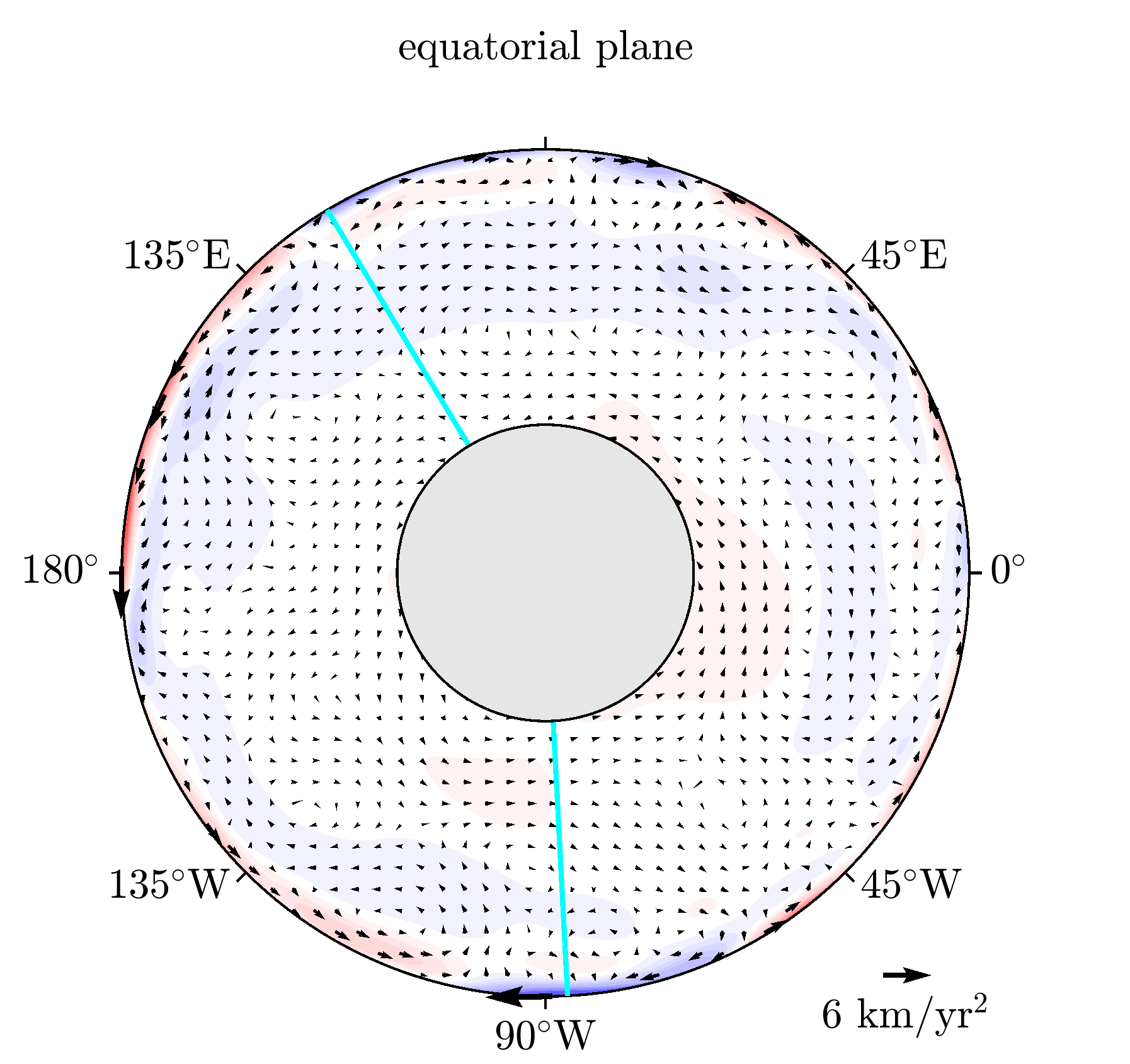}
	\includegraphics[width=10.5cm]{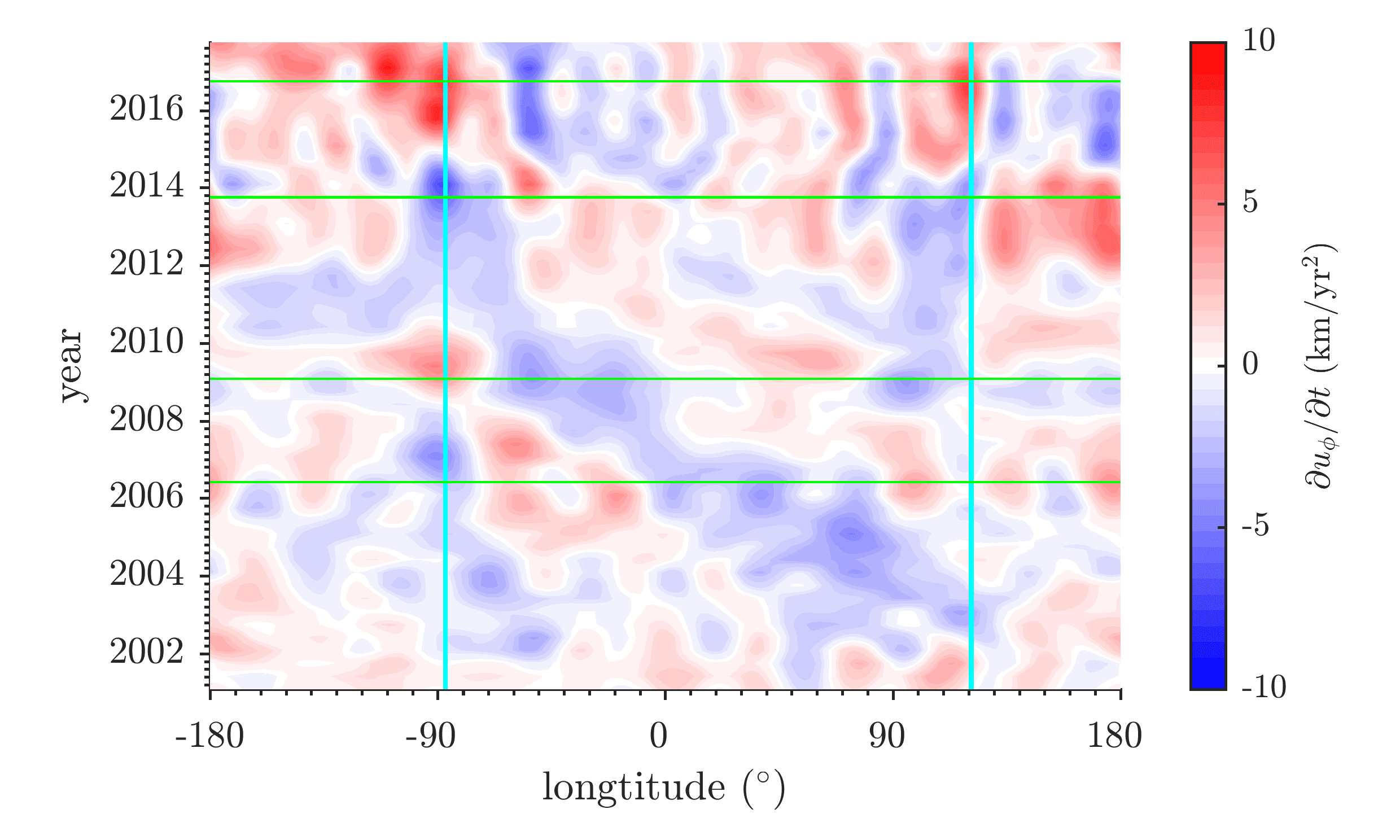}
	\caption{Flow acceleration at the core surface (top), on two meridional planes and the equatorial plane (middle) cutting across the core region. The cyan lines at $87^\circ$W and $121^\circ$E indicate where the meridional planes intersect with the equatorial plane. The color scale shows the strength of the azimuthal flow acceleration, while the grey patch masks the area of the inner core. We also plot the azimuthal flow acceleration averaged over $15^\circ\text{N, S}$ latitude as a function of time and longitude (bottom). The green lines coincide with the time of increased SA as shown in Fig.~\ref{fig:flow_acc}.}
	\label{fig:flow_3D_acc}
\end{figure*}

Fig.~\ref{fig:flow_acc} and \ref{fig:flow_3D_acc}  illustrate how SA pulses in {\it CoreFlo-LL.1} result from localised accelerations of the non-zonal azimuthal flow.  In 2006 the strongest SA  at low latitudes was situated under northeastern South America and under the equatorial Atlantic.  At this time in {\it CoreFlo-LL.1}  there was a generally eastward acceleration under northern South America and westward flow acceleration under western equatorial Africa. In 2009 the strongest SA feature was again under northeastern South America, slightly to the west of the location of the maximum amplitude feature in 2006 and opposite in sign.  In {\it CoreFlo-LL.1} this is associated with a strong westwards acceleration under northern South America. The SA patterns under South America alternated in sign for the subsequent pulses in 2013 and 2016, as did the direction of the non-zonal azimuthal flow acceleration in this location. There is a clear divergence of azimuthal flow acceleration under northern South America in 2013. In 2016, when there is excellent {\it Swarm} data available for a year on either side, the flow accelerations in this region had reversed direction to consist of a regions of strong convergence. 

The spatial structure of the non-zonal azimuthal flow acceleration at the core surface in 2013 is more clearly seen in  Fig.~\ref{fig:flow_3D_acc} where the azimuthal flow acceleration is contoured. The low latitude flow acceleration is directed from regions of acceleration divergence towards regions of acceleration convergence.  In order to satisfy conservation of mass the former must be related with accelerations in upwelling and the latter to regions of acceleration in downwelling. The relevance of flow acceleration upwellings and downwellings to understanding rapid field accelerations has previously been highlighted by \cite[]{olsen2008rapidly}.  The bottom panel in  Fig.~\ref{fig:flow_3D_acc} shows a time longitude plot of the azimuthal flow accelerations in  {\it CoreFlo-LL.1}, averaged $15^\circ$N and S of the equator.   Times of SA acceleration pulses, marked by the green lines, correspond to times when there were strong azimuthal flow accelerations at specific locations.  The longitude sectors from $60^\circ$W to $100^\circ$W, and from $80^\circ$E to $130^\circ$E, have experienced a succession of strong azimuthal flow acceleration features since 2000 with alternating sign.  The Pacific region, from $130^\circ$E to $150^\circ$W, has also experienced strong azimuthal flow accelerations since 2012, with a change in sign from eastward to westward acceleration during 2014.

Sign changes in the azimuthal flow acceleration are apparent at times and locations where geomagnetic jerks have been reported, for example in the Pacific region during 2014 \cite[]{torta2015evidence}.   Sign changes in the azimuthal flow acceleration correspond to times when the azimuthal flow acceleration is changing rapidly; the time between strongly positive and negative accelerations can be short, of order one year or less.  Under the equatorial Pacific in 2014, as seen by comparing Figs.~\ref{fig:flow_3D_acc} and \ref{fig:flow_tdep_av_eqline}, the sign change in the acceleration occurred as the time-varying azimuthal flow reached its highest amplitude, in this case a strong eastward flow. Not all zero crossings of the non-zonal azimuthal flow acceleration (or extrema in the time-varying, non-zonal, azimuthal flow) are associated with jerk-type events at Earth's surface.  In order for jerks to have a significant effect at the Earth's surface, the flow acceleration sign changes must involve sufficiently large-scale flow structures.  Following this logic, strong global jerk events seen in historical records (e.g. in 1969, 1978) could be the result of sign changes in the non-zonal azimuthal flow acceleration that are of sufficiently large amplitude and spatial scale.  More localised jerk events can in this case be due to changes in the sign of the non-zonal azimuthal flow acceleration of smaller lengthscale.  Overall, a spectrum of jerk events of different sizes and amplitudes is therefore expected, depending on the size and amplitude of the changes in the underlying non-zonal azimuthal flow acceleration. 

In  Fig.~\ref{fig:flow_3D_acc} we also explore the 3D structure of the flow accelerations, as captured by {\it CoreFlo-LL.1}. Since the modes on which the model is based are three dimensional, and mode amplitudes have been inferred from the observations,  we can present the flow and its accelerations throughout the core volume.   At the core surface, the non-zonal azimuthal accelerations required to explain the observations are found to be strongly localised in an equatorial belt within $15^\circ$ of the equator and consist of successive regions of convergence and divergence, with enhanced amplitude at specific longitudes, for example under the region west of northern South America in 2013.  The flow acceleration structures of largest scale in 2013 are under the Pacific region, close to where there was a change in the sign of the acceleration in 2014.  Low latitude azimuthal flow acceleration is not present at all longitudes, it is weaker under the central Pacific and west of Africa in 2013. We find that the acceleration changes sign as we move away from the equator.  For example under the Pacific region from $150^\circ$E to $180^\circ$E, the azimuthal acceleration is eastwards on the equator in 2013, but westwards at plus and minus $15^\circ$ latitude. The time-dependent modes in {\it CoreFlo-LL.1} are primarily equatorially symmetric and nearly axially invariant, hence meridional sections of the azimuthal flow acceleration show essentially columnar structures, with highest amplitudes close to the core surface at the equator. The sign change of the acceleration moving away from the equator at the core surface is seen in the meridional sections to be due a change in sign of the columnar structures of the azimuthal flow acceleration in the cylindrical radial direction.  Strong azimuthal flow accelerations are  focused in a very small part of the volume of the outer core, close to the core surface, as can be seen in the equatorial section. Considering a time-longitude plot of the azimuthal flow acceleration, oscillation in the polarity of the non-zonal azimuthal flow acceleration are seen for example close to $90^\circ$W and $120^\circ$E (cyan lines); local intensifications of the azimuthal flow acceleration are also evident at times of SA pulses (green lines).



\afterpage{\clearpage}  

\subsection{Experiments with alternative flow regularizations and parameterizations}
In order to address the question of the robustness of the time-varying flow in  {\it CoreFlo-LL.1}, and the associated accelerations, we have carried out a number of experiments with different choices of flow parameterization and model regularization, as summarized in Table \ref{tab:statistics}, see the related discussion in section \ref{sec:results} for a description of the experimental models.  We examined in detail results for the azimuthal flow accelerations in 2013 for these experimental flows, similar in form to  Fig.~\ref{fig:flow_3D_acc}. Regarding the time-longitude plots, for both \textit{model 1b and model 1c}, we can identify many similar features as in Fig.~\ref{fig:flow_3D_acc}. In particular, around $90^\circ$W we see unambiguous alternation in the direction of the azimuthal flow acceleration, and find that the intense eastward and westward accelerations again occur in similar pairs of converging and diverging accelerations. The amplitudes of the accelerations are however higher for \textit{model 1b} and there are more small-scale structures compared with {\it CoreFlo-LL.1}, as might be expected for a model constructed with equatorial symmetry of flow acceleration imposed. Nevertheless it consists of the same essential details: the non-zonal azimuthal flow acceleration alternates in time, for example around $90^\circ$W, or under the Pacific after 2012, and is organized around centers of flow acceleration convergence and divergence. In both \textit{model 1b} and \textit{model 1c} the azimuthal flow accelerations remain localised close to the core surface, and they vary little in the axial direction.  Their time-longitude plots also suggests a coherent westward movement of flow acceleration features from $90^\circ$W to $90^\circ$E between 2003 and 2014. 

Overall we find that flows regularized in different ways, or with different assumptions made regarding the equatorial symmetry of the time-varying flow, produce oscillating SA pulses in a similar manner to  {\it CoreFlo-LL.1}, via alternating non-zonal azimuthal flow acceleration at low latitudes.  Although the basic patterns of azimuthal flow acceleration are similar across the investigated flows, if one is willing to accept time-dependent cross-equatorial flow, as in {\it CoreFlo-LL.1}, then less vigorous accelerations of flow convergence and divergence, and smaller amplitude flow accelerations are required.  Further comparisons with numerical dynamo models in a regime appropriate for studying rapid dynamics are required to determine whether or not such transient $\EA$ flow accelerations are physically reasonable.




\subsection{Relation to previous studies}

Patterns of time-varying core flow at low latitudes have been examined in a number of previous studies.   \cite{Jackson1997} mentioned the presence of oscillatory motions at low latitudes in his tangentially-geostrophic flows, but did not analyse the motions in detail. \cite{Gillet2015a} inverted the COV-OBS field model for QG core flows, and found vigorous azimuthal jets in the equatorial belt below $10^\circ$ latitude; these were particularly intense after 2000 when satellite observations contributed to the field model. \cite{Gillet2015a}  argued that the dynamics in the equatorial region was different to that occurring at mid latitudes, though coupled via meridional flows arriving or leaving at longitudes of 130$^\circ$E and 60$^\circ$W. They found flow interannual variations of the equatorial jets, for example at 85$^\circ$W. \cite{Finlay2016} carried out a similar inversion of the higher resolution CHAOS-6 model, that included {\it Swarm} data, and found similar oscillating, non-zonal, azimuthal jets in the equatorial region, for example at 40$^\circ$W.  They noted that pulses of SA coincided with times of large azimuthal flow acceleration, between maxima and minima of the time varying azimuthal flow.  Our results are in good accord with these previous QG studies, despite our flows being parameterized and regularized in a very different way, and being based directly on ground and observatory data rather than on field models.  Our flow \textit{model 1c} is most similar to that of  \cite{Gillet2015a} and \cite{Finlay2016} since it has purely equatorially-symmetric time-dependent flows and a spatial regularization applied to its poloidal and toroidal flow components.

\cite{barrois2017contributions} recently introduced a new Kalman filter approach to core flow estimation, taking output from a numerical dynamo simulation as prior information, and the COV-OBS.x1 field model as input. \cite{Barrois2018} have extended this approach to work with ground and satellite datasets similar to those employed here.  We have compared our flow results with those of \cite{Barrois2018} as well as the corrected flows reported by \cite{Barrois2018a}, producing the same time-longitude plots of azimuthal flow acceleration; we find Barrois's flows show similar large-scale patterns of azimuthal flow acceleration to {\it CoreFlo-LL.1} , with regions of azimuthal flow acceleration convergence and divergence alternating on interannual timescales.  The amplitudes of the accelerations in the models  of Barrois are however smaller than those found in {\it CoreFlo-LL.1}; this may be in part due to their reliance on a dynamo prior which may lack realistic rapid dynamics at low latitudes  \cite[]{aubert2018geomagnetic}. It is possible that magnetic diffusion, due to field gradients being built up by the time-varying core flow, could also make a contribution to the observed secular acceleration.  This could conceivably be accounted for following the approach described by \cite{amit2008accounting} where diffusion is correlated with horizontal divergence at the core surface.  Note however that in this case diffusion is enslaved to the flow and cannot drive field changes on its own.

Another approach to studying the recent time-varying core flow has been independently pursued by \cite{Whaler2016}.  They inverted ground observatory data alone, applying a strong norm on the spatial complexity of the flow (varying with $n^5$, forcing the flow to be large scale), but without enforcing any symmetry constraint or any requirement that the flow be quasi or tangentially geostrophic.  Like us, they also minimized a measure of the flow acceleration.  \cite{Whaler2016} obtained flows that were primarily toroidal, equatorially-symmetric and tangentially geostrophic, with the time-averaged part explaining much of the variance, and being dominated by an eccentric planetary gyre.  The details of their time-varying flows are however sensitive to the choice of flow parameterization and amount of applied temporal regularization.   Nonetheless, they reported that their time-dependent flow had a larger component of poloidal energy than the time averaged flow, as is the case for the largest scale part of our flows, see Fig.~\ref{fig:torpol_spectrum}.  They also highlighted an anti-clockwise rotation of the time-varying flow under the equatorial western Pacific between 1997 and 2015.  In  {\it CoreFlo-LL.1} we find the azimuthal flow direction changes from westward to eastward around 2012 in this region, and this occurs via an anti-clockwise rotation.   \cite{Whaler2016} also found peaks of poloidal flow acceleration in the equatorial region under the Western Atlantic and the eastern Americas with opposite signs in 2006 and 2009; we find similar features in our flows, although the accelerations are generally stronger.    They discussed a sign change in their non-zonal azimuthal flow at 85$^\circ$W around 2006; we find the same type of change in our time-varying azimuthal flow from eastward to westward at this location in 2007 (see Fig.~\ref{fig:flow_tdep_av_eqline}).   To summarize, the azimuthal flow variations found by \cite{Whaler2016} are generally consistent with our results, but our flows have greater power at small scales, and the amplitude of our time-varying flows are much larger. Given the restriction of our observations to mid and low latitudes, and because we base our flow parameterization on inertial modes in a full sphere that are not suitable for studying flow close to and inside the tangent cylinder, we are unable to comment on the acceleration of high latitude flows recently discussed by \citet{Livermore2017} and \citet{Barrois2018}.  

The data employed, the choice of flow parameterization, and the applied regularization all affect the details of the inferred time-varying core flow.  Nevertheless, there now seems to be some agreement across the above studies that non-zonal azimuthal flow acceleration at low latitudes are required by the observations. Localised peaks in the azimuthal flow acceleration, associated with flow acceleration convergence/divergences,  give rise to SA pulses with alternating polarity.  The meridional part of the time-varying low latitude flows is less well constrained by the data alone.  Purely $\ES$ time-varying flows can account for the secular variation observations, but allowing cross-equatorial flow enables a similar fit to the data with simpler flows. 

\subsection{Possible implications for core dynamics}
{\it CoreFlo-LL.1} supports the hypothesis that the equatorial region beneath the core-mantle boundary is a place of vigorous localised fluid motions, with oscillations on interannual timescales involving strong acceleration of the non-zonal azimuthal flow.    \cite{chulliat2015fast} proposed that these may be related to waves that could propagate in a stratified layer beneath the core-mantle boundary at low latitudes. \cite{knezek2018influence} have developed detailed numerical models of how non-zonal MAC waves could be trapped near the equator, for plausible configurations of the background magnetic field. Such waves propagate in the azimuthal direction, and \cite{knezek2018influence} find they travel predominantly eastward for plausible field strengths.  We do not find compelling evidence in our flow models for eastward propagation of waves close to the core surface.  Rather we find strong oscillation in the non-zonal flow at particular locations, and perhaps tentative indications of westward propagation of disturbances under the Atlantic sector, see  Figs.~\ref{fig:flow_3D_acc} as previously noted by \cite{chulliat2015fast}.
 
An alternative source for low latitude flow accelerations is the arrival at the core-mantle boundary of QG Alfv\'{e}n waves triggered by buoyancy release deep in the core  \cite[][]{aubert2018geomagnetic}.  The energy of these waves is focused by the geometry of the magnetic field within the core, and their arrival at the core surface is characterized by non-zonal azimuthal flow accelerations of alternating sign, with a sub-decadal timescale related to that of the period of the underlying Alfv\'{e}n waves.  This scenario seems to be in good agreement with many of the details of our  time varying flows.  The apparent westward motion of azimuthal flow acceleration features under the Atlantic may in this case perhaps be a signature of a QG  Alfv\'{e}n wave front arriving first under western Africa and then progressively later at locations to the west.  \cite{Aubert2018b} have argued that such waves may indeed provide an explanation for geomagnetic jerks. In our flows, jerk events are associated with a rapid change in the sign of the low latitude, non-zonal, azimuthal flow acceleration pattern.

\section{Conclusions}
\label{sec:conc}
We have derived a new model describing the time-varying flow at low latitudes in the Earth's outer core, called {\it CoreFlo-LL.1}.  It is based on ground and satellite magnetic observations and assumes that the time-varying flow can be parameterized primarily in terms of geostrophic and  QG modes.  We find evidence for strong non-zonal azimuthal flow accelerations in the equatorial region, that change polarity on an interannual timescale. SA pulses at the core surface are associated with localised extrema of these azimuthal flow accelerations.  We find that geomagnetic jerks may occur when the non-zonal azimuthal flow acceleration rapidly changes sign, for example if the time-varying azimuthal flow reaches a peak in its amplitude.  The 2014 jerk that was seen most clearly in the Pacific region involved a large-scale non-zonal azimuthal flow acceleration structure that changed sign in this fashion.  One explanation for such an event could be a QG  Alfv\'{e}n wave, triggered deep within the core, arriving at the core surface  \cite[][]{Aubert2018b}.

We find that flows based on geostrophic and inertial mode spatial structures are able to provide an adequate description of geomagnetic observations, with time-averaged flows dominated by a planetary scale eccentric gyre.  Compared to previous studies, {\it CoreFlo-LL.1} shows more intense time-varying flows, and small lengthscale structures, localised in the equatorial region.  It is remarkable that field accelerations at both mid and low latitudes at Earth's surface and at satellite altitude can be well fit by time-varying flows focused at the equatorial region on the core surface. The meridional component of our time-varying flows  are however strongly dependent on a-priori assumptions made concerning the  flow structure, in particular whether or not time-dependent cross-equatorial flow is permissible.  

In this work we have neglected the role of magnetic diffusion. Though not driving field acceleration, it should be included in future studies, for example following the approach of \cite{amit2008accounting}. Moving forward it will also be important to use more complete data error covariance descriptions, for example taking into account the correlations between errors on neighbouring ground and virtual observatories. A fundamental limitation of this study is that it uses the free oscillation modes of a rotating sphere, rather than a spherical shell, motivated by our focus on low latitudes and the existence of analytic expression for the full sphere modes. One could also in principle solve a numerical eigenvalue problem and obtain correct modes for a spherical shell, and taking into account finite dissipation; these could then instead be used these as the basis for an inversion.  Hypotheses testing for example exploring different choices of the background density profile and core magnetic field structure could then be undertaken. 

A more complete understanding of the physical processes underlying time-varying low latitude core flows will ultimately require fitting of observations by full dynamic models, rather than the kinematic descriptions we have employed epoch by epoch.  As an initial step, such models should describe non-zonal QG  Alfv\'{e}n waves, as these appear to be of interest in explaining SA pulses and geomagnetic jerks.  The lengthening high resolution data series being generated by the {\it Swarm} satellite mission will provide increasingly powerful observational tests of such ideas.    
 

\begin{acknowledgments}
We thank ESA for providing L1b magnetic field data from the {\it Swarm} satellite mission that were crucial to this study.  The staff of the geomagnetic observatories and INTERMAGNET are thanked for supplying high-quality observatory data, and BGS are thanked for providing us with checked and corrected observatory hourly mean values. The support of the CHAMP mission by the German Aerospace Center (DLR) and the Federal Ministry of Education and Research is gratefully acknowledged.  Nicolas Gillet, Olivier Barrois and Julien Aubert are thanked for helpful discussions and suggestions. CK and CCF were partly supported by the European Research Council (ERC) under the European Union's Horizon 2020 research and innovation programme (grant agreement No 772561).  Ciaran Beggan and an anonymous reviewer are thanked for constructive comments that improved the manuscript.
\end{acknowledgments}

\bibliography{references}{}
\bibliographystyle{gji}

\appendix

\section{Explicit expressions of modes}
\label{app:mode_expressions}

\subsection{Geostrophic mode}
\label{app:geostrophic_mode}

The unnormalized geostrophic polynomials with $k\geq 1$ are defined by \cite[]{Zhang2017}
\begin{equation}
G_{2k-1}(s)=\sum_{j=1}^{k} C^\mathrm{G}_{k;j}s^{2j-1},
\end{equation}
with $s=r\sin\theta$ and
\begin{equation}
	C^\mathrm{G}_{k;j} = \frac{(-1)^{k-j}[2(k+j)-1]!!}{2^{k-1}(k-j)!(j-1)!(2j)!!}
\end{equation}
In this case, the mean over the full sphere of the squared absolute value becomes
\begin{equation}
\frac{3}{4\uppi}\int_\mathcal{V}\abs{G_{2k-1}}^2\mathrm{d}\mathcal{V}=\frac{3(2k+1)!!(2k-1)!!}{(4k+1)(2k)!!(2k-2)!!},
\end{equation}
which is used to normalize the polynomials. The enstrophy, as defined in Eq.~\eqref{eq:enstrophy} of the unnormalized geostrophic polynomials is
\begin{equation}
\label{eq:geostrophic_vorticity}
	Q^2\left( G_{2k-1}\ephi\right) =3\sum_{jl}^kC^\mathrm{G}_{k;j}C^\mathrm{G}_{k;l}2^{j+l}jl\frac{[j+l-2]!}{[2(j+l)-1]!!}.
\end{equation}

\subsection{Symmetric inertial modes}
\label{app:inertial_mode_ES}

Following closely \cite{zhang2001inertial} and \cite{Zhang2017}, the unnormalized and complex-valued three components of both non-axisymmetric and axisymmetric modes are given in spherical coordinates by
\begin{subequations}
	\label{eq:symmetric_modes}
	\begin{align}
	\begin{split}
	\er\cdot \mathbf{u}_{mnk}^\mathrm{S} =
	-\frac{\I}{2}\sum_{i=0}^{k}\sum_{j=0}^{k-i}&C^\mathrm{S}_{mk;ij}\sigma^{2i-1}(1-\sigma^2)^{j-1}
	r^{m+2(i+j)-1}\sin^{m+2j}\theta\cos^{2i}\theta\e^{\I m\phi}\\
	&\cdot[\sigma(m + m\sigma + 2j\sigma)-2i(1-\sigma^2)]
	\end{split}\\
	\begin{split}
	\etheta\cdot \mathbf{u}_{mnk}^\mathrm{S}=
	-\frac{\I}{2}\sum_{i=0}^{k}\sum_{j=0}^{k-i}&C^\mathrm{S}_{mk;ij}\sigma^{2i-1}(1-\sigma^2)^{j-1}
	r^{m+2(i+j)-1}\sin^{m+2j-1}\theta\cos^{2i-1}\theta\e^{\I m\phi}\\
	&\cdot[\sigma(m + m\sigma + 2j\sigma)\cos^2\theta+2i(1-\sigma^2)\sin^2\theta]
	\end{split}\\
	\begin{split}
	\ephi\cdot \mathbf{u}_{mnk}^\mathrm{S}=
	\frac{1}{2}\sum_{i=0}^{k}\sum_{j=0}^{k-i}&C^\mathrm{S}_{mk;ij}\sigma^{2i}(1-\sigma^2)^{j-1}
	r^{m+2(i+j)-1}\sin^{m+2j-1}\theta \cos^{2i}\theta\e^{\I m\phi}\\
	&\cdot(m + m\sigma + 2j),
	\end{split}
	\end{align}
\end{subequations}
where the coefficients $C^\mathrm{S}_{mk;ij}$ are defined as
\begin{equation}
C^\mathrm{S}_{mk;ij}=\frac{(-1)^{i+j}[2(k+m+i+j)-1]!!}{2^{j+1}(2i-1)!!(k-i-j)!i!j!(m+j)!}.
\end{equation}
The subscripts of $\sigma_{mnk}$ have been omitted to make the equations easier to read. The normalization of the modes is given by
\begin{equation}
\label{eq:symmetric_normalization}
\begin{aligned}
\frac{3}{4\uppi}\int_{\mathcal{V}}\abs{\mathbf{u}^\mathrm{S}_{mnk}}^2\mathrm{d}\mathcal{V}=\sum_{i=0}^{k}\sum_{j=0}^{k-i}\sum_{q=0}^{k}\sum_{l=0}^{k-q}&C^\mathrm{S}_{mk;ij}C^\mathrm{S}_{mk;ql}
\frac{3\cdot 2^{m+j+l-3}}{[2(m+i+j+q+l)+1]!!}\sigma^{2(i+q)}(1-\sigma^2)^{j+l}\\
&\cdot\bigg(\big[(m+m\sigma+2j)(m+m\sigma+2l)
+(m+m\sigma+2j\sigma)(m+m\sigma+2l\sigma)\big]\\
&\phantom{\cdot\bigg(}\cdot(m+j+l-1)!\frac{[2(i+q)-1]!!}{(1-\sigma^2)^2}
+8iq(m+j+l)!\frac{[2(i+q)-3]!!}{\sigma^2}\bigg).
\end{aligned}
\end{equation}
Given the components in Eq.~\eqref{eq:symmetric_modes}, one can also compute the inner product of the vorticity of two unnormalized $\ES$ modes $\mathbf{u}_\alpha^\mathrm{S}$ and $\mathbf{u}_\beta^\mathrm{S}$ with
\begin{equation}
\label{eq:symmetric_vorticity}
\begin{aligned}
\frac{3}{4\pi}\int_\mathcal{V}(\nabla\times\mathbf{u}_\alpha^\mathrm{S})\cdot(\nabla&\times\mathbf{u}_\beta^\mathrm{S})^*\mathrm{d}\mathcal{V}=\\
\sum_{i=0}^{k_\alpha}\sum_{j=0}^{k_\alpha-i}\sum_{q=0}^{k_\beta}\sum_{l=0}^{k_\beta-q}&
C^\mathrm{S}_{mk_\alpha;ij} C^\mathrm{S}_{mk_\beta;ql}\frac{3\cdot 2^{j+l+m-1}iq}{[2(i+j+q+l+m)-1]!!}
\sigma_\alpha^{2i-2}\sigma_\beta^{2q-2}(1-\sigma_\alpha^2)^{j-1}(1-\sigma_\beta^2 )^{l-1}\\
&\cdot\bigg(\sigma_\alpha \sigma_\beta[l+j+m-1]![2(i+q)-3]!\\
&\phantom{\cdot\bigg(}\cdot \big[(2j\sigma_\alpha+m+m\sigma_\alpha)(2l\sigma_\beta+m+m\sigma_\beta)+(2j+m+m\sigma_\alpha)(2l+m+m\sigma_\beta)\big]\\
&\phantom{\cdot\bigg(}+2(2i-1)(2q-1)(1-\sigma_\alpha^2)(1-\sigma_\beta^2)[l+j+m]![2(i+q)-5]!!\bigg),
\end{aligned}
\end{equation}
for $m_\alpha=m_\beta=m$ and zero otherwise. The asterisk denotes the complex conjugate. In this paper, we only used terms with $\alpha=\beta$. 

\subsection{Antisymmetric inertial modes}
\label{app:inertial_mode_EA}

Again following closely \cite{zhang2001inertial} and \cite{Zhang2017} the three unnormalized and complex-valued velocity components in spherical polar coordinates are
\begin{subequations}
	\label{eq:antisymmetric_modes}
	\begin{align}
	\begin{split}
	\er\cdot\mathbf{u}^\mathrm{A}_{mnk}=-\frac{\I}{2}\sum_{i=0}^{k}\sum_{j=0}^{k-i}&C^\mathrm{A}_{mk;ij}\sigma^{2i-1}(1-\sigma^2)^{j-1}
	r^{m+2(i+j)}\sin^{m+2j}\theta\cos^{2i+1}\theta\e^{\I m\phi}\\
	&\cdot[\sigma(m + m\sigma + 2j\sigma)-(2i+1)(1-\sigma^2)]
	\end{split}\\
	\begin{split}
	\etheta\cdot\mathbf{u}^\mathrm{A}_{mnk}=-\frac{\I}{2}\sum_{i=0}^{k}\sum_{j=0}^{k-i}&C^\mathrm{A}_{mk;ij}\sigma^{2i-1}(1-\sigma^2)^{j-1}
	r^{m+2(i+j)}\sin^{m+2j-1}\theta\cos^{2i}\theta\e^{\I m\phi}\\
	&\cdot[\sigma(m + m\sigma + 2j\sigma)\cos^2\theta+(2i+1)(1-\sigma^2)\sin^2\theta]
	\end{split}\\
	\begin{split}
	\ephi\cdot\mathbf{u}^\mathrm{A}_{mnk}=\frac{1}{2}\sum_{i=0}^{k}\sum_{j=0}^{k-i}&C^\mathrm{A}_{mk;ij}\sigma^{2i}(1-\sigma^2)^{j-1}
	r^{m+2(i+j)}\sin^{m+2j-1}\theta \cos^{2i+1}\theta\e^{\I m\phi}\\
	&\cdot (m + m\sigma + 2j),
	\end{split}
	\end{align}
\end{subequations}
where the coefficients $C^\mathrm{A}_{mk;ij}$ are defined as
\begin{equation}
C^\mathrm{A}_{ij;km}=\frac{(-1)^{i+j}[2(k+m+i+j)+1]!!}{2^{j+1}(2i+1)!!(k-i-j)!i!j!(m+j)!}.
\end{equation}
The normalization\footnote{The normalization is different here from the one given in \cite{Zhang2017}, where +3 is +1 in the first denominator with the double factorial. We believe this is a misprint and should be +3 as given here.} of the components in this definition is
\begin{equation}
\label{eq:antisymmetric_normalization}
\begin{aligned}
\frac{3}{4\uppi}\int_{\mathcal{V}}\abs{\mathbf{u}^\mathrm{A}_{mnk}}^2\mathrm{d}\mathcal{V}=\sum_{i=0}^{k}\sum_{j=0}^{k-i}\sum_{q=0}^{k}\sum_{l=0}^{k-q}&C^\mathrm{A}_{mk;ij}C^\mathrm{A}_{mk;ql}
\frac{3\cdot 2^{m+j+l-3}}{[2(m+i+j+q+l)+3]!!}\sigma^{2(i+q)}(1-\sigma^2)^{j+l}\\
&\cdot\bigg([(m+m\sigma+2j)(m+m\sigma+2l)+(m+m\sigma+2j\sigma)(m+m\sigma+2l\sigma)]\\
&\phantom{\cdot\bigg(}\cdot(m+j+l-1)!\frac{[2(i+q)+1]!!}{(1-\sigma^2)^2}+2(2i+1)(2q+1)(m+j+l)!\\
&\phantom{\cdot\bigg(}\cdot\frac{[2(i+q)-1]!!}{\sigma^2}\bigg).
\end{aligned}
\end{equation}
The inner product of the vorticity of two unnormalized modes becomes
\begin{equation}
\label{eq:antisymmetric_vorticity}
\begin{aligned}
\frac{3}{4\pi}\int_\mathcal{V}(\nabla\times\mathbf{u}_\alpha^\mathrm{A})\cdot(\nabla&\times\mathbf{u}_\beta^\mathrm{A})^*\mathrm{d}\mathcal{V}=\\
\sum_{i=0}^{k_\alpha}\sum_{j=0}^{k_\alpha-i}\sum_{q=0}^{k_\beta}\sum_{l=0}^{k_\beta-q}&
C^\mathrm{A}_{mk_\alpha;ij} C^\mathrm{A}_{mk_\beta;ql}\frac{3\cdot 2^{j+l+m-3}(2i+1)(2q+1)}{[2(i+j+q+l+m)+1]!!}
\sigma_\alpha^{2i-2}\sigma_\beta^{2q-2}(1-\sigma_\alpha^2)^{j-1}(1-\sigma_\beta^2 )^{l-1}\\
&\cdot\bigg(\sigma_\alpha \sigma_\beta[l+j+m-1]![2(i+q)-1]!\bigg.\\
&\phantom{\cdot\bigg(}\cdot\big[(2j\sigma_\alpha+m+m\sigma_\alpha)(2l\sigma_\beta+m+m\sigma_\beta)+(2j+m+m\sigma_\alpha)(2l+m+m\sigma_\beta)\big]\\
&\phantom{\cdot\bigg(}+8iq(1-\sigma_\alpha^2)(1-\sigma_\beta^2)[l+j+m]![2(i+q)-3]!!\bigg),
\end{aligned}
\end{equation}
for $m_\alpha=m_\beta=m$ and zero otherwise. In this paper, we only used terms with $\alpha=\beta$.

\section{Details on small-scale field covariance}
\label{app:details_covariance}

For $\mathbf{C}^\mathrm{\vphantom{T}}_{\tilde{b}}$, we assume at all epochs that the expansion of the unresolved small-scale core field $\mathbf{\tilde{b}}_p$ has zero mean and variance $\sigma^2_{\tilde{b}}$, and are correlated in time according to a correlation function $\rho$ that only depends on the SH degree $n$. We estimate the variance by extending the spectrum of the CHAOS-6-x7 model at the surface up to degree $N_{\tilde{b}} = 30$ in the year 2015
\begin{equation}
\label{eq:power_spectrum}
W_b(n)=(n+1)\sum_{m=1}^n\big(g_n^m(t)^2+h_n^m(t)^2\big)_{t=2015}
\end{equation}
with an exponential fit to the spectrum for degrees \mbox{$n\in[2,13]$}. The variance of the SH expansion, which we assume to be independent of order $m$ and suitable for all epochs, is then
\begin{equation}
\sigma^2_{\tilde{b}}(n)=\frac{1.10\cdot 10^9}{(n+1)(2n+1)}\exp(-1.26n),
\end{equation}
with \mbox{$N_b<n\leq N_{\tilde{b}}$}. The correlation time in years can also be derived from the CHAOS-6 model with \cite[]{Gillet2013}
\begin{equation}
\tau_\mathrm{c}(n)=\sqrt{3\frac{W_b(n)}{W_{\dot{b}}(n)}},
\end{equation}
where $W_{\dot{b}}$ is the power spectrum of the SV and given by the expression in Eq.~\eqref{eq:power_spectrum} by replacing the SH coefficients with the respective time derivatives $\dot{g}_n^m$ and $\dot{h}_n^m$. An extension towards the degrees $N_b<n\leq N_{\tilde{b}}$ is achieved by fitting an exponential model for degrees $n\in[2,10]$\label{lastpage}
\begin{equation}
\tau_\mathrm{c}(n) = 5.00\cdot 10^2\exp(-0.23n).
\end{equation}







\end{document}